\documentclass{article}
\usepackage{arxiv}
\usepackage{amssymb}
\usepackage{amsmath}
\usepackage{bm}
\usepackage{caption}
\usepackage{subcaption}
\usepackage[utf8]{inputenc} 
\usepackage[T1]{fontenc}    
\usepackage{hyperref}       
\usepackage{url}            
\usepackage{booktabs}       
\usepackage{amsfonts}       
\usepackage{nicefrac}       
\usepackage{microtype}      
\usepackage{lipsum}		
\usepackage{graphicx}
\usepackage{doi}

\title{Thermodynamics of Encoding and Encoders}
\author{ 
Yang Tian \\
	Department of Psychology \& Tsinghua Laboratory of Brain and Intelligence\\ Tsinghua University\\ Beijing, 100084, China.\\
	\texttt{tiany20@mails.tsinghua.edu.cn} \\
	\And
	Pei Sun \\
	Department of Psychology \& Tsinghua Laboratory of Brain and Intelligence\\ Tsinghua University\\ Beijing, 100084, China.\\
	\texttt{peisun@tsinghua.edu.cn}
}

\renewcommand{\shorttitle}{Thermodynamics of Encoding and Encoders}

\begin{document}
\maketitle

\begin{abstract}
Non-isolated systems have diverse coupling relations with the external environment. These relations generate complex thermodynamics and information transmission between the system and its environment. The framework depicted in the current research attempts to glance at the critical role of the internal orders inside the non-isolated system in shaping the information thermodynamics coupling. We characterize the coupling as a generalized encoding process, where the system acts as an information thermodynamics encoder to encode the external information based on thermodynamics. We formalize the encoding process in the context of the nonequilibrium second law of thermodynamics, revealing an intrinsic difference in information thermodynamics characteristics between information thermodynamics encoders with and without internal correlations. During the information encoding process of an external source $\mathsf{Y}$, specific sub-systems in an encoder $\mathsf{X}$ with internal correlations can exceed the information thermodynamics bound on $\left(\mathsf{X},\mathsf{Y}\right)$ and encode more information than system $\mathsf{X}$ works as a whole. We computationally verify this theoretical finding in an Ising model with a random external field and a neural data set of the human brain during visual perception and recognition. Our analysis demonstrates that the stronger internal correlation inside these systems implies a higher possibility for specific sub-systems to encode more information than the global one. These findings may suggest a new perspective in studying information thermodynamics in diverse physical and biological systems.
\end{abstract}

\begin{quotation}
\textbf{Perhaps the connection between thermodynamics and information is one of the most intriguing relations discovered in physics. It is impressive to see the ubiquitous and critical roles of information thermodynamics links in diverse physical, chemical, and biological systems. Nowadays, emerging interests have arisen in studying information thermodynamics in complex systems (e.g., the human brain), whose complexity originates from heterogeneous elements and intricate internal correlations. An open challenge relates to the elusive roles of internal orders inside a complex system in influencing information thermodynamics. This research attempts to lay a cornerstone of this rising direction. We present a theory of the thermodynamics of encoding, where a non-isolated system $\mathsf{X}$ acts as an information thermodynamics encoder to encode the information of an external source $\mathsf{Y}$ through thermodynamics coupling. This theory reveals an intrinsic difference between the non-isolated system with intra-system coupling and the non-isolated system of independent elements in information thermodynamics characteristics. It demonstrates that a stronger internal correlation inside system $\mathsf{X}$ implies a higher possibility for specific sub-systems of $\mathsf{X}$ to encode more information of $\mathsf{Y}$ than $\mathsf{X}$ itself. We further verify the ubiquity of these findings in representative non-isolated systems in physics and biology, such as the Ising model with a random external field and the human brain, an ultra-complex system of neurons, during perception. These validated theoretical results may deepen our understanding of physical and biological information processing systems and inspire advanced computing architecture, storage infrastructure, and artificial neural network designs.}
\end{quotation}

\section{Introduction}
As you are reading this sentence, synergistic neural dynamics emerges in your brain (a non-isolated system of neurons) to encode the text information \cite{churchland2010stimulus,dayan2001theoretical}, changing the thermodynamic state of your brain continuously \cite{collell2015brain}. This is an instance of the coupling between a non-isolated system and its external environment. Similar instances is common in classic \cite{schwarz1995thermodynamics,kuhl2003mass,talkner2020colloquium}, quantum \cite{talkner2020colloquium,campisi2009thermodynamics,brandner2016periodic}, biological \cite{katchalsky1962thermodynamics,qian2005nonequilibrium,haynie2001biological}, ecological \cite{jorgensen2004towards,svirezhev2000thermodynamics,ruth2013integrating}, and social \cite{ruth2013integrating,dopfer2007general,sawyer2005social} systems. The pursuit to understand this coupling plays a pivotal role in statistical physics. Distinguished from the macroscopic systems studied by classic thermodynamics \cite{callen1998thermodynamics}, the microscopic and mesoscopic systems in stochastic thermodynamics \cite{seifert2012stochastic,sekimoto2010stochastic} usually feature non-negligible system-environment coupling \cite{jarzynski2017stochastic}. Over decades, the study of the thermodynamic attributes of these systems has become a rapidly emerging direction \cite{jarzynski2004nonequilibrium,gelin2009thermodynamics,esposito2010entropy,pucci2013entropy,seifert2016first,philbin2016thermal,talkner2016open}.

Apart from the above progress, a frequently neglected perspective is that the coupling can also be a process for the system to encode the external environment information. At first glance, this information-theoretical aspect seems to focus on quantifying the information processed by a population of elements about external sources (e.g., with multiple mutual information \cite{mcgill1954multivariate}, total correlation \cite{watanabe1960information}, or connected information \cite{schneidman2003network}), irrelevant to the thermodynamics analysis. However, as suggested by Szilard's \cite{szilard1929entropieverminderung} and Landauer's \cite{landauer1991information,landauer1996physical} classical works, information is physical. Historically, the information-thermodynamics connection is first discovered in the paradox of the Maxwell demon \cite{leff2014maxwell}. Recently, more fundamental connections have been identified in reset \cite{leff2014maxwell,sagawa2009minimal,sagawa2014thermodynamic}, measurement \cite{horowitz2013imitating,sagawa2009minimal,granger2011thermodynamic,sagawa2013role,hasegawa2010generalization,esposito2011second}, and feedback \cite{horowitz2013imitating,sagawa2013role,sagawa2009minimal,abreu2011extracting} processes and have been verified experimentally \cite{berut2012experimental,berut2013detailed,jun2014high,toyabe2010experimental}. The idea to consider thermodynamics from the information aspect (or vice versa) has been proven effective in identifying potential links between physics and information quantities \cite{kawai2007dissipation,sagawa2010generalized,still2012thermodynamics,ito2013information,ito2018stochastic,parrondo2015thermodynamics,hasegawa2010generalization,esposito2011second}. Meanwhile, these intriguing connections have inspired widespread explorations on the possibility for the information-carrying capacity of a memory device to function as a thermodynamic fuel \cite{boyd2017correlation,boyd2016maxwell,strasberg2013thermodynamics,barato2013autonomous,hoppenau2014energetics,um2015total,merhav2015sequence}, leading to the remarkable development of designing nanoscale autonomous machines that are fueled by information (referred to as information engines) \cite{diana2013finite,mandal2013maxwell,mandal2012work,lu2014engineering,chapman2015autonomous}. These recent advances in physics have further stimulated growing attention from other fields such as biology \cite{cao2015thermodynamics,collell2015brain,andrieux2008nonequilibrium,jarzynski2008thermodynamics,barato2014efficiency,sartori2014thermodynamic} and computer science \cite{hylton2021vision}. 

Till now, it is unclear how internal orders inside a non-isolated system affect information thermodynamics. The necessity to study this problem emerges especially for the complex systems (e.g., the brain), whose complexity originates from heterogeneous elements and intricate internal correlations. Our primary objective is to seek a systematic framework that formalizes the information thermodynamics coupling between a non-isolated system $\mathsf{X}$ and an external source $\mathsf{Y}$ and helps understand the critical roles of the existing order inside the system $\mathsf{X}$ on this coupling relation. Another objective is to explore potential diversities of information thermodynamics characteristics determined by intra-system factors and verify these diversities in concrete non-isolated systems.  

The rest of our paper is organized as follows. Sec. \ref{Sec2} depicts the theoretical framework of our research. We suggest an original perspective to characterize the non-isolated system $\mathsf{X}$ as an information thermodynamics encoder, which encodes the information of a coupled external source $\mathsf{Y}$ through thermodynamics. After contextualizing this perspective with existing theories, we formalize the definition of encoding in the context of the nonequilibrium second law of thermodynamics. Based on this formalization, Sec. \ref{Sec3} goes deep into the information thermodynamics of different non-isolated systems. We unveil how the nature of order inside the non-isolated system determines the information thermodynamics characteristics. An intrinsic difference in information thermodynamics is revealed between the encoders with and without intra-system coupling---unlike those with independent elements, an encoder $\mathsf{X}$ with intra-system coupling allows the encoded information of $\mathsf{Y}$ in its sub-systems to exceed the information thermodynamics bound on the joint system $\left(\mathsf{X},\mathsf{Y}\right)$. In other words, every time a particular amount of irreversible work is extracted or dispensed from the joint system $\left(\mathsf{X},\mathsf{Y}\right)$, specific sub-systems of $\mathsf{X}$ may be able to encode more information of $\mathsf{Y}$ than system $\mathsf{X}$ itself. This possibility originates from the internal correlation inside encoder $\mathsf{X}$. In Sec. \ref{Sec4}, we have computationally verified our theory in an Ising model with random external field and a real data set of the human brain during the perception process. 

\section{Information thermodynamics encoder and encoding}\label{Sec2}
\subsection{Definition of information thermodynamics encoder}\label{Sec2a}
Let us consider a non-isolated system $\mathsf{X}=\left(\lbrace X_{i}\rbrace,\mathcal{R}\right)$. Here we represent it as a graph. Set $\lbrace X_{i}\rbrace$ includes $n$ elements (vertices), where each $X_{i}$ has $m$ states $\Omega\left(X_{i}\right)=\lbrace X_{i}^{j}\rbrace$ (here $\Omega\left(\cdot\right)$ defines the sample space). Set $\mathcal{R}\subseteq\lbrace X_{i}\rbrace\times\lbrace X_{i}\rbrace$ describes the intra-system coupling relations (edges). Note that the coupling is a kind of equivalence relation, making every connected component of the graph be a clique (see Fig. \ref{G1}). To function as an information processor, system $\mathsf{X}$ must have multiple distinguishable states for information storage \cite{parrondo2015thermodynamics}. Therefore, we always consider the cases with $\sum_{i}\vert \Omega\left(X_{i}\right)\vert>n$ to guarantee $\vert \Omega\left(\mathsf{X}\right)\vert>1$. 

We refer to $\mathsf{X}$ as an information thermodynamics encoder when it couples with an external source $\mathsf{Y}=\lbrace Y_{i}\rbrace$ (see Fig. \ref{G1} for an instance). Following the idea in \cite{granger2011thermodynamic,esposito2011second}, the joint system $\left(\mathsf{X},\mathsf{Y}\right)$ is in contact
with a heat bath $\mathsf{HB}$ and the total system $\left[\left(\mathsf{X},\mathsf{Y}\right),\mathsf{HB}\right]$ is assumed as isolated. 

On the one hand, the coupling between $\mathsf{X}$ and $\mathsf{Y}$ allows the transformation of heat, energy or substances. On the other hand, this coupling ensures the existence of $\mathcal{P}\left(\mathsf{X}\mid\mathsf{Y}\right)$ (here $\mathcal{P}\left(\cdot\right)$ denotes the probability), establishing an information channel between $\mathsf{X}$ and $\mathsf{Y}$. Such a channel supports the information-theoretical analysis of the coupling relation in the context of that $\mathsf{X}$ encodes the information of $\mathsf{Y}$. Taken together, the dual attributes of this coupling allow us implement a unified analysis of information and thermodynamics. 

\subsection{The second law of thermodynamics}\label{Sec2b}
To embed our analysis with solid physics backgrounds, we begin with a discussion on the second law of thermodynamics. 

 \begin{figure}[b!]
\centering
\includegraphics[width=0.7\columnwidth]{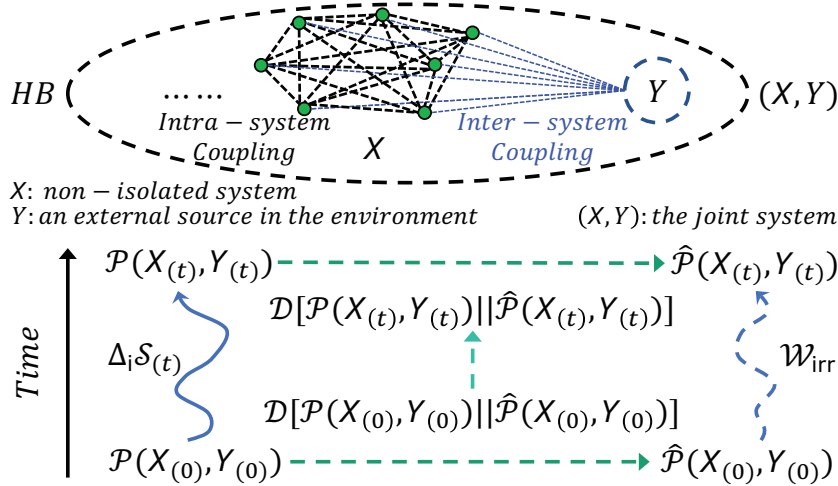}
\caption{\label{G1} A schematic diagram of the coupled system $\left(\mathsf{X},\mathsf{Y}\right)$ and the illustration of equality (\ref{EQ1}).}
 \end{figure}

Let us consider a general case where $\mathsf{X}$ will change if it interacts with $\mathsf{Y}$ during an interval $\left[0,t\right]$. The joint system $\left(\mathsf{X},\mathsf{Y}\right)$ varies due to this interaction and the heat bath $\mathsf{HB}$ accounts for supplying necessary heat. In \cite{sagawa2013role,sagawa2012fluctuation}, the time evolution of $\mathsf{X}$ under the influence of $\mathsf{Y}$ has been previously used for deriving the nonequilibrium second law of thermodynamics from the information perspective. 

Here we offer a more fundamental derivation from the thermodynamics perspective. System $\left(\mathsf{X},\mathsf{Y}\right)$ varies during $\left[0,t\right]$, accompanied by a nonequilibrium entropy change $\Delta\mathcal{S}$ and a nonequilibrium free energy change $\Delta\mathcal{F}$. By decomposing $\Delta\mathcal{S}$ as $\Delta\mathcal{S}=\Delta_{e}\mathcal{S}+\Delta_{i}\mathcal{S}$, one can distinguish between the reversible contribution $\Delta_{e}\mathcal{S}=QT^{-1}$ due to the heat flow (heat $Q$ comes from the heat bath $\mathsf{HB}$ and $T$ is the temperature) and the irreversible contribution $\Delta_{i}\mathcal{S}\geq 0$. As demonstrated in \cite{esposito2011second}, an equivalent formulation of the nonequilibrium second law can be obtained by $T\Delta_{i}\mathcal{S}=\mathcal{W}-\Delta\mathcal{F}\geq 0$, where $\mathcal{W}$ is the work performed on system $\left(\mathsf{X},\mathsf{Y}\right)$. 

Based on the knowledge of free energy \cite{donald1987free,qian2001relative}, the actual free energy $\mathcal{F}_{\left(\tau\right)}$ at moment $\tau\in\left[0,t\right]$ is associated with the equilibrium free energy $\widehat{\mathcal{F}}_{\left(\tau\right)}$ as well as the relative entropy between the actual state $\mathcal{P}\left(\mathsf{X}_{\left(\tau\right)},\mathsf{Y}_{\left(\tau\right)}\right)$ and the equilibrium state $\widehat{\mathcal{P}}\left(\mathsf{X}_{\left(\tau\right)},\mathsf{Y}_{\left(\tau\right)}\right)$ (here notion $\mathsf{X}_{\left(\tau\right)}$ represents the state of $\mathsf{X}$ at moment $\tau$). More specifically, one can know $\mathcal{F}_{\left(\tau\right)}=\widehat{\mathcal{F}}_{\left(\tau\right)}+T\mathcal{D}\big[\mathcal{P}\left(\mathsf{X}_{\left(\tau\right)},\mathsf{Y}_{\left(\tau\right)}\right)\big\Vert\widehat{\mathcal{P}}\left(\mathsf{X}_{\left(\tau\right)},\mathsf{Y}_{\left(\tau\right)}\right)\big]$, where $\mathcal{D}\left[\cdot\right]$ denotes the relative entropy \cite{cover1999elements}. 

Therefore, the nonequilibrium second law during $\left[0,t\right]$ can be specified as $T\Delta_{i}\mathcal{S}_{\left(t\right)}=\mathcal{W}-\left(\mathcal{F}_{\left(t\right)}-\mathcal{F}_{\left(0\right)}\right)=\mathcal{W}-\left(\widehat{\mathcal{F}}_{\left(t\right)}-\widehat{\mathcal{F}}_{\left(0\right)}\right)-T\mathcal{D}\big[\mathcal{P}\left(\mathsf{X}_{\left(t\right)},\mathsf{Y}_{\left(t\right)}\right)\big\Vert\widehat{\mathcal{P}}\left(\mathsf{X}_{\left(t\right)},\mathsf{Y}_{\left(t\right)}\right)\big]+T\mathcal{D}\big[\mathcal{P}\left(\mathsf{X}_{\left(0\right)},\mathsf{Y}_{\left(0\right)}\right)\big\Vert\widehat{\mathcal{P}}\left(\mathsf{X}_{\left(0\right)},\mathsf{Y}_{\left(0\right)}\right)\big]$, where the term $\mathcal{W}-\left(\widehat{\mathcal{F}}_{\left(t\right)}-\widehat{\mathcal{F}}_{\left(0\right)}\right)=\mathcal{W}-\Delta\widehat{\mathcal{F}}=\mathcal{W}_{\mathsf{irr}}$ is referred to as the irreversible work \cite{esposito2011second}. Taking these derivations together, the nonequilibrium second law can be reformulated as following \cite{esposito2011second} (please see Fig. \ref{G1} for an illustration)
\begin{align}
 \frac{\mathcal{W}_{\mathsf{irr}}}{T}=\mathcal{D}\big[\mathcal{P}\left(\mathsf{X}_{\left(t\right)},\mathsf{Y}_{\left(t\right)}\right)\big\Vert\widehat{\mathcal{P}}\left(\mathsf{X}_{\left(t\right)},\mathsf{Y}_{\left(t\right)}\right)\big]-\mathcal{D}\big[\mathcal{P}\left(\mathsf{X}_{\left(0\right)},\mathsf{Y}_{\left(0\right)}\right)\big\Vert\widehat{\mathcal{P}}\left(\mathsf{X}_{\left(0\right)},\mathsf{Y}_{\left(0\right)}\right)\big]+\Delta_{i}\mathcal{S}_{\left(t\right)}. \label{EQ1}
\end{align}
Here the term $\mathcal{D}\big[\mathcal{P}\left(\mathsf{X}_{\left(t\right)},\mathsf{Y}_{\left(t\right)}\right)\big\Vert\widehat{\mathcal{P}}\left(\mathsf{X}_{\left(t\right)},\mathsf{Y}_{\left(t\right)}\right)\big]-\mathcal{D}\big[\mathcal{P}\left(\mathsf{X}_{\left(0\right)},\mathsf{Y}_{\left(0\right)}\right)\big\Vert\widehat{\mathcal{P}}\left(\mathsf{X}_{\left(0\right)},\mathsf{Y}_{\left(0\right)}\right)\big]$ can be treated as information gain if it is positive, otherwise it is information reduction \cite{esposito2011second}. Accordingly, the irreversible work $\mathcal{W}_{\mathsf{irr}}$ in (\ref{EQ1}) can be either negative when the information reduction is greater than the entropy production, or non-negative in other cases \cite{esposito2011second}.

\subsection{Thermodynamics and information}\label{Sec2c}
Given the second law of thermodynamics in (\ref{EQ1}), one might be curious about its connection to the above mentioned information-theoretical perspective. Let us consider the information amount of $\mathsf{Y}$ can be encoded in $\mathsf{X}$ \cite{cover1999elements}. This aspect is studied in the term of measurement in statistical physics \cite{parrondo2015thermodynamics,horowitz2013imitating,sagawa2009minimal,granger2011thermodynamic,sagawa2013role,hasegawa2010generalization,esposito2011second} and substantial progress has been made in its application in neuroscience \cite{collell2015brain,dayan2001theoretical,yarrow2012fisher}. 

 \begin{figure}[b!]
\centering
\includegraphics[width=0.7\columnwidth]{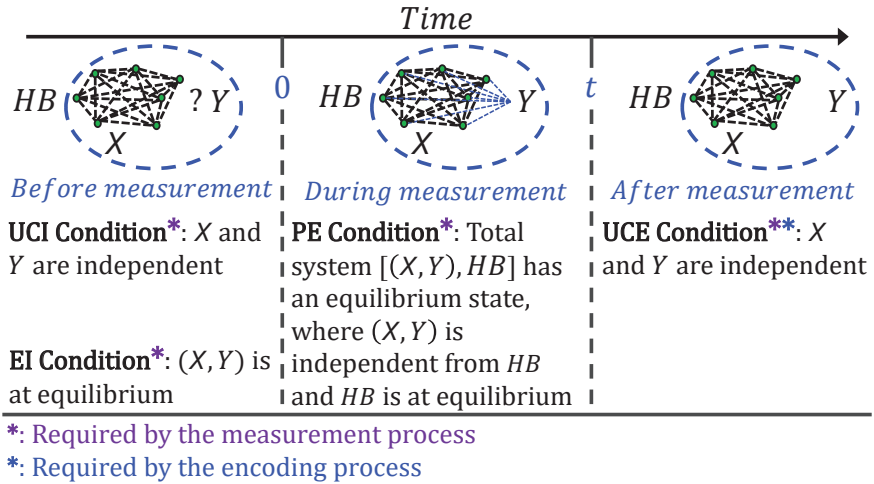}
\caption{\label{G2} The difference between measurement and encoding across time sections.}
 \end{figure}

To organize our derivation, we divide the timeline into 3 sections, respectively corresponding to the period before, during, and after the measurement process.
\begin{itemize}
    \item\;Before the measurement (before moment $0$), there is no restriction on $\mathsf{X}$ and $\mathsf{Y}$. They can be either independent or coupled. Specially, we refer to the case where $\mathsf{X}$ and $\mathsf{Y}$ are originally independent until the measurement begins as the \textbf{UCI Condition} (Uncoupled initialization condition), namely that $\widehat{\mathcal{P}}\left(\mathsf{X}_{\left(0\right)},\mathsf{Y}_{\left(0\right)}\right)= \widehat{\mathcal{P}}\left(\mathsf{X}_{\left(0\right)}\right)\widehat{\mathcal{P}}\left(\mathsf{Y}_{\left(0\right)}\right)$ in (\ref{EQ1}). Moreover, for the special case where the joint system $\left(\mathsf{X},\mathsf{Y}\right)$ is at equilibrium when the measurement begins, we refer to it as the \textbf{EI Condition} (Equilibrium initialization condition), namely that $\mathcal{P}\left(\mathsf{X}_{\left(0\right)},\mathsf{Y}_{\left(0\right)}\right)= \widehat{\mathcal{P}}\left(\mathsf{X}_{\left(0\right)},\mathsf{Y}_{\left(0\right)}\right)$;
    \item\;During the measurement (during $\left[0,t\right]$), it is necessary for $\mathsf{X}$ and $\mathsf{Y}$ to be coupled with each other. Specially, for the case where the total system $\left[\left(\mathsf{X}_{\left(\tau\right)},\mathsf{Y}_{\left(\tau\right)}\right),\mathsf{HB}_{\left(\tau\right)}\right]$ (here $\tau\in\left[0,t\right]$) always features a equilibrium state where system $\left(\mathsf{X},\mathsf{Y}\right)$ and $\mathsf{HB}$ are independent and $\mathsf{HB}$ is at equilibrium (there is no restriction on the state of system $\left(\mathsf{X},\mathsf{Y}\right)$), we refer to it as the \textbf{PE Condition} (Partial equilibrium condition), namely that $\widehat{\mathcal{P}}\left[\left(\mathsf{X}_{\left(\tau\right)},\mathsf{Y}_{\left(\tau\right)}\right),\mathsf{HB}_{\left(\tau\right)}\right]=\mathcal{P}\left(\mathsf{X}_{\left(\tau\right)},\mathsf{Y}_{\left(\tau\right)}\right)\widehat{\mathcal{P}}\left(\mathsf{HB}_{\left(\tau\right)}\right)$ holds for any $\tau\in\left[0,t\right]$;
    \item\;After the measurement (after moment $t$), there is no restriction on $\mathsf{X}$ and $\mathsf{Y}$ as well. For the case where the coupling between $\mathsf{X}$ and $\mathsf{Y}$ is turned off after the measurement ends, we refer to it as the \textbf{UCE Condition} (Uncoupled ending condition). Under this condition, we have $\widehat{\mathcal{P}}\left(\mathsf{X}_{\left(t\right)},\mathsf{Y}_{\left(t\right)}\right)= \widehat{\mathcal{P}}\left(\mathsf{X}_{\left(t\right)}\right)\widehat{\mathcal{P}}\left(\mathsf{Y}_{\left(t\right)}\right)$ in (\ref{EQ1}).
\end{itemize}
 
As suggested by \cite{esposito2011second,esposito2010entropy,jarzynski1999microscopic}, the \textbf{PE Condition} is necessary for defining the entropy production in (\ref{EQ1})
\begin{align}
   \forall\tau,\;\Delta_{i}\mathcal{S}_{\left(\tau\right)}=\mathcal{D}\big\{\mathcal{P}\left[\left(\mathsf{X}_{\left(\tau\right)},\mathsf{Y}_{\left(\tau\right)}\right),\mathsf{HB}_{\left(\tau\right)}\right]\big\Vert\mathcal{P}\left(\mathsf{X}_{\left(\tau\right)},\mathsf{Y}_{\left(\tau\right)}\right)\widehat{\mathcal{P}}\left(\mathsf{HB}_{\left(\tau\right)}\right)\big\}, \label{EQ2}
\end{align}
where the entropy production is defined as the irreversible contribution to the entropy change of system $\left(\mathsf{X},\mathsf{Y}\right)$ when it is coupled with $\mathsf{HB}$. For the \textbf{PE Condition} and (\ref{EQ2}), one can see their detailed derivation in \cite{esposito2010entropy} and a summary in \cite{esposito2011second}. Moreover, an equivalent vision of the \textbf{PE Condition} can be seen in \cite{granger2011thermodynamic}.

When the \textbf{UCE Condition} is satisfied in (\ref{EQ1}), we can know that $\mathcal{D}\big[\mathcal{P}\left(\mathsf{X}_{\left(t\right)},\mathsf{Y}_{\left(t\right)}\right)\big\Vert\widehat{\mathcal{P}}\left(\mathsf{X}_{\left(t\right)},\mathsf{Y}_{\left(t\right)}\right)\big]=\mathcal{D}\big[\mathcal{P}\left(\mathsf{X}_{\left(t\right)},\mathsf{Y}_{\left(t\right)}\right)\big\Vert\widehat{\mathcal{P}}\left(\mathsf{X}_{\left(t\right)}\right)\widehat{\mathcal{P}}\left(\mathsf{Y}_{\left(t\right)}\right)\big]$. In previous studies, the term $\mathcal{D}\big[\mathcal{P}\left(\mathsf{X}_{\left(t\right)},\mathsf{Y}_{\left(t\right)}\right)\big\Vert\widehat{\mathcal{P}}\left(\mathsf{X}_{\left(t\right)}\right)\widehat{\mathcal{P}}\left(\mathsf{Y}_{\left(t\right)}\right)\big]$ is treated as the mutual information $\mathcal{I}\left(\mathsf{X}_{\left(t\right)},\mathsf{Y}_{\left(t\right)}\right)$ between systems $\mathsf{X}$ and $\mathsf{Y}$ (e.g., see \cite{esposito2011second}). Considering the non-negative entropy production $\Delta_{i}\mathcal{S}_{\left(t\right)}\geq 0$, one can derive the following inequality
\begin{align}
    \frac{\mathcal{W}_{\mathsf{irr}}}{T}+\mathcal{D}\big[\mathcal{P}\left(\mathsf{X}_{\left(0\right)},\mathsf{Y}_{\left(0\right)}\right)\big\Vert\widehat{\mathcal{P}}\left(\mathsf{X}_{\left(0\right)},\mathsf{Y}_{\left(0\right)}\right)\big]\geq\mathcal{I}\big(\mathsf{X}_{\left(t\right)};\mathsf{Y}_{\left(t\right)}\big). \label{EQ3}
\end{align}
In the meanwhile, one can obtain a special case 
\begin{align}
    \frac{\mathcal{W}_{\mathsf{irr}}}{T}\geq\mathcal{I}\big(\mathsf{X}_{\left(t\right)},\mathsf{Y}_{\left(t\right)}\big) \label{EQ4}
\end{align}
if the \textbf{EI Condition} is satisfied as well. (\ref{EQ4}) is the usual form of the thermodynamics cost of the measurement process \cite{granger2011thermodynamic,hasegawa2010generalization,parrondo2015thermodynamics}. Although the \textbf{UCI Condition} is not necessarily met in the above derivation, the energy definition of $\left(\mathsf{X},\mathsf{Y}\right)$ can be calculated in a simple form under this condition \cite{esposito2011second}. In \cite{granger2011thermodynamic}, this condition is set for the measurement process as well. In a summary, the definition of measurement essentially requires the above four conditions to be met (see Fig. \ref{G2}).

\subsection{Thermodynamics of encoding}
Given the summary above, the connection between the second law in (\ref{EQ1}) and the information has been revealed from the aspect of measurement. Then, we show how to generalize the concept of measurement to encoding. 

For encoding, we only require the \textbf{UCE Condition} to be satisfied. Therefore, the measurement process studied by previous researches \cite{parrondo2015thermodynamics,horowitz2013imitating,sagawa2009minimal,granger2011thermodynamic,sagawa2013role,hasegawa2010generalization,esposito2011second} is a particular case of encoding. In Fig. \ref{G2}, we summarize the difference between the measurement process and encoding.

\begin{figure}[t!]
\centering
\includegraphics[width=0.7\columnwidth]{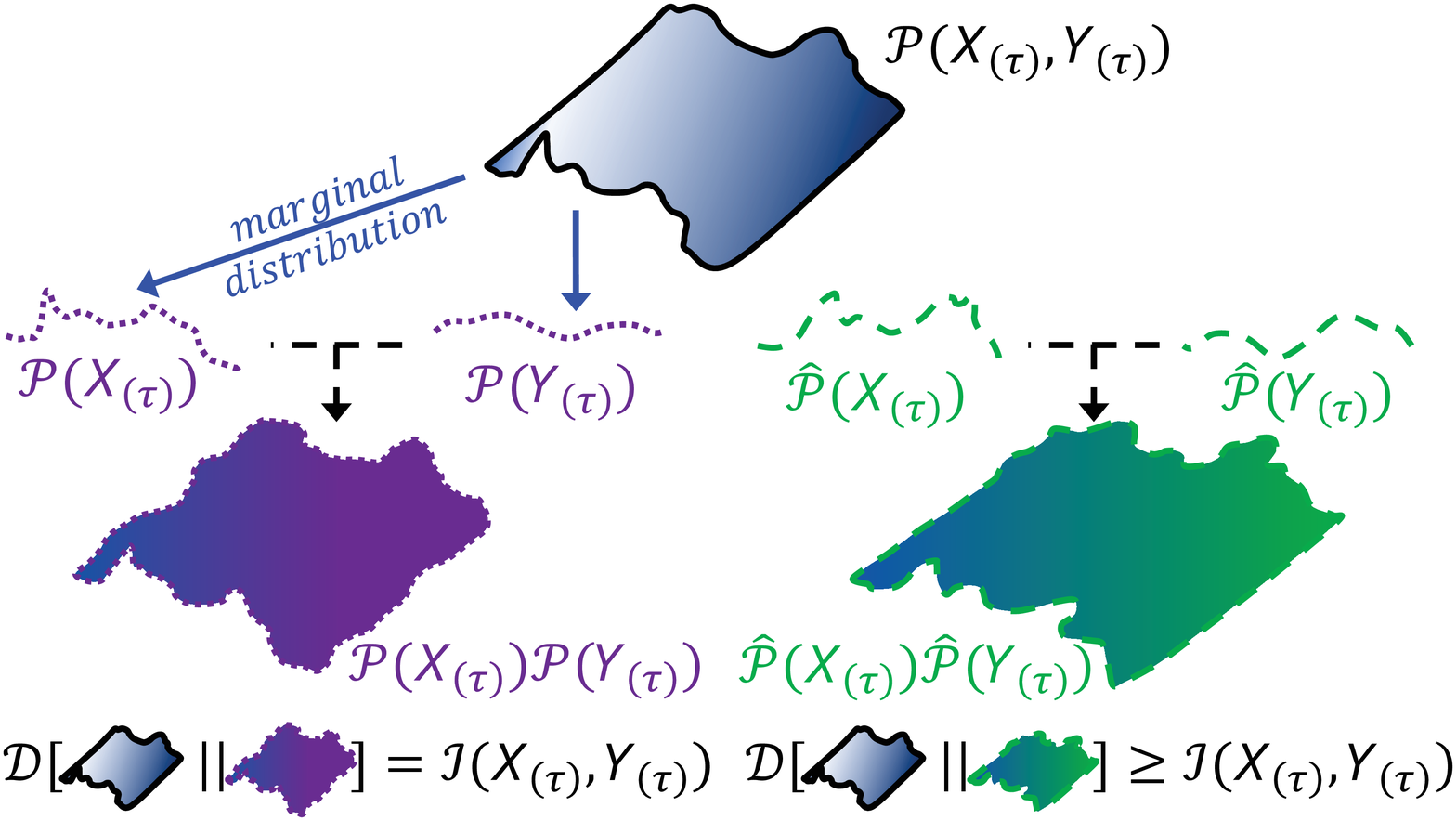}
\caption{\label{G3} The relation between $\mathcal{I}\left(\mathsf{X}_{\left(\tau\right)},\mathsf{Y}_{\left(\tau\right)}\right)$ and $\mathcal{D}\big[\mathcal{P}\left(\mathsf{X}_{\left(\tau\right)},\mathsf{Y}_{\left(\tau\right)}\right)\big\Vert\widehat{\mathcal{P}}\left(\mathsf{X}_{\left(\tau\right)}\right)\widehat{\mathcal{P}}\left(\mathsf{Y}_{\left(\tau\right)}\right)\big]$.}
 \end{figure}

To formalize the thermodynamics of encoding, we need to note that $\mathcal{D}\big[\mathcal{P}\left(\mathsf{X}_{\left(\tau\right)},\mathsf{Y}_{\left(\tau\right)}\right)\big\Vert\widehat{\mathcal{P}}\left(\mathsf{X}_{\left(\tau\right)}\right)\widehat{\mathcal{P}}\left(\mathsf{Y}_{\left(\tau\right)}\right)\big]$ (here $\tau\in\left[0,t\right]$) is relevant with rather than equivalent to the mutual information $\mathcal{I}\left(\mathsf{X}_{\left(\tau\right)},\mathsf{Y}_{\left(\tau\right)}\right)$ (please see Fig. \ref{G3}). Although this nominal equivalence has been used to derive (\ref{EQ3}-\ref{EQ4}) in previous studies \cite{esposito2011second,granger2011thermodynamic}, one can still verify that
\begin{align}
    \mathcal{D}\big[\mathcal{P}\left(\mathsf{X}_{\left(\tau\right)},\mathsf{Y}_{\left(\tau\right)}\right)\big\Vert\widehat{\mathcal{P}}\left(\mathsf{X}_{\left(\tau\right)}\right)\widehat{\mathcal{P}}\left(\mathsf{Y}_{\left(\tau\right)}\right)\big]&=\mathcal{I}\left(\mathsf{X}_{\left(\tau\right)},\mathsf{Y}_{\left(\tau\right)}\right)+\notag\\&\sum_{\Omega\left(\mathsf{X}\right)\times\Omega\left(\mathsf{Y}\right)}\mathcal{P}\left(\mathsf{X}_{\left(\tau\right)},\mathsf{Y}_{\left(\tau\right)}\right)\log\left(\frac{\mathcal{P}\left(\mathsf{X}_{\left(\tau\right)}\right)\mathcal{P}\left(\mathsf{Y}_{\left(\tau\right)}\right)}{\widehat{\mathcal{P}}\left(\mathsf{X}_{\left(\tau\right)}\right)\widehat{\mathcal{P}}\left(\mathsf{Y}_{\left(\tau\right)}\right)}\right),\label{EQ5}
\end{align}
where $\mathcal{D}\big[\mathcal{P}\left(\mathsf{X}_{\left(\tau\right)},\mathsf{Y}_{\left(\tau\right)}\right)\big\Vert\widehat{\mathcal{P}}\left(\mathsf{X}_{\left(\tau\right)}\right)\widehat{\mathcal{P}}\left(\mathsf{Y}_{\left(\tau\right)}\right)\big]$ consists of not only the mutual information $\mathcal{I}\left(\mathsf{X}_{\left(\tau\right)},\mathsf{Y}_{\left(\tau\right)}\right)$ but also a compound term. One can find the detailed derivation of (\ref{EQ5}) in (\ref{A1}-\ref{A3}) in appendix \ref{ASec1}. Luckily, the non-equivalence suggested in (\ref{EQ5}) does not contradict with (\ref{EQ3}) or (\ref{EQ4}), since
\begin{align}
    \sum_{\Omega\left(\mathsf{X}\right)\times\Omega\left(\mathsf{Y}\right)}\mathcal{P}\left(\mathsf{X}_{\left(\tau\right)},\mathsf{Y}_{\left(\tau\right)}\right)\log\left(\frac{\mathcal{P}\left(\mathsf{X}_{\left(\tau\right)}\right)\mathcal{P}\left(\mathsf{Y}_{\left(\tau\right)}\right)}{\widehat{\mathcal{P}}\left(\mathsf{X}_{\left(\tau\right)}\right)\widehat{\mathcal{P}}\left(\mathsf{Y}_{\left(\tau\right)}\right)}\right)\geq 0\label{EQ6}
\end{align}
can be proven (see (\ref{A3}-\ref{A15}) in appendix \ref{ASec1}). 

Given (\ref{EQ6}), we can formalize the thermodynamics of encoding in the form of (\ref{EQ3}). Under the \textbf{UCE Condition}, the irreversible work $\mathcal{W}_{\mathsf{irr}}$ that can be extracted or dispensed from system $\left(\mathsf{X},\mathsf{Y}\right)$ satisfies
\begin{align}
    &\frac{\mathcal{W}_{\mathsf{irr}}}{T}+\mathcal{D}\big[\mathcal{P}\left(\mathsf{X}_{\left(0\right)},\mathsf{Y}_{\left(0\right)}\right)\big\Vert\widehat{\mathcal{P}}\left(\mathsf{X}_{\left(0\right)},\mathsf{Y}_{\left(0\right)}\right)\big]-\Delta_{i}\mathcal{S}_{\left(t\right)}\notag\\
    =&\mathcal{D}\big[\mathcal{P}\left(\mathsf{X}_{\left(t\right)},\mathsf{Y}_{\left(t\right)}\right)\big\Vert\widehat{\mathcal{P}}\left(\mathsf{X}_{\left(t\right)}\right)\widehat{\mathcal{P}}\left(\mathsf{Y}_{\left(t\right)}\right)\big] \label{EQ7}\\
    \geq&\mathcal{I}\big(\mathsf{X}_{\left(t\right)};\mathsf{Y}_{\left(t\right)}\big), \label{EQ8}
\end{align}
where the equality is satisfied only when $\mathcal{P}\left(\mathsf{X}_{\left(t\right)}\right)\mathcal{P}\left(\mathsf{Y}_{\left(t\right)}\right)=\widehat{\mathcal{P}}\left(\mathsf{X}_{\left(t\right)}\right)\widehat{\mathcal{P}}\left(\mathsf{Y}_{\left(t\right)}\right)$. Moreover, we prove that the \textbf{UCE Condition} is necessary for deriving (\ref{EQ6}) in appendix \ref{ASec2}. Therefore, the thermodynamics of encoding described in (\ref{EQ7}-\ref{EQ8}) can not be independent of the \textbf{UCE Condition}.

Before we move on, let us rethink about the \textbf{UCE Condition}. A possible question might be about the physical ground underlying its definition: $\widehat{\mathcal{P}}\left(\mathsf{X}_{\left(t\right)},\mathsf{Y}_{\left(t\right)}\right)= \widehat{\mathcal{P}}\left(\mathsf{X}_{\left(t\right)}\right)\widehat{\mathcal{P}}\left(\mathsf{Y}_{\left(t\right)}\right)$. Here we emphasize that any thermodynamics variation costs time. Although the coupling between $\mathsf{X}$ and $\mathsf{Y}$ is closed at moment $t$, one can not expect that $\mathcal{P}\left(\mathsf{X}_{\left(t\right)},\mathsf{Y}_{\left(t\right)}\right)$ jump to the independent state $\mathcal{P}\left(\mathsf{X}_{\left(t\right)}\right)\mathcal{P}\left(\mathsf{Y}_{\left(t\right)}\right)$ without time delay. Therefore, the approaching of $\left(\mathsf{X}_{\left(t\right)},\mathsf{Y}_{\left(t\right)}\right)$ to independence can only be treated as an internal tendency at moment $t$.

\section{Information thermodynamics during encoding} \label{Sec3}
\subsection{Problem setup}
Different from the classic ideas, our research concentrates on a new aspect of encoding. We notice that $\mathsf{X}$ consists of multiple sub-systems, and every sub-system involves encoding as well. An exploration of the relative role of each sub-system (or individual element) in information encoding (system coupling) helps reveal the intrinsic order of the system \cite{schneidman2003network}. The significance of such an analysis has been discussed in information theory \cite{watanabe1960information}, statistical physics \cite{schneidman2003network}, and neuroscience \cite{palm1988significance,abeles1991corticonics,schnitzer2003multineuronal}. Combine this new aspect with the classic ideas, an important question is whether (\ref{EQ8}) can be generalized to these sub-systems. Specifically, we wonder if the encoded information of $\mathsf{Y}$ in an arbitrary sub-system of $\mathsf{X}$ is bounded by the irreversible work $\mathcal{W}_{\mathsf{irr}}$ from the joint system $\left(\mathsf{X},\mathsf{Y}\right)$ following (\ref{EQ8}) as well.

\begin{figure}[t!]
\centering
\includegraphics[width=0.7\columnwidth]{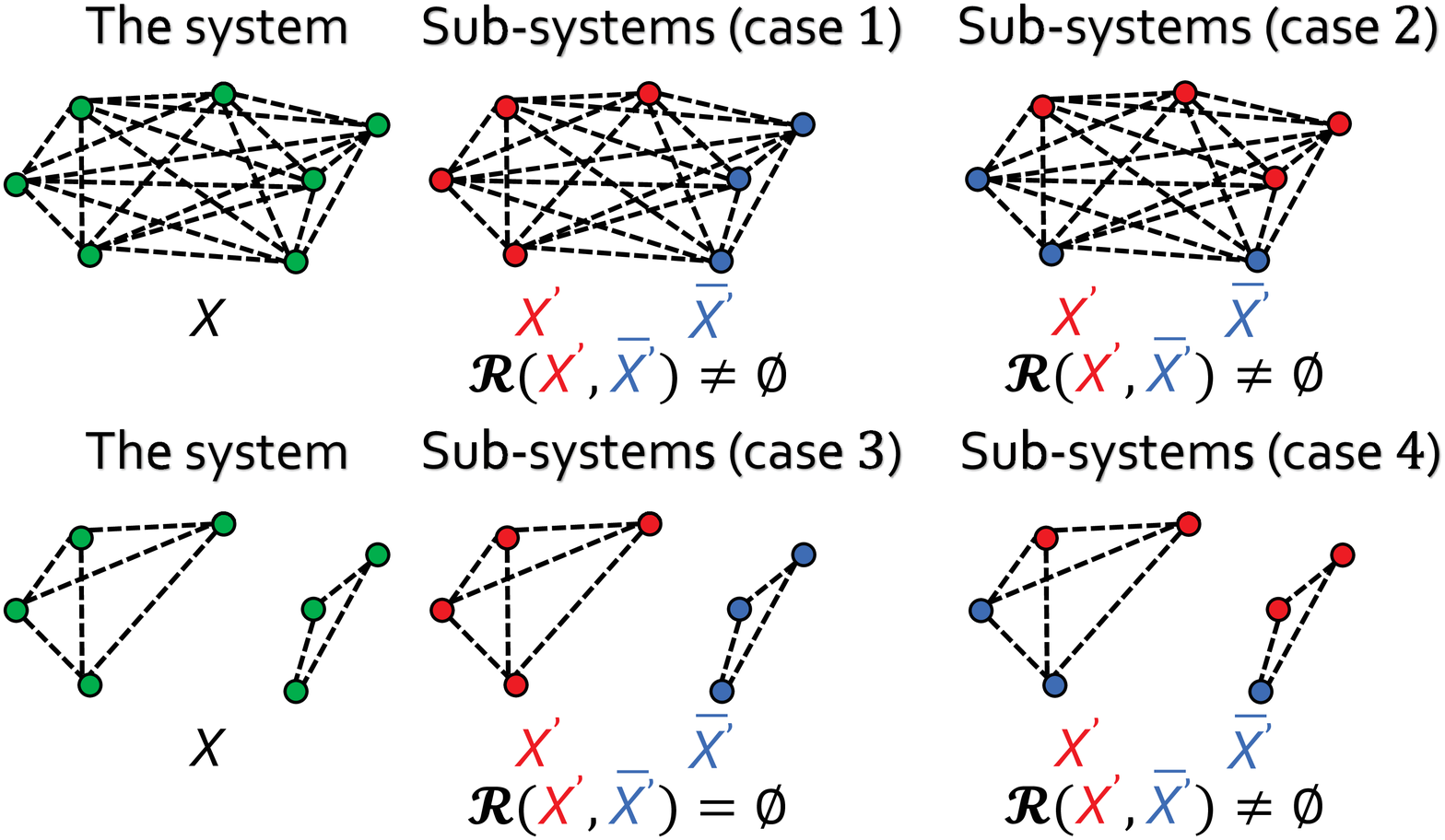}
\caption{\label{G4} Examples of the sub-system selection with different attributes of $\mathcal{R}\left(\mathsf{X}_{\left(t\right)}^{\prime},\overline{\mathsf{X}^{\prime}}_{\left(t\right)}\right)$.}
 \end{figure}

Consider any sub-system $\mathsf{X}^{\prime}=\left(\lbrace X_{O}\rbrace,\mathcal{R}^{\prime}\right)$ of $\mathsf{X}$ (here $O\subseteq\lbrace 1,\ldots,n\rbrace$ is the index set), we mark a corresponding sub-system $\overline{\mathsf{X}^{\prime}}=\left(\lbrace X_{i}\rbrace-\lbrace X_{O}\rbrace,\overline{\mathcal{R}^{\prime}}\right)$. We denote the relations between $\mathsf{X}^{\prime}$ and $\overline{\mathsf{X}^{\prime}}$ utilizing $\mathcal{R}\left(\mathsf{X}_{\left(t\right)}^{\prime},\overline{\mathsf{X}^{\prime}}_{\left(t\right)}\right)$. Then, we can rewrite the whole system as a tuple $\mathsf{X}=\left(\left(\mathsf{X}_{\left(t\right)}^{\prime},\overline{\mathsf{X}^{\prime}}_{\left(t\right)}\right),\mathcal{R}\left(\mathsf{X}_{\left(t\right)}^{\prime},\overline{\mathsf{X}^{\prime}}_{\left(t\right)}\right)\right)$. Depending on the nature of $\mathcal{R}\left(\mathsf{X}_{\left(t\right)}^{\prime},\overline{\mathsf{X}^{\prime}}_{\left(t\right)}\right)$, we can distinguish between two cases of system $\mathsf{X}$ where $\mathcal{R}\left(\mathsf{X}_{\left(t\right)}^{\prime},\overline{\mathsf{X}^{\prime}}_{\left(t\right)}\right)\equiv\emptyset$ (one can see the case 3 in Fig. \ref{G4}) or $\mathcal{R}\left(\mathsf{X}_{\left(t\right)}^{\prime},\overline{\mathsf{X}^{\prime}}_{\left(t\right)}\right)\neq\emptyset$ (cases 1, 2, and 4 in Fig. \ref{G4}). These definitions allow reformulating (\ref{EQ1}) under the \textbf{UCE Condition} as \begin{align}
    \frac{\mathcal{W}_{\mathsf{irr}}}{T}\geq\mathcal{I}\left[\left(\mathsf{X}_{\left(t\right)}^{\prime},\overline{\mathsf{X}^{\prime}}_{\left(t\right)}\right);\mathsf{Y}_{\left(t\right)}\right]-\mathcal{D}\big[\mathcal{P}\left(\mathsf{X}_{\left(0\right)},\mathsf{Y}_{\left(0\right)}\right)\big\Vert\widehat{\mathcal{P}}\left(\mathsf{X}_{\left(0\right)},\mathsf{Y}_{\left(0\right)}\right)\big]+\Delta_{i}\mathcal{S}_{\left(t\right)}. \label{EQ9}
\end{align}
Given (\ref{EQ9}), a natural thought for solving our question on the generalization of (\ref{EQ8}) is to compare $\mathcal{I}\left[\left(\mathsf{X}_{\left(t\right)}^{\prime},\overline{\mathsf{X}^{\prime}}_{\left(t\right)}\right);\mathsf{Y}_{\left(t\right)}\right]$ and $\mathcal{I}\left(\mathsf{X}_{\left(t\right)}^{\prime};\mathsf{Y}_{\left(t\right)}\right)$. Below, we demonstrate that the solution to this question can be developed based on fundamental information theory tools and further leads to an interesting view regarding non-isolated systems.

\subsection{Encoding without intra-system coupling}
\begin{figure}[t!]
\centering
\includegraphics[width=0.7\columnwidth]{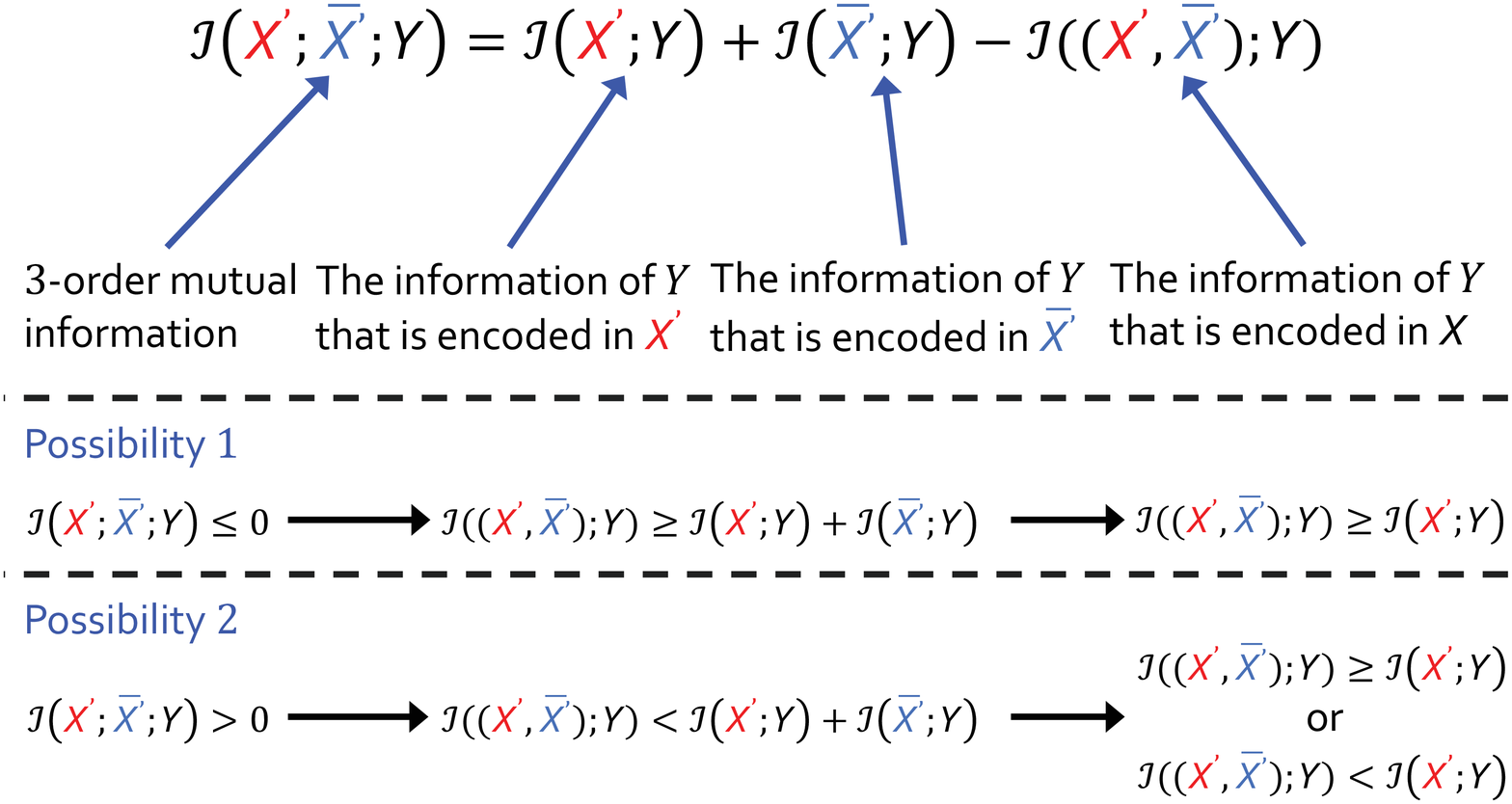}
\caption{\label{G5} A schematic diagram of the 3-order mutual information defined in (\ref{EQ10}) and our idea underlying inequality (\ref{EQ11}), suggesting the necessity to verify if the 3-order mutual information can be positive.}
 \end{figure}

\begin{figure}[t!]
\centering
\includegraphics[width=0.7\columnwidth]{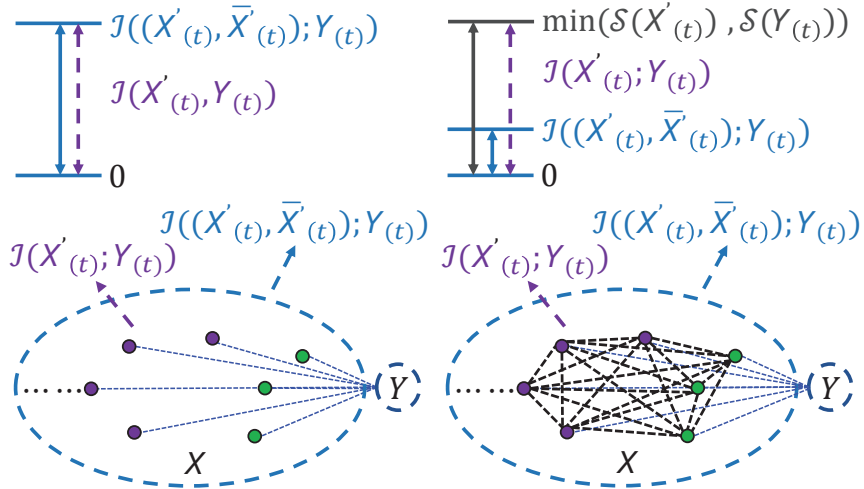}
\caption{\label{G6} The encoding processes with (right panel) or without (left panel) intra-system coupling. An arbitrary sub-system $\mathsf{X}^{\prime}$ is marked as purple and the corresponding sub-system $\overline{\mathsf{X}^{\prime}}$ is marked as green.}
 \end{figure}

We start the comparison in a simple case where $\mathcal{R}\left(\mathsf{X}_{\left(t\right)}^{\prime},\overline{\mathsf{X}^{\prime}}_{\left(t\right)}\right)\equiv\emptyset$, namely that $\mathsf{X}$ can always be divided into two uncorrelated sub-systems (see the left panel of Fig. \ref{G6}). In this case, we know that $\mathcal{I}\Big(\mathsf{X}_{\left(t\right)}^{\prime};\overline{\mathsf{X}^{\prime}}_{\left(t\right)}\Big)\equiv0$ is ensured by the independence between $\mathsf{X}^{\prime}_{\left(t\right)}$ and $\overline{\mathsf{X}^{\prime}}_{\left(t\right)}$. Moreover, we can find the item $\mathcal{I}\left(\left(\mathsf{X}_{\left(t\right)}^{\prime},\overline{\mathsf{X}^{\prime}}_{\left(t\right)}\right);\mathsf{Y}_{\left(t\right)}\right)$ in the $3$-order multiple mutual information \cite{mcgill1954multivariate,schneidman2003network}
\begin{align}
    \mathcal{I}\left(\mathsf{X}_{\left(t\right)}^{\prime};\overline{\mathsf{X}^{\prime}}_{\left(t\right)};\mathsf{Y}_{\left(t\right)}\right)=&\mathcal{I}\left(\mathsf{X}_{\left(t\right)}^{\prime};\mathsf{Y}_{\left(t\right)}\right)+\mathcal{I}\Big(\overline{\mathsf{X}^{\prime}}_{\left(t\right)};\mathsf{Y}_{\left(t\right)}\Big)-\notag\\&\mathcal{I}\left[\left(\mathsf{X}_{\left(t\right)}^{\prime},\overline{\mathsf{X}^{\prime}}_{\left(t\right)}\right);\mathsf{Y}_{\left(t\right)}\right]. \label{EQ10}
\end{align}
Such an equality helps connect between $\mathcal{I}\left(\mathsf{X}_{\left(t\right)}^{\prime};\mathsf{Y}_{\left(t\right)}\right)$ and $\mathcal{I}\left(\left(\mathsf{X}_{\left(t\right)}^{\prime},\overline{\mathsf{X}^{\prime}}_{\left(t\right)}\right);\mathsf{Y}_{\left(t\right)}\right)$ (see Fig. \ref{G5}). We need to remind that equality (\ref{EQ4}) (or an equivalent formulation of it proposed by \cite{ting1962amount}) can not be calculated directly with our general settings. An estimation for it is necessary. By applying Yeung's inequality \cite{yeung1991new} on (\ref{EQ10}), we can derive
\begin{align}
&\mathcal{I}\left(\mathsf{X}_{\left(t\right)}^{\prime};\mathsf{Y}_{\left(t\right)}\right)+\mathcal{I}\Big(\overline{\mathsf{X}^{\prime}}_{\left(t\right)};\mathsf{Y}_{\left(t\right)}\Big)-\mathcal{I}\left[\left(\mathsf{X}_{\left(t\right)}^{\prime},\overline{\mathsf{X}^{\prime}}_{\left(t\right)}\right);\mathsf{Y}_{\left(t\right)}\right]\notag\\
    \leq&\min\Big\lbrace\mathcal{I}\left(\mathsf{X}_{\left(t\right)}^{\prime};\mathsf{Y}_{\left(t\right)}\right),\mathcal{I}\Big(\overline{\mathsf{X}^{\prime}}_{\left(t\right)};\mathsf{Y}_{\left(t\right)}\Big),\mathcal{I}\Big(\mathsf{X}_{\left(t\right)}^{\prime};\overline{\mathsf{X}^{\prime}}_{\left(t\right)}\Big)\Big\rbrace\notag\\\leq&0. \label{EQ11}
\end{align}
Based on (\ref{EQ11}), we know that $\mathcal{I}\left[\left(\mathsf{X}_{\left(t\right)}^{\prime},\overline{\mathsf{X}^{\prime}}_{\left(t\right)}\right);\mathsf{Y}_{\left(t\right)}\right]\geq\max\big\lbrace\mathcal{I}\big(\mathsf{X}_{\left(t\right)}^{\prime};\mathsf{Y}_{\left(t\right)}\big),\mathcal{I}\big(\overline{\mathsf{X}^{\prime}}_{\left(t\right)};\mathsf{Y}_{\left(t\right)}\big)\big\rbrace$ (see Fig. \ref{G5} and the left panel of Fig. \ref{G6}). Therefore, one can immediately obtain
\begin{align}
    \frac{\mathcal{W}_{\mathsf{irr}}}{T}+\mathcal{D}\big[\mathcal{P}\left(\mathsf{X}_{\left(0\right)},\mathsf{Y}_{\left(0\right)}\right)\big\Vert\widehat{\mathcal{P}}\left(\mathsf{X}_{\left(0\right)},\mathsf{Y}_{\left(0\right)}\right)\big]\geq\max\big\lbrace\mathcal{I}\big(\mathsf{X}_{\left(t\right)}^{\prime};\mathsf{Y}_{\left(t\right)}\big),\mathcal{I}\big(\overline{\mathsf{X}^{\prime}}_{\left(t\right)};\mathsf{Y}_{\left(t\right)}\big)\big\rbrace, \label{EQ12}
\end{align}
proving that (\ref{EQ8}) holds for sub-systems in this case. 

Let us suspend the analysis and review the above derivation. For the encoding without intra-system coupling, the item $\mathcal{I}\left[\left(\mathsf{X}_{\left(t\right)}^{\prime},\overline{\mathsf{X}^{\prime}}_{\left(t\right)}\right);\mathsf{Y}_{\left(t\right)}\right]$ in (\ref{EQ3}) can be connected with $\mathcal{I}\left(\mathsf{X}_{\left(t\right)}^{\prime};\mathsf{Y}_{\left(t\right)}\right)$ by the three order multiple mutual information \cite{mcgill1954multivariate,schneidman2003network} in (\ref{EQ10}). Utilizing Yeung's inequality \cite{yeung1991new} in (\ref{EQ11}), we can implement the comparison in (\ref{EQ6}) and demonstrate that (\ref{EQ8}) can be generalized to sub-systems in this case. Considering the arbitrary selection of $\mathsf{X}^{\prime}$, the condition $\mathcal{R}\left(\mathsf{X}_{\left(t\right)}^{\prime},\overline{\mathsf{X}^{\prime}}_{\left(t\right)}\right)\equiv\emptyset$ means that each element in $\mathsf{X}$ is independent. An intuitive discovery can be identified in such a case, suggesting that the information that can be encoded by any sub-system is bounded by the encoded information base on the whole system. 

Furthermore, one can obtain more knowledge about the above discovery after applying the theory of information synergy and redundancy \cite{mcgill1954multivariate,schneidman2003network}. More specifically, there is synergy between $\mathsf{X}_{\left(t\right)}^{\prime}$ and $\overline{\mathsf{X}^{\prime}}_{\left(t\right)}$ if (\ref{EQ10}) is negative, namely that $\mathsf{X}_{\left(t\right)}^{\prime}$ and $\overline{\mathsf{X}^{\prime}}_{\left(t\right)}$ taken together (denoted by $\left(\mathsf{X}_{\left(t\right)}^{\prime},\overline{\mathsf{X}^{\prime}}_{\left(t\right)}\right)$ in (\ref{EQ10})) can encode more information of $\mathsf{Y}_{\left(t\right)}$ than the case when they are taken separately. If (\ref{EQ10}) is positive, then the pair
$\left(\mathsf{X}_{\left(t\right)}^{\prime},\overline{\mathsf{X}^{\prime}}_{\left(t\right)}\right)$ is redundant in the information encoding about $\mathsf{Y}_{\left(t\right)}$. Based on (\ref{EQ11}), one can find that there is no information redundancy in the system of independent elements. Moreover, one can easily prove that the equality in (\ref{EQ11}) can be satisfied if and only if $\mathcal{I}\left(\mathsf{X}_{\left(t\right)}^{\prime};\overline{\mathsf{X}^{\prime}}_{\left(t\right)}\mid\mathsf{Y}_{\left(t\right)}\right)=0$, because $\mathcal{I}\left(\mathsf{X}_{\left(t\right)}^{\prime};\overline{\mathsf{X}^{\prime}}_{\left(t\right)};\mathsf{Y}_{\left(t\right)}\right)=\mathcal{I}\Big(\mathsf{X}_{\left(t\right)}^{\prime};\overline{\mathsf{X}^{\prime}}_{\left(t\right)}\Big)-\mathcal{I}\left(\mathsf{X}_{\left(t\right)}^{\prime};\overline{\mathsf{X}^{\prime}}_{\left(t\right)}\mid\mathsf{Y}_{\left(t\right)}\right)$ \cite{cover1999elements} and $\mathcal{I}\Big(\mathsf{X}_{\left(t\right)}^{\prime};\overline{\mathsf{X}^{\prime}}_{\left(t\right)}\Big)=0$. Applying the definition of conditional mutual information \cite{cover1999elements}, it can be known that $\mathcal{I}\left(\mathsf{X}_{\left(t\right)}^{\prime};\overline{\mathsf{X}^{\prime}}_{\left(t\right)}\mid\mathsf{Y}_{\left(t\right)}\right)=0$ is equivalent to the conditionally independence between $\mathsf{X}_{\left(t\right)}^{\prime}$ and $\overline{\mathsf{X}^{\prime}}_{\left(t\right)}$ given $\mathsf{Y}_{\left(t\right)}$
\begin{equation}
    \mathcal{P}\left(\mathsf{X}_{\left(t\right)}^{\prime};\overline{\mathsf{X}^{\prime}}_{\left(t\right)}\mid\mathsf{Y}_{\left(t\right)}\right)=\mathcal{P}\left(\mathsf{X}_{\left(t\right)}^{\prime}\mid\mathsf{Y}_{\left(t\right)}\right)\mathcal{P}\left(\overline{\mathsf{X}^{\prime}}_{\left(t\right)}\mid\mathsf{Y}_{\left(t\right)}\right). \label{EQ13}
\end{equation}
There exists information synergy in system $\mathsf{X}$ unless (\ref{EQ13}) is satisfied. Considering that (\ref{EQ13}) is a strong condition, one can generally treat it as a rare case for a system of independent elements to have no information synergy (see the top panel of Fig. \ref{G7a}).

\subsection{Encoding with intra-system coupling}
We turn to the comparison in a more complex case where $\mathcal{R}\left(\mathsf{X}_{\left(t\right)}^{\prime},\overline{\mathsf{X}^{\prime}}_{\left(t\right)}\right)\neq\emptyset$, namely that $\mathsf{X}$ is a system with non-negligible intra-system coupling (the right panel of Fig. \ref{G6}). In this case, we can see $\mathcal{I}\Big(\mathsf{X}_{\left(t\right)}^{\prime};\overline{\mathsf{X}^{\prime}}_{\left(t\right)}\Big)\geq0$. Following the same idea that has been shown in (\ref{EQ10}-\ref{EQ12}), we know that $\mathcal{I}\left(\mathsf{X}_{\left(t\right)}^{\prime};\mathsf{Y}_{\left(t\right)}\right)+\mathcal{I}\Big(\overline{\mathsf{X}^{\prime}}_{\left(t\right)};\mathsf{Y}_{\left(t\right)}\Big)-\mathcal{I}\left[\left(\mathsf{X}_{\left(t\right)}^{\prime},\overline{\mathsf{X}^{\prime}}_{\left(t\right)}\right);\mathsf{Y}_{\left(t\right)}\right]
    \leq\min\Big\lbrace\mathcal{I}\left(\mathsf{X}_{\left(t\right)}^{\prime};\mathsf{Y}_{\left(t\right)}\right),\mathcal{I}\Big(\overline{\mathsf{X}^{\prime}}_{\left(t\right)};\mathsf{Y}_{\left(t\right)}\Big),\mathcal{I}\Big(\mathsf{X}_{\left(t\right)}^{\prime};\overline{\mathsf{X}^{\prime}}_{\left(t\right)}\Big)\Big\rbrace$ is ensured by Yeung's inequality \cite{yeung1991new}. Note that one can easily find that the right side of the inequality is either zero or positive in this case. Therefore, the upper bound in Yeung's inequality fails in determining the sign of item $\mathcal{I}\left(\mathsf{X}_{\left(t\right)}^{\prime};\mathsf{Y}_{\left(t\right)}\right)+\mathcal{I}\Big(\overline{\mathsf{X}^{\prime}}_{\left(t\right)};\mathsf{Y}_{\left(t\right)}\Big)-\mathcal{I}\left[\left(\mathsf{X}_{\left(t\right)}^{\prime},\overline{\mathsf{X}^{\prime}}_{\left(t\right)}\right);\mathsf{Y}_{\left(t\right)}\right]$. Combining this finding with (\ref{EQ9}), one can prove that the bound in (\ref{EQ8}) does not always hold for sub-systems $\mathsf{X}^{\prime}$ and $\overline{\mathsf{X}^{\prime}}$. Although the encoded information of $\mathsf{Y}$ in system $\mathsf{X}$ is bounded by the irreversible work $\mathcal{W}_{\mathsf{irr}}$ from the joint system $\left(\mathsf{X},\mathsf{Y}\right)$ following (\ref{EQ8}), an arbitrary sub-system $\mathsf{X}^{\prime}$ (or $\overline{\mathsf{X}^{\prime}}$) might disobey this bound by encoding more information than $\mathsf{X}$ (see the right panel of Fig. \ref{G6}). We should emphasize that this does not disobey the second law of thermodynamics. The encoded information in sub-system $\mathsf{X}^{\prime}$ is still bound by the irreversible work $\mathcal{W}_{\mathsf{irr}}^{\prime}$ from $\left(\mathsf{X}^{\prime},\mathsf{Y}\right)$ following $\frac{\mathcal{W}^{\prime}_{\mathsf{irr}}}{kT}+\mathcal{D}\big[\mathcal{P}\left(\mathsf{X}^{\prime}_{\left(0\right)},\mathsf{Y}_{\left(0\right)}\right)\big\Vert\widehat{\mathcal{P}}\left(\mathsf{X}^{\prime}_{\left(0\right)},\mathsf{Y}_{\left(0\right)}\right)\big]\geq\mathcal{I}\big(\mathsf{X}^{\prime}_{\left(t\right)};\mathsf{Y}_{\left(t\right)}\big)$. Moreover, one can measure the maximum extent that an arbitrary sub-system $\mathsf{X}^{\prime}$ can disobey the bound offered by the irreversible work $\mathcal{W}_{\mathsf{irr}}$. An elementary bound 
    \begin{align}
        \mathcal{I}\left(\mathsf{X}_{\left(t\right)}^{\prime};\mathsf{Y}_{\left(t\right)}\right)-\mathcal{I}\left[\left(\mathsf{X}_{\left(t\right)}^{\prime},\overline{\mathsf{X}^{\prime}}_{\left(t\right)}\right);\mathsf{Y}_{\left(t\right)}\right]\leq\min\lbrace\mathcal{S}\left(\mathsf{X}_{\left(t\right)}^{\prime}\right),\mathcal{S}\left(\mathsf{Y}_{\left(t\right)}\right)\rbrace\label{EQ14}
    \end{align}
    can be derived from Shannon entropy $\mathcal{S}$ (the right panel of Fig. \ref{G6}). Nevertheless, the non-trivial bound remains elusive and requires more explorations in future.

\begin{figure}[b!]
     \centering
     \begin{subfigure}[b]{0.48\columnwidth}
         \includegraphics[width=\columnwidth]{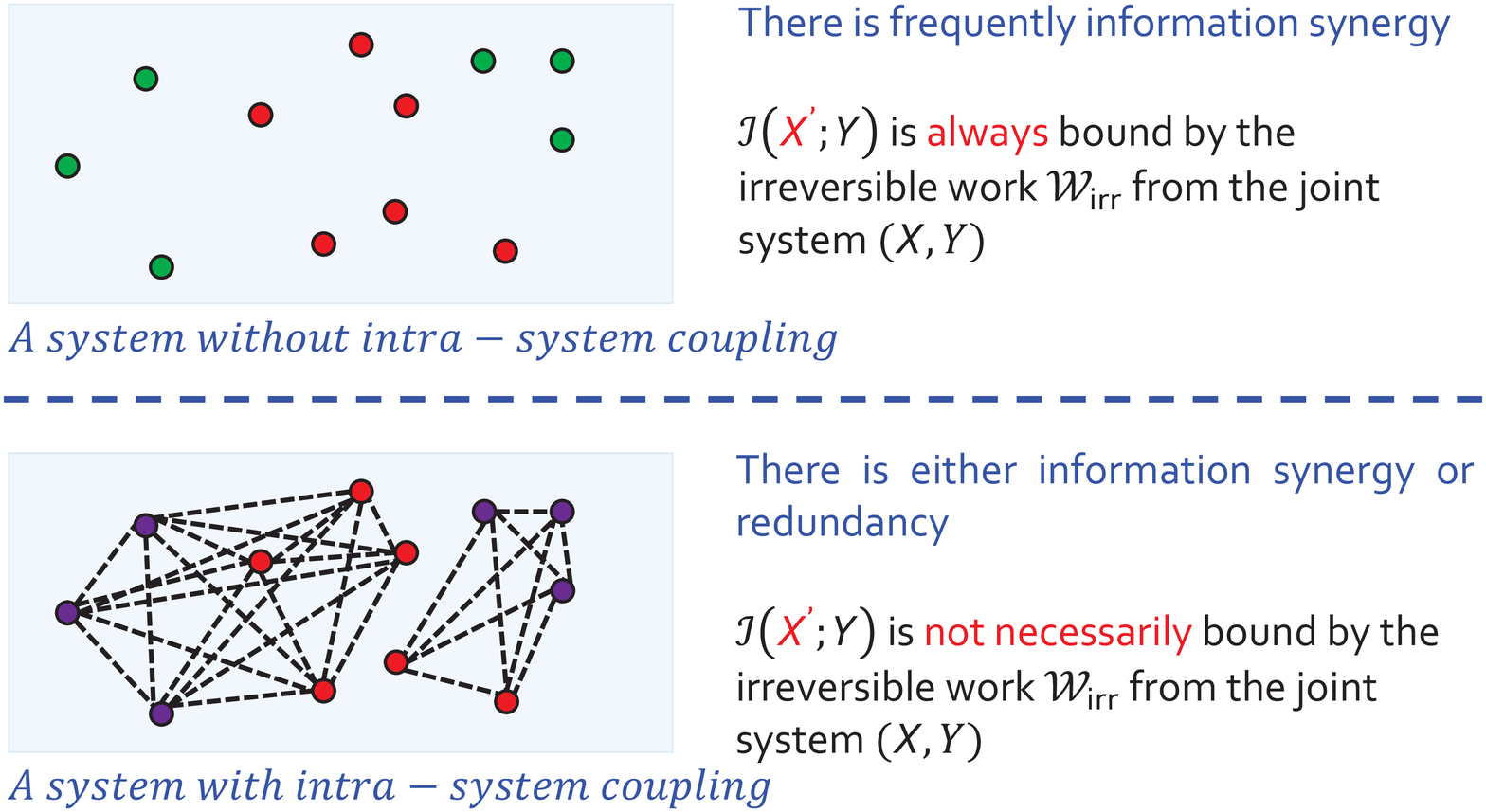}
        \caption{\label{G7a}}
     \end{subfigure}
     \hfill
     \begin{subfigure}[b]{0.48\columnwidth}
         \includegraphics[width=\columnwidth]{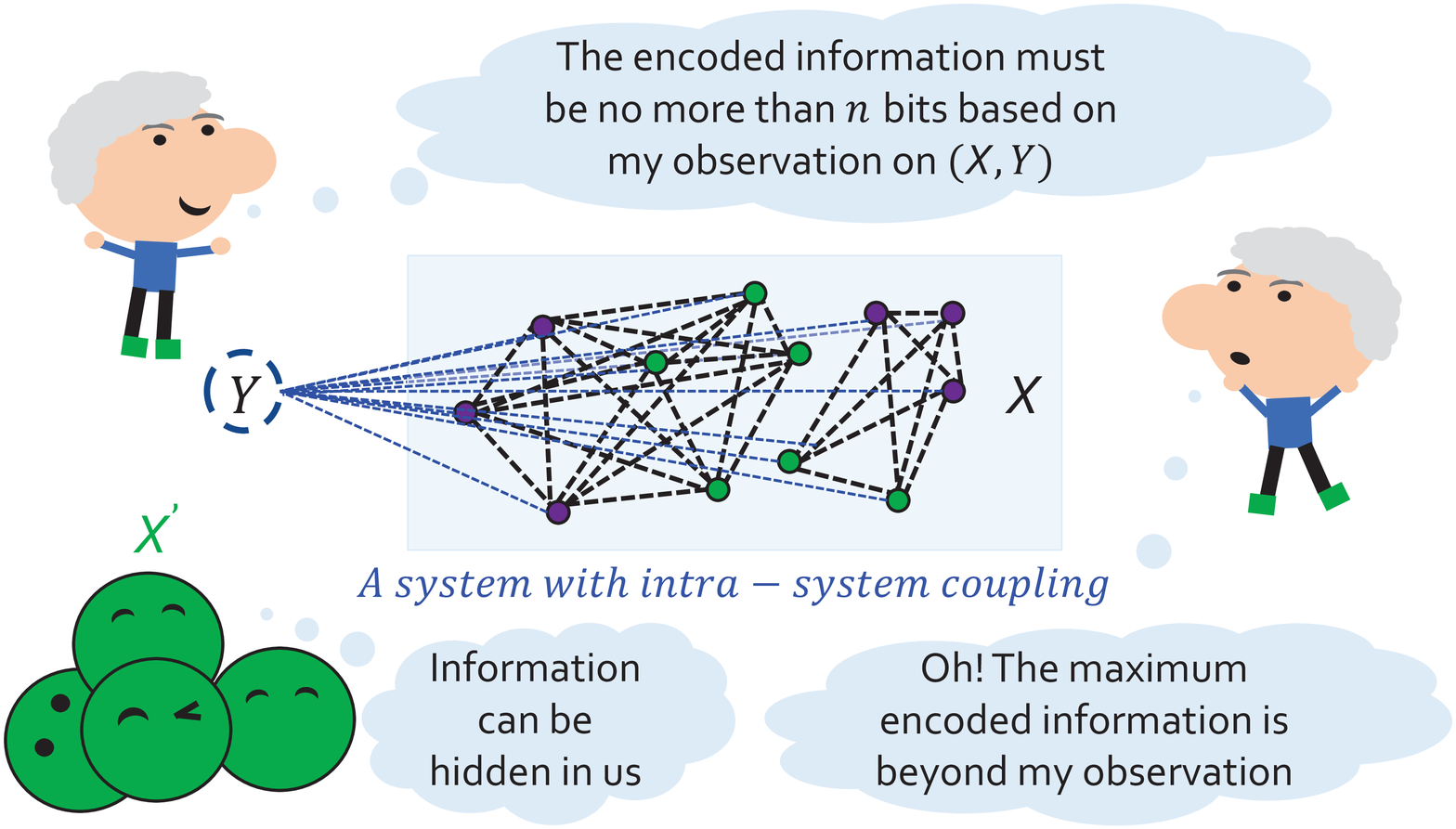}
     \caption{\label{G7b}}
     \end{subfigure}
     \hfill
        \caption{\label{G7} Summary of theoretical findings. (a) The differences between the encoding processes in systems with/without intra-system coupling. (b) The illustration of hiding information in non-isolated system without intra-system coupling.}
\end{figure}

   One can see that the encoding by a system with non-negligible intra-system coupling might involve both information synergy and redundancy. The synergy case is in accord with our common sense of the relation between a system and its sub-systems. The redundancy case (where the bound offered by the irreversible work $\mathcal{W}_{\mathsf{irr}}$ is disobeyed by sub-systems) corresponds to the situation where an arbitrary sub-system encodes more information than the whole system. Such a case results from the in-harmony between sub-systems in encoding (see the bottom panel of Fig. \ref{G7a}). 
    
    To this point, we have proposed our answer to the question on the generalization of (\ref{EQ8}). We demonstrate that the irreversible work $\mathcal{W}_{\mathsf{irr}}$ from $\left(\mathsf{X},\mathsf{Y}\right)$ can bound the encoded information of $\mathsf{Y}$ in any sub-system of $\mathsf{X}$ if there is no intra-system coupling. Otherwise, this bound might be disobeyed by the sub-systems of $\mathsf{X}$ (see Fig. \ref{G7}). We have discussed these findings from the aspect of information synergy and redundancy \cite{mcgill1954multivariate,schneidman2003network}. These discussions lead to an interesting view that a non-isolated system without intra-system coupling frequently involves information synergy, in comparison, the system with intra-system coupling can involve either information synergy or redundancy (see Fig. \ref{G7}).
    
    Rather than limit ourselves in treating the redundancy as specific information inefficiency, we suggest a new perspective that the redundancy means a possibility to hide information in a non-isolated system. By observing the irreversible work $\mathcal{W}_{\mathsf{irr}}$ from a joint system $\left(\mathsf{X},\mathsf{Y}\right)$, one might underestimate the information of $\mathsf{Y}$ encoded in $\mathsf{X}$ (underestimate the inter-system coupling strength) if $\mathsf{X}$ consists of correlated elements. In such a case, more information of $\mathsf{Y}$ can be hidden in the sub-systems of $\mathsf{X}$ (see an illustration in Fig. \ref{G7b}). 
    
    Below, we implement our analysis of this perspective by applying our theoretical framework to concrete non-isolated systems. These systems come from statistical physics and neuroscience, helping demonstrate the ubiquity of the discussed phenomenon above.
\section{Computational verification in non-isolated systems}\label{Sec4}
    \subsection{Ising model with random external field} \label{Sec4a}
    We begin our demonstration with the Ising model \cite{cipra1987introduction}, one of the significant themes in physics. Specifically, we consider a 1-dimensional Ising model with random external field, whose Hamiltonian is
    \begin{align}
        \mathbf{H}\left(\bm{\sigma},h\right)=-J\sum_{\langle i,j\rangle}\sigma_{i}\sigma_{j}-h\sum_{i}\sigma_{i}. \label{EQ15}
    \end{align}
    Here $\bm{\sigma}$ is a shorthand for a specification of the spin $\sigma_{i}$ on each site. The notion $J$ denotes the coupling strength between spins (the intra-system coupling). The external field $h\in \left(0,m\right]\cap\mathbb{Z}$ is defined as a discrete random variable (the inter-system coupling)
    \begin{align}
    \mathcal{P}\left(h\right)=\frac{1}{m}.\label{EQ16}  
    \end{align}
    To this point, a spin system $\bm{\sigma}$ and an external source $h$ have been characterized, functioning as the systems $\mathsf{X}$ and $\mathsf{Y}$ in our previous discussion, respectively. Here we define a random external field $h$ rather than a fixed one since it is more natural to consider a random variable $h$ in our subsequent encoding analysis. One can find other similar models in \cite{acharyya1998nonequilibrium,zhang2016critical}.
    
    Given a $h$, we know that system $\bm{\sigma}$ always has a unique equilibrium Boltzman distribution at a constant temperature $T$ \cite{ruelle1972use,lebowitz1972uniqueness}
    \begin{align}
       \widehat{\mathcal{P}}\left(\bm{\sigma}\mid h\right)=\frac{1}{Z\left(h\right)}\exp\left[-\frac{1}{kT}\mathbf{H}\left(\bm{\sigma},h\right)\right], \label{EQ17} 
    \end{align}
    where $k$ is the Boltzmann constant, and $Z\left(h\right)$ denotes the normalization term (partition function)
     \begin{align}
       Z\left(h\right)&=\sum_{\bm{\sigma}}\exp\left[-\frac{1}{kT}\mathbf{H}\left(\bm{\sigma},h\right)\right]\notag\\
       &=\exp\left(\frac{nJ}{kT}\right)\left[\cosh\frac{h}{kT}+\sqrt{\sinh^{2}\frac{h}{kT}+\exp(-4\frac{J}{kT})}\right]^{n}\notag\\&\;\;\;\;+\exp\left(\frac{nJ}{kT}\right)\left[\cosh\frac{h}{kT}-\sqrt{\sinh^{2}\frac{h}{kT}+\exp(-4\frac{J}{kT})}\right]^{n}, \label{EQ18} 
    \end{align}
    where $n$ is the number of spins. Assuming (\ref{EQ16}) is the equilibrium distribution $\widehat{\mathcal{P}}\left(h\right)$ of $h$, we can calculate the equilibrium distribution of joint system $\left(\bm{\sigma},h\right)$
    \begin{align}
       \widehat{\mathcal{P}}\left(\bm{\sigma},h\right)=\frac{\widehat{\mathcal{P}}\left(h\right)}{Z\left(h\right)}\exp\left[-\frac{1}{kT}\mathbf{H}\left(\bm{\sigma},h\right)\right].  \label{EQ19} 
    \end{align}
    
    Now, let us consider the thermodynamics of encoding in the joint system $\left(\bm{\sigma},h\right)$ during $\left[0,t\right]$ (e.g., try to calculate inequality (\ref{EQ1})). We treat system $\bm{\sigma}$ as an information thermodynamics encoder to encode the information of $h$. Given the above review, we can calculate the equilibrium distribution $\widehat{\mathcal{P}}\left(\bm{\sigma}_{\left(\tau\right)}, h_{\left(\tau\right)}\right)$ at any moment $\tau\in\left[0,t\right]$. When $\bm{\sigma}$ is coupled with an external field $h\in\left(0,m\right]$, the equilibrium distribution of $\left(\bm{\sigma},h\right)$ can be worked out following (\ref{EQ19}). When the \textbf{UCE Condition} is met at moment $t$, one can see that $\widehat{\mathcal{P}}\left(\bm{\sigma}_{\left(t\right)}\right)=\widehat{\mathcal{P}}\left(\bm{\sigma}_{\left(t\right)}\mid 0\right)$, where $\widehat{\mathcal{P}}\left(\bm{\sigma}_{\left(t\right)}\mid 0\right)$ measures the equilibrium distribution of $\bm{\sigma}$ independent from the external field $h$. Thus,
    \begin{align}
        \widehat{\mathcal{P}}\left(\bm{\sigma}_{\left(t\right)}, h_{\left(t\right)}\right)=\widehat{\mathcal{P}}\left(\bm{\sigma}_{\left(t\right)}\mid 0\right)\widehat{\mathcal{P}}\left(h_{\left(t\right)}\right). \label{EQ20}
    \end{align}
    As for the work $\mathcal{W}$ performed on system $\left(\bm{\sigma},h\right)$, it can be calculated under the framework of stochastic thermodynamics \cite{seifert2012stochastic}
         \begin{align}
            \dot{\mathcal{W}}=&\sum_{h}\sum_{\bm{\sigma}}\mathcal{P}\left(\bm{\sigma}\mid h\right)\mathcal{P}\left(h\right)\dot{\mathbf{H}}\left(\bm{\sigma},h\right).\label{EQ21} 
        \end{align}
   For convenience, we implement the experiment in a special case where $h_{\left(t\right)}=h_{\left(0\right)}$ and the \textbf{UCI Condition} is satisfied. One can know that $\widehat{\mathcal{P}}\left(\bm{\sigma}_{\left(0\right)}, h_{\left(0\right)}\right)=\widehat{\mathcal{P}}\left(\bm{\sigma}_{\left(t\right)}, h_{\left(t\right)}\right)$ and therefore the change in equilibrium free energy $\Delta\widehat{\mathcal{F}}=0$. Under this condition, one can derive $\mathcal{W}_{irr}=\mathcal{W}$. Taken together, the most part of (\ref{EQ1}) has been calculated to this point.
    
    However, one can immediately find it non-trivial to define the time-dependent distribution $\mathcal{P}\left(\bm{\sigma}_{\left(\tau\right)}, h_{\left(\tau\right)}\right)$ analytically. To overcome this challenge, we turn to the computational implementation of (\ref{EQ1}).
        
    Computationally, we generate $m\times k_{J}\times k_{T}\times k_{\theta}$ random sequences of the external field, where each random sequence $\left(h_{\left(0\right)},\ldots,h_{\left(t\right)}\right)$ corresponds to an experiment condition. The condition includes a possible variation process of external field that begins and ends with a unique $h_{\left(0\right)}=h_{\left(t\right)}\in \left(0,m\right]\cap\mathbb{Z}$. This sequence features a certain transition rate $\theta^{-1}$ (there are $k_{\theta}$ transition rates that quantify the variation speed of the external field). Meanwhile, each condition corresponds to a coupling strength $J$ and temperature $T$ (there are $k_{J}$ and $k_{T}$ kinds of coupling strength and temperature quantities, respectively). Under each experiment condition, the simulation experiment is implemented as following
        \begin{itemize}
            \item\;We randomly initialize a specification $\bm{\sigma}_{\left(0\right)}$ at moment $0$;
            \item\;We perform the continuous-time Monte Carlo simulation utilizing the Metropolis algorithm $\left(\bm{\sigma}_{\left(0\right)},\ldots,\bm{\sigma}_{\left(t\right)}\right)=\mathsf{Metropolis}\left[\bm{\sigma}_{\left(0\right)},\left(h_{\left(0\right)},\ldots,h_{\left(t\right)}\right)\right]$ \cite{bhanot1988metropolis}. The simulation lasts for a duration $\left[0,t\right]$. During the simulation, we use the update of the external field $h$ to drive the Metropolis algorithm. The work is initialized as $\mathcal{W}=0$. Every time when the spin specification changes from $\bm{\sigma}_{\left(\tau\right)}$ to $\bm{\sigma}^{\prime}_{\left(\tau+1\right)}$, we add $\frac{\mathbf{H}\left(\bm{\sigma}_{\left(\tau\right)},h_{\left(\tau\right)}\right)-\mathbf{H}\left(\bm{\sigma}_{\left(\tau+1\right)}^{\prime},h_{\left(\tau+1\right)}\right)}{T}$ to $\frac{\mathcal{W}}{T}$;
            \item\;The Metropolis algorithm $\left(\bm{\sigma}_{\left(0\right)},\ldots,\bm{\sigma}_{\left(t\right)}\right)=\mathsf{Metropolis}\left[\bm{\sigma}_{\left(0\right)},\left(h_{\left(0\right)},\ldots,h_{\left(t\right)}\right)\right]$ runs $\xi$ times in total. The $i$-th simulation corresponds to a final state $\bm{\sigma}_{\left(t\right)}^{i}$. Then, we estimate $\mathcal{P}\left(\bm{\sigma}_{\left(t\right)}\mid h_{\left(t\right)}\right)$ based on the frequency distribution on $\lbrace\bm{\sigma}_{\left(t\right)}^{1},\ldots,\bm{\sigma}_{\left(t\right)}^{\xi}\rbrace$.
        \end{itemize}
    Based on the above procedure, we can computationally obtain the probability distribution $\mathcal{P}\left(\bm{\sigma}_{\left(t\right)}\mid h_{\left(t\right)}\right)$ and the work $\mathcal{W}$ (or the irreversible work $\mathcal{W}_{\mathsf{irr}}$). Moreover, we can further calculate the term $\mathcal{D}\big[\mathcal{P}\left(\bm{\sigma}_{\left(t\right)},h_{\left(t\right)}\right)\big\Vert\widehat{\mathcal{P}}\left(\bm{\sigma}_{\left(t\right)},h_{\left(t\right)}\right)\big]$ following (\ref{EQ20}). As for $\mathcal{P}\left(\bm{\sigma}_{\left(0\right)}, h_{\left(0\right)}\right)$, we assume it is uniformly distributed on the sample space $\Omega\left(\bm{\sigma}\right)\times\Omega\left(h\right)$. Given this assumption, we can calculate the term $\mathcal{D}\big[\mathcal{P}\left(\bm{\sigma}_{\left(0\right)},h_{\left(0\right)}\right)\big\Vert\widehat{\mathcal{P}}\left(\bm{\sigma}_{\left(0\right)},h_{\left(0\right)}\right)\big]$ under the \textbf{UCI Condition}. Finally, we can directly obtain the entropy production $\Delta_{i}\mathcal{S}_{\left(t\right)}$ since other parts of (\ref{EQ1}) have been given. We implement the computational experiment using the parameter settings in appendix \ref{ASec3}. The spin system $\bm{\sigma}$ is set with a small size, allowing us to do a relatively exhaustive sampling.
    
    \begin{figure}[t!]
     \centering
     \begin{subfigure}[b]{0.48\columnwidth}
         \includegraphics[width=\columnwidth]{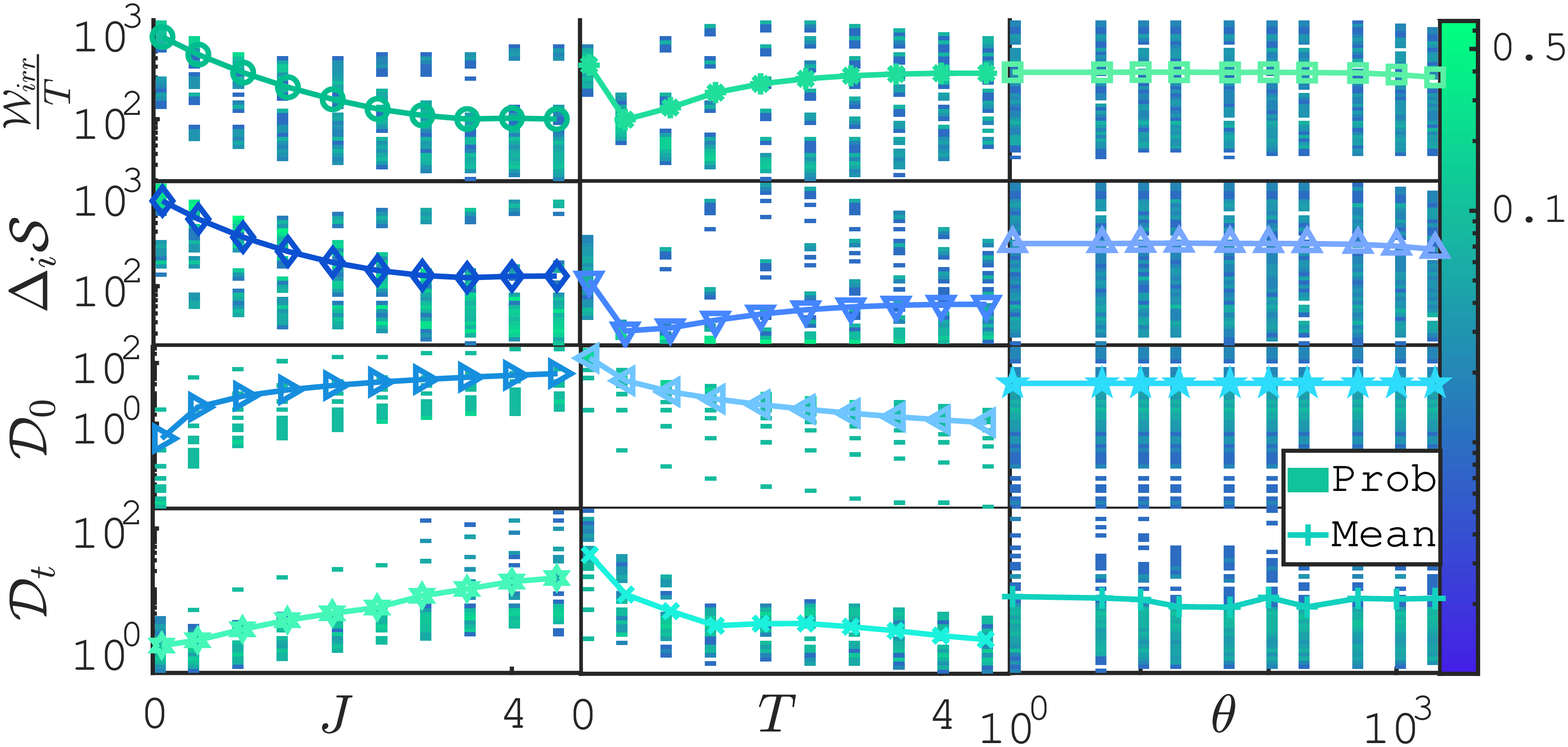}
        \caption{\label{G8a}}
     \end{subfigure}
     \hfill
     \begin{subfigure}[b]{0.48\columnwidth}
         \includegraphics[width=\columnwidth]{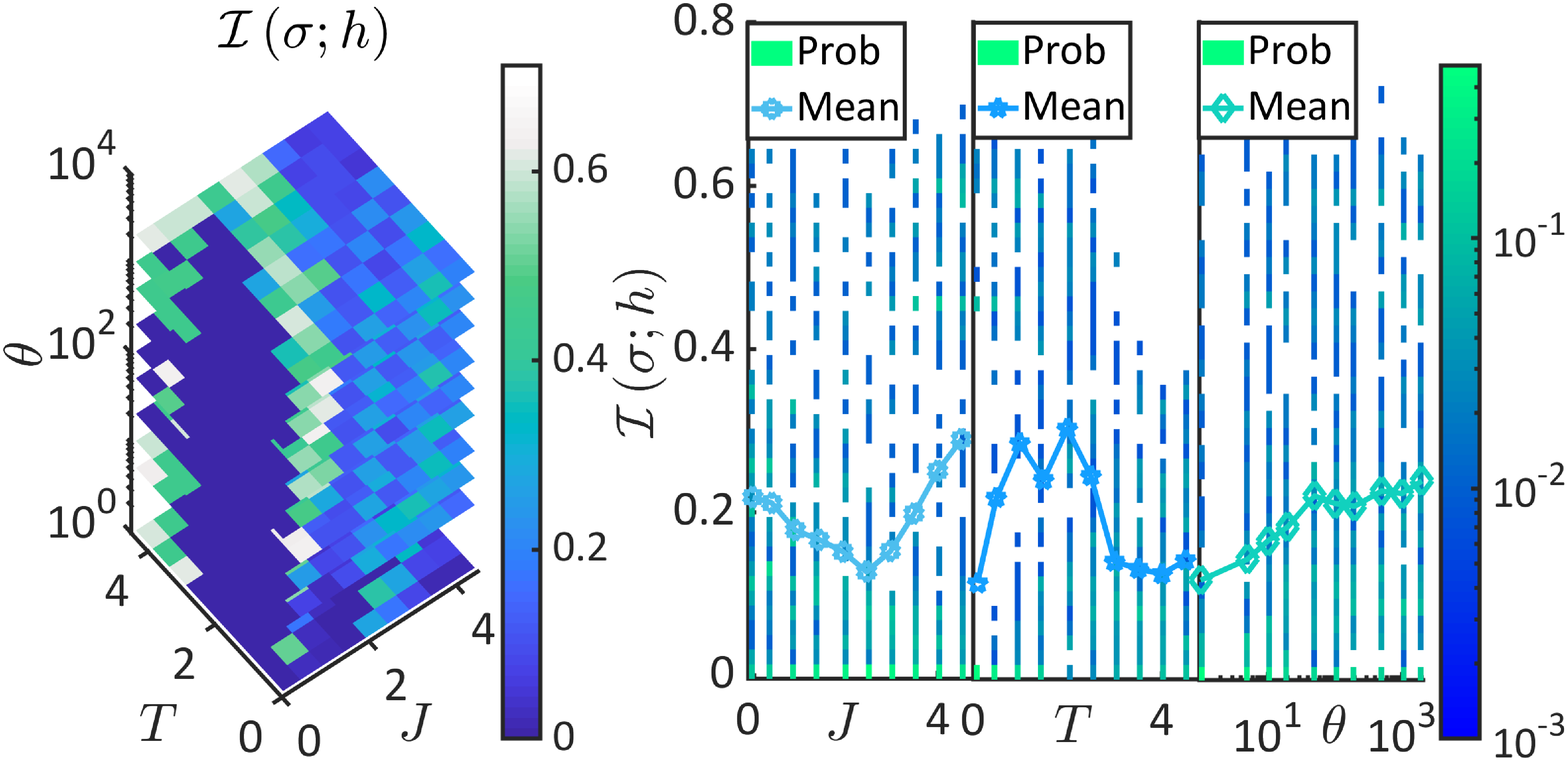}
     \caption{\label{G8b}}
     \end{subfigure}
     \hfill
     \begin{subfigure}[b]{0.48\columnwidth}
         \includegraphics[width=\columnwidth]{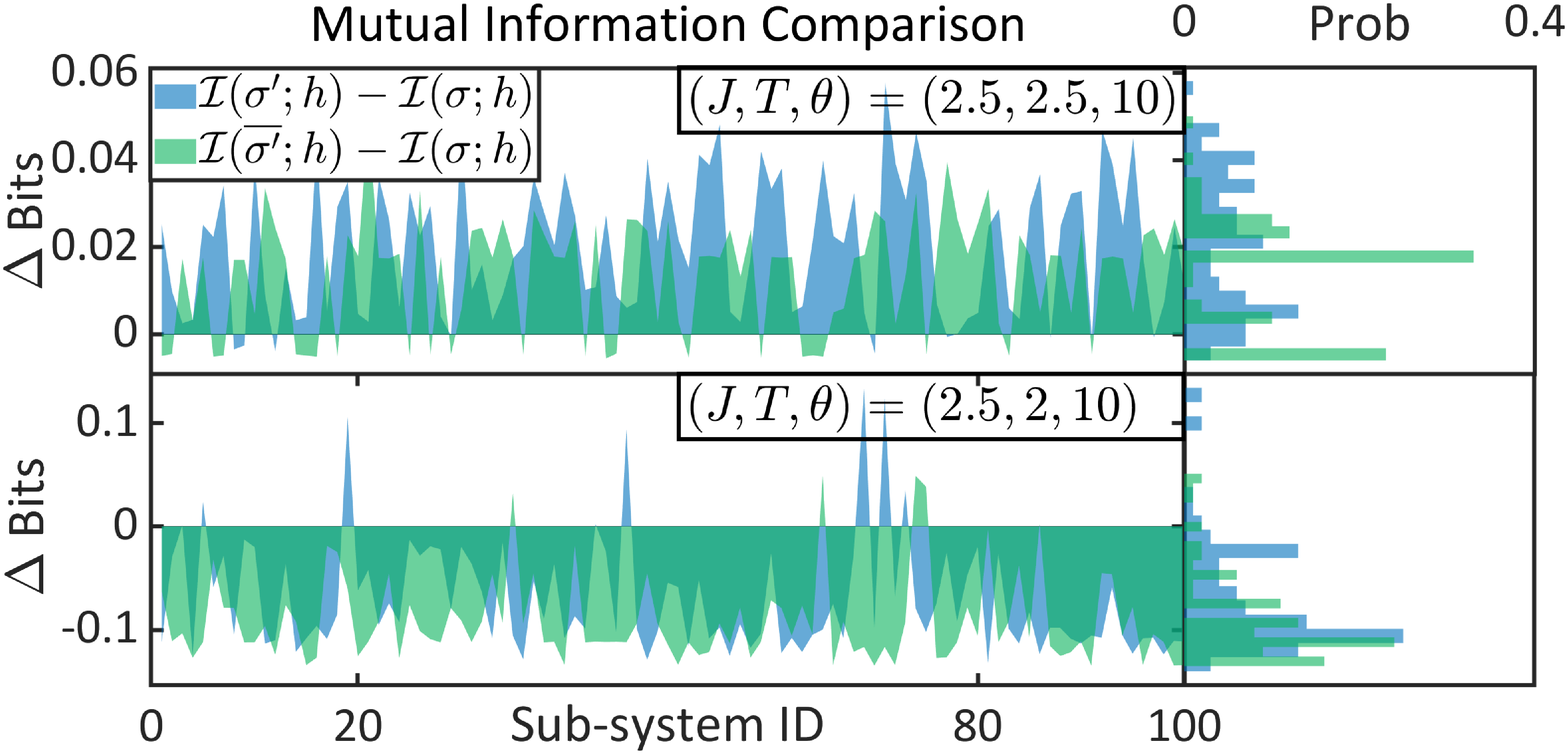}
     \caption{\label{G8c}}
     \end{subfigure}
     \hfill
     \begin{subfigure}[b]{0.48\columnwidth}
         \includegraphics[width=\columnwidth]{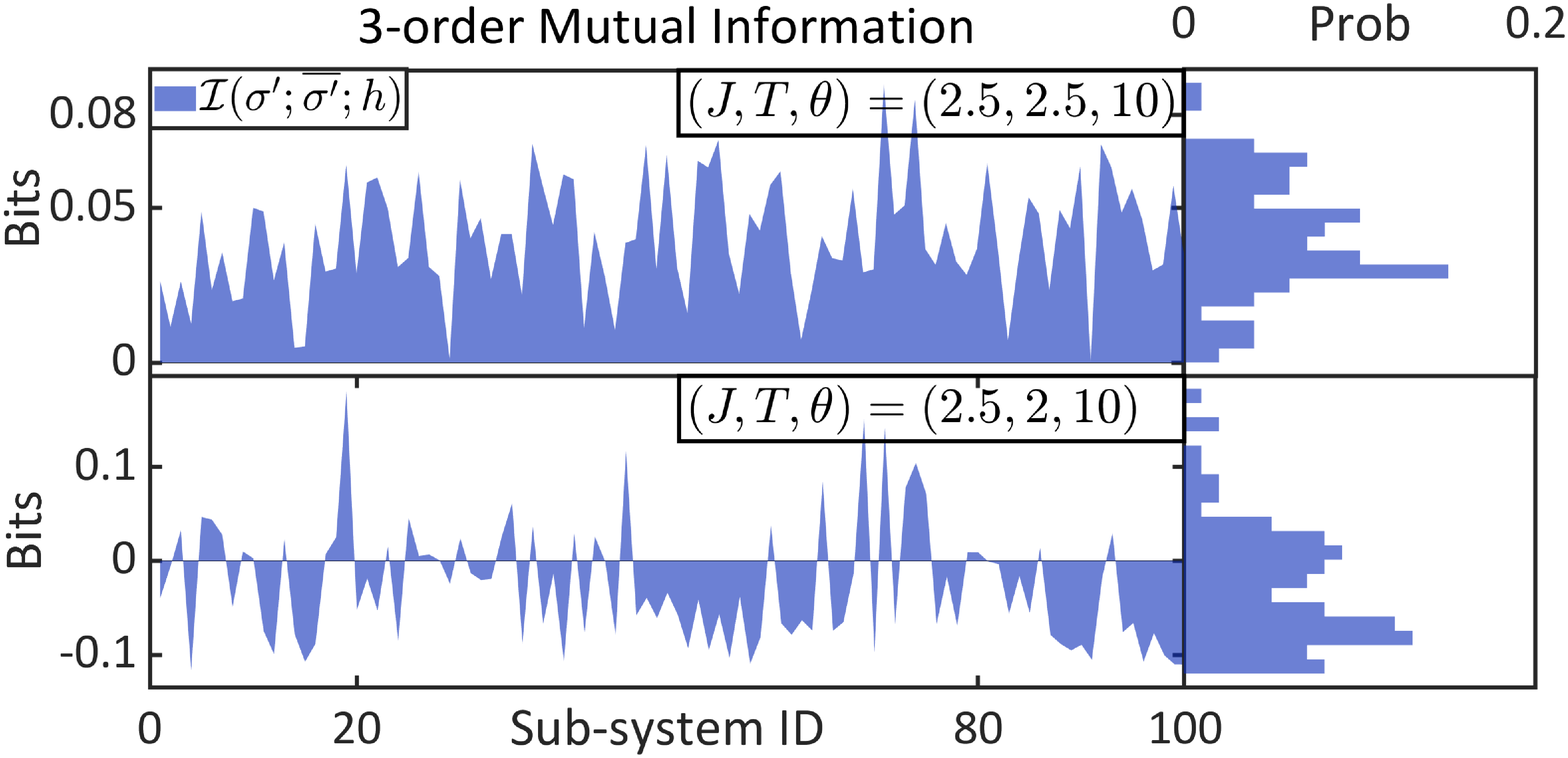}
     \caption{\label{G8d}}
     \end{subfigure}
     \hfill
     \begin{subfigure}[b]{0.48\columnwidth}
         \includegraphics[width=\columnwidth]{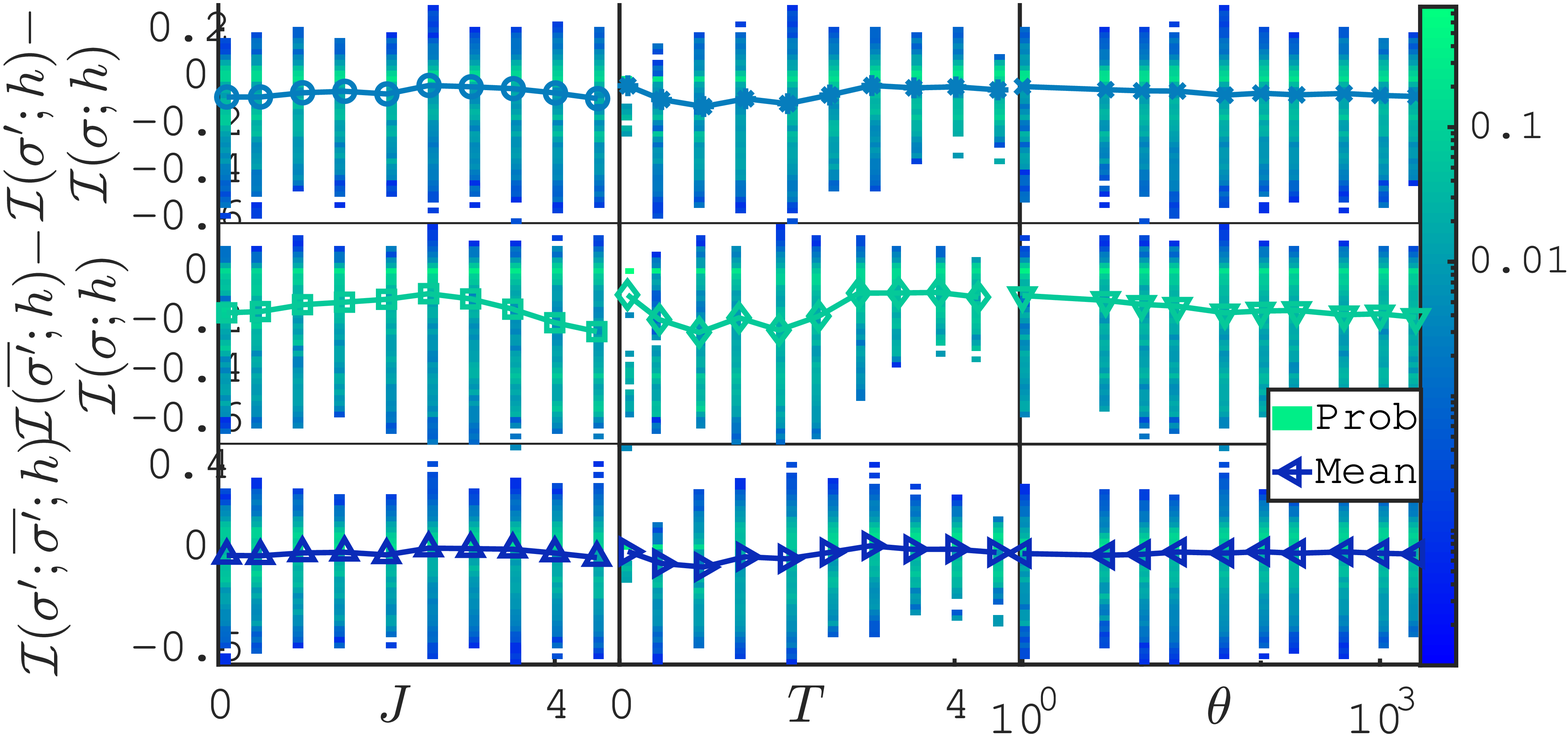}
     \caption{\label{G8e}}
     \end{subfigure}
     \hfill
     \begin{subfigure}[b]{0.48\columnwidth}
         \includegraphics[width=\columnwidth]{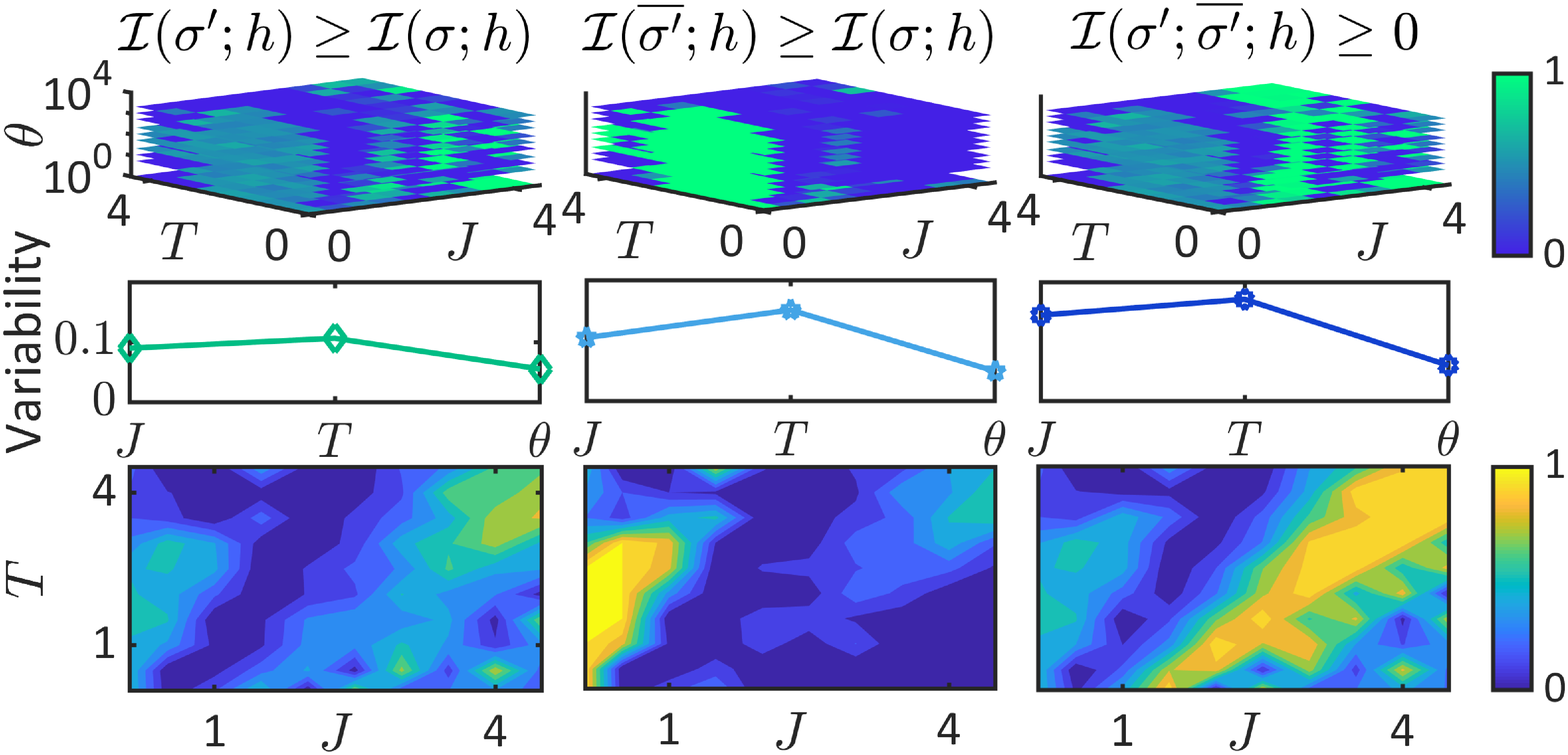}
     \caption{\label{G8f}}
     \end{subfigure}
     \hfill
        \caption{\label{G8} Information thermodynamics in the Ising model. (a) The calculated terms (mean value and probability distributions) of (\ref{EQ1}) are shown as the functions of $J$, $T$, and $\theta$. While analyzing each parameter direction among $J$, $T$, and $\theta$, the other two directions are averaged. (b) The calculated mutual information quantities (raw data distributions, mean value, and probability distributions) act as the functions of $J$, $T$, and $\theta$. (c-d) Two representative examples of the mutual information comparison and their corresponding $3$-order mutual information. (e) The mutual information comparison result and the corresponding $3$-order mutual information are analyzed as the functions of $J$, $T$, and $\theta$. (f) The probability distributions of the cases where $\mathcal{I}\left(\bm{\sigma}_{\left(t\right)}^{\prime};h_{\left(t\right)}\right)\geq\mathcal{I}\left(\bm{\sigma}_{\left(t\right)};h_{\left(t\right)}\right)$, $\mathcal{I}\left(\overline{\bm{\sigma}^{\prime}}_{\left(t\right)};h_{\left(t\right)}\right)\geq\mathcal{I}\left(\bm{\sigma}_{\left(t\right)};h_{\left(t\right)}\right)$, or $\mathcal{I}\left(\bm{\sigma}_{\left(t\right)}^{\prime};\overline{\bm{\sigma}^{\prime}}_{\left(t\right)};h_{\left(t\right)}\right)\geq 0$ are shown as the functions of $J$, $T$, and $\theta$ (upper line). We also compare the probability variability of these cases in each parameter direction (middle line). Then, we analyze the probability distributions of these cases in the directions of $J$ and $T$ due to the high variability (bottom line).}
\end{figure}

Fig. \ref{G8a} shows the variation trends of every term of (\ref{EQ1}) along with the increasing $J$, $T$, or $\theta$ conditions. The corresponding mutual information calculated based on (\ref{EQ5}) is illustrated as the function of $J$, $T$, or $\theta$ in Fig. \ref{G8b}. One can see that the mutual information (mean value) features a positive correlation with the inverse transition rate $\theta$ (logarithmic value). The correlation is strong with a Pearson coefficient $R=0.9429$ and significant at $p<10^{-5}$, suggesting that the spin system $\bm{\sigma}$ encodes more information about the external field $h$ when it has more relaxation time between any two times of external field variations (larger $\theta$). Apart of that, more complex variation trends of the mutual information can be found along with the coupling strength $J$ and the temperature $T$. 

Featuring intra-system coupling relations (the interactions between different spins), the spin system $\bm{\sigma}$ is expected to allow the encoded information by its sub-systems to exceed the encoded information by itself. To verify this speculation, we randomly select sub-systems $\bm{\sigma}^{\prime}$ (and the corresponding $\overline{\bm{\sigma}^{\prime}}$) 100 times under each experiment conditions. As shown by the examples in Fig. \ref{G8c}, one can see that the information quantity differences $\mathcal{I}\left(\bm{\sigma}_{\left(t\right)}^{\prime};h_{\left(t\right)}\right)-\mathcal{I}\left(\bm{\sigma}_{\left(t\right)};h_{\left(t\right)}\right)$ and $\mathcal{I}\left(\overline{\bm{\sigma}^{\prime}}_{\left(t\right)};h_{\left(t\right)}\right)-\mathcal{I}\left(\bm{\sigma}_{\left(t\right)};h_{\left(t\right)}\right)$ can be either non-positive (meaning that the bound (\ref{EQ8}) is followed by sub-systems) or positive (the bound (\ref{EQ8}) may be disobeyed by sub-systems). Moreover, one can find the corresponding positive $3$-order mutual information $\mathcal{I}\left(\bm{\sigma}_{\left(t\right)}^{\prime};\overline{\bm{\sigma}^{\prime}}_{\left(t\right)};h_{\left(t\right)}\right)$ in Fig. \ref{G8d}, which suggests the existence of information redundancy or synergy. 

Although the experiment condition $\left(J,T,\theta\right)$ has significant effects on $\mathcal{I}\left(\bm{\sigma}_{\left(t\right)}^{\prime};h_{\left(t\right)}\right)-\mathcal{I}\left(\bm{\sigma}_{\left(t\right)};h_{\left(t\right)}\right)$, $\mathcal{I}\left(\bm{\sigma}_{\left(t\right)}^{\prime};h_{\left(t\right)}\right)-\mathcal{I}\left(\bm{\sigma}_{\left(t\right)};h_{\left(t\right)}\right)$, and $\mathcal{I}\left(\bm{\sigma}_{\left(t\right)}^{\prime};\overline{\bm{\sigma}^{\prime}}_{\left(t\right)};h_{\left(t\right)}\right)$ in Fig. \ref{G8c}-\ref{G8d}, we suggest that these effects do not show clear trends if they are attributed to a single parameter among $J$, $T$, and $\theta$ in Fig. \ref{G8e}. Therefore, we turn to search for underlying patterns from the probabilistic perspective and implement a unified analysis. In Fig. \ref{G8f}, we investigate the probability distributions of $\mathcal{I}\left(\bm{\sigma}_{\left(t\right)}^{\prime};h_{\left(t\right)}\right)\geq\mathcal{I}\left(\bm{\sigma}_{\left(t\right)};h_{\left(t\right)}\right)$, $\mathcal{I}\left(\overline{\bm{\sigma}^{\prime}}_{\left(t\right)};h_{\left(t\right)}\right)\geq\mathcal{I}\left(\bm{\sigma}_{\left(t\right)};h_{\left(t\right)}\right)$, and $\mathcal{I}\left(\bm{\sigma}_{\left(t\right)}^{\prime};\overline{\bm{\sigma}^{\prime}}_{\left(t\right)};h_{\left(t\right)}\right)\geq 0$ as the functions of $\left(J,T,\theta\right)$. After measuring the mean variability of these probability distributions in each parameter direction (e.g., the probability variability in the direction of $J$ at $\left(J,T,\theta\right)$ is defined as the variance $\operatorname*{Var}_{J}\left[\mathcal{P}\left(J,T,\theta\right)\right]$. Then the mean variability is quantified as $\operatorname*{\mathbb{E}}_{T,\theta}\lbrace\operatorname*{Var}_{J}\left[\mathcal{P}\left(J,T,\theta\right)\right]\rbrace$), we confirm that the variability of these probability distributions can be better explained by $\left(J,T\right)$ rather than $\theta$. This finding inspires us to concentrate on the effects of $\left(J,T\right)$ on the probability distributions. The bottom panel of Fig. \ref{G8f} demonstrates that the probability densities of $\mathcal{I}\left(\bm{\sigma}_{\left(t\right)}^{\prime};h_{\left(t\right)}\right)\geq\mathcal{I}\left(\bm{\sigma}_{\left(t\right)};h_{\left(t\right)}\right)$, $\mathcal{I}\left(\overline{\bm{\sigma}^{\prime}}_{\left(t\right)};h_{\left(t\right)}\right)\geq\mathcal{I}\left(\bm{\sigma}_{\left(t\right)};h_{\left(t\right)}\right)$, and $\mathcal{I}\left(\bm{\sigma}_{\left(t\right)}^{\prime};\overline{\bm{\sigma}^{\prime}}_{\left(t\right)};h_{\left(t\right)}\right)\geq 0$ relatively concentrate on the sub-region where $J\sim T$ (or equivalently, $\min\left(\frac{J}{T},\frac{T}{J}\right)\sim 1$). Because the temperature $T$ is not an inherent characteristic of the spin system $\bm{\sigma}$, our research focuses on the coupling strength $J$. As the temperature $T$ increases, the coupling strength $J$ where these probability densities concentrate will also increase. In other words, the cases where the encoded information of $h$ by a sub-system $\bm{\sigma}^{\prime}$ (or $\overline{\bm{\sigma}^{\prime}}$) exceeds $\mathcal{I}\left(\bm{\sigma};h\right)$ and the cases with information redundancy frequently emerge when the coupling strength $J$ (the intra-system coupling) is relatively proportional to the temperature $T$.

    Assuming there is an observer who pursues to estimate the maximum encoded information of $h$ in $\bm{\sigma}$. Given our findings above, the encoded information might be underestimated if this observer implements the observation outside the system $\left(\bm{\sigma},h\right)$ (e.g., the observer estimates the encoded information utilizing the irreversible work $\mathcal{W}_{\mathsf{irr}}$ from system $\left(\bm{\sigma},h\right)$ and (\ref{EQ8})). When the encoded information $\mathcal{I}\left(\bm{\sigma}^{\prime};h\right)$ by a sub-system $\bm{\sigma}^{\prime}$ (or $\overline{\bm{\sigma}^{\prime}}$) can exceed $\mathcal{I}\left(\bm{\sigma};h\right)$, a certain amount of encoded information (measured as $\mathcal{I}\left(\bm{\sigma}^{\prime};h\right)-\mathcal{I}\left(\bm{\sigma};h\right)$ or $\mathcal{I}\left(\overline{\bm{\sigma}^{\prime}};h\right)-\mathcal{I}\left(\bm{\sigma};h\right)$) is hidden from the observer. 
    
    \subsection{Real data of the human brain during the perception process}\label{Sec4b}
Perhaps the brain is the most natural example of information thermodynamics encoder with intra-system coupling. We choose the brain as our second demonstration since neuroscience studies can offer vast amounts of data for $\mathsf{X}$ (neural data) and a clear characterization for $\mathsf{Y}$ (stimulus data). Nevertheless, we need to note that the temperature $T$ and the irreversible work $\mathcal{W}_{\mathsf{irr}}$ are not well-defined concepts for the brain and can not be calculated directly. Therefore, the analysis can only be performed indirectly.
   
   Specifically, we implement our analysis on an open-source functional magnetic resonance imaging (fMRI) data set obtained from the visual object recognition experiment, where the random stimulus sequence consists of 8 kinds of objects \cite{ds000105}. The fMRI technique measures neural activities through the blood-oxygen-level-dependent (BOLD) contrast in the magnetic field \cite{heeger2002does}. This data set includes the fMRI signals of 6 subjects from a 3 Tesla scanner, covering a high-resolution whole-brain region of 163840 voxels (small neural clusters) for each subject. Each voxel corresponds to a time series of the activities of a neural cluster. In the experiment, 12 pairs of (neural data, stimulus data) are obtained from each subject, and we arrange these pairs into a single matrix. Thus, every subject corresponds to a joint system $\left(\mathsf{X},\mathsf{Y}\right)$. The validity of the data has been verified previously \cite{haxby2001distributed,hanson2004combinatorial,o2005partially}, ensuring the reproducibility of our analysis. Please see appendix \ref{ASec4} for more details.
    
    To measure information quantities in real data set, we estimate the mutual information $\widehat{\mathcal{I}}$ following the computational approach proposed by Gao et al. \cite{gao2017estimating}
    \begin{align}
    \widehat{\mathcal{I}}\left(\mathsf{X},\mathsf{Y}\right)=\mathbb{E}_{\tau}\left[\psi\left(\widehat{k}_{\left(\tau\right)}\right)- \log \left(\mathcal{C}_{\mathsf{X}}\left(\tau\right)+1\right)\left(\mathcal{C}_{\mathsf{Y}}\left(\tau\right)+1\right)\right]+\log t, \label{EQXX}
    \end{align}
    where $\psi\left(\cdot\right)$ denotes the digamma function. The estimation is implemented based on $k$-nearest neighbor (KNN) method, where each $\tau\in\left[0,t\right]$ corresponds to a sample (here $t$ is the length of data). Let $\epsilon_{\left(\tau\right)}$ be the distance from $\left(\mathsf{X}_{\left(\tau\right)},\mathsf{Y}_{\left(\tau\right)}\right)$ to its $k$-th neighbor with a given parameter $k$ (in our research, we set $k=5$). One have $\widehat{k}_{\left(\tau\right)}=k$ if $\epsilon_{\left(\tau\right)}>0$. Otherwise, one need to measure $\widehat{k}_{\left(\tau\right)}=\vert\lbrace\left(\mathsf{X}_{\left(\tau^{\prime}\right)},\mathsf{Y}_{\left(\tau^{\prime}\right)}\right)\mid\left(\mathsf{X}_{\left(\tau^{\prime}\right)},\mathsf{Y}_{\left(\tau^{\prime}\right)}\right)=\left(\mathsf{X}_{\left(\tau\right)},\mathsf{Y}_{\left(\tau\right)}\right)\rbrace\vert$. In the sub-space of $\mathsf{X}$, one can count the number $\mathcal{C}_{\mathsf{X}}\left(\tau\right)$ of the sample whose distance from $\mathsf{X}_{\left(\tau\right)}$ is no more than $\epsilon_{\left(\tau\right)}$. Similarily, one can measure $\mathcal{C}_{\mathsf{Y}}\left(\tau\right)$ in the sub-space of $\mathsf{Y}$. In \cite{gao2017estimating}, the distance is defined utilizing the maximum norm. Unlike the classic estimator designed with $3\mathcal{H}$-principle that $\widehat{\mathcal{I}}\left(\mathsf{X},\mathsf{Y}\right)=\widehat{\mathcal{H}}\left(\mathsf{X}\right)+\widehat{\mathcal{H}}\left(\mathsf{Y}\right)-\widehat{\mathcal{H}}\left(\mathsf{X},\mathsf{Y}\right)$ (e.g., see \cite{kraskov2004estimating}), this estimation is more efficient in controlling systematic errors and dealing with mixture spaces \cite{gao2017estimating}. 

\begin{figure}[t!]
     \centering
     \begin{subfigure}[b]{0.48\columnwidth}
         \includegraphics[width=\columnwidth]{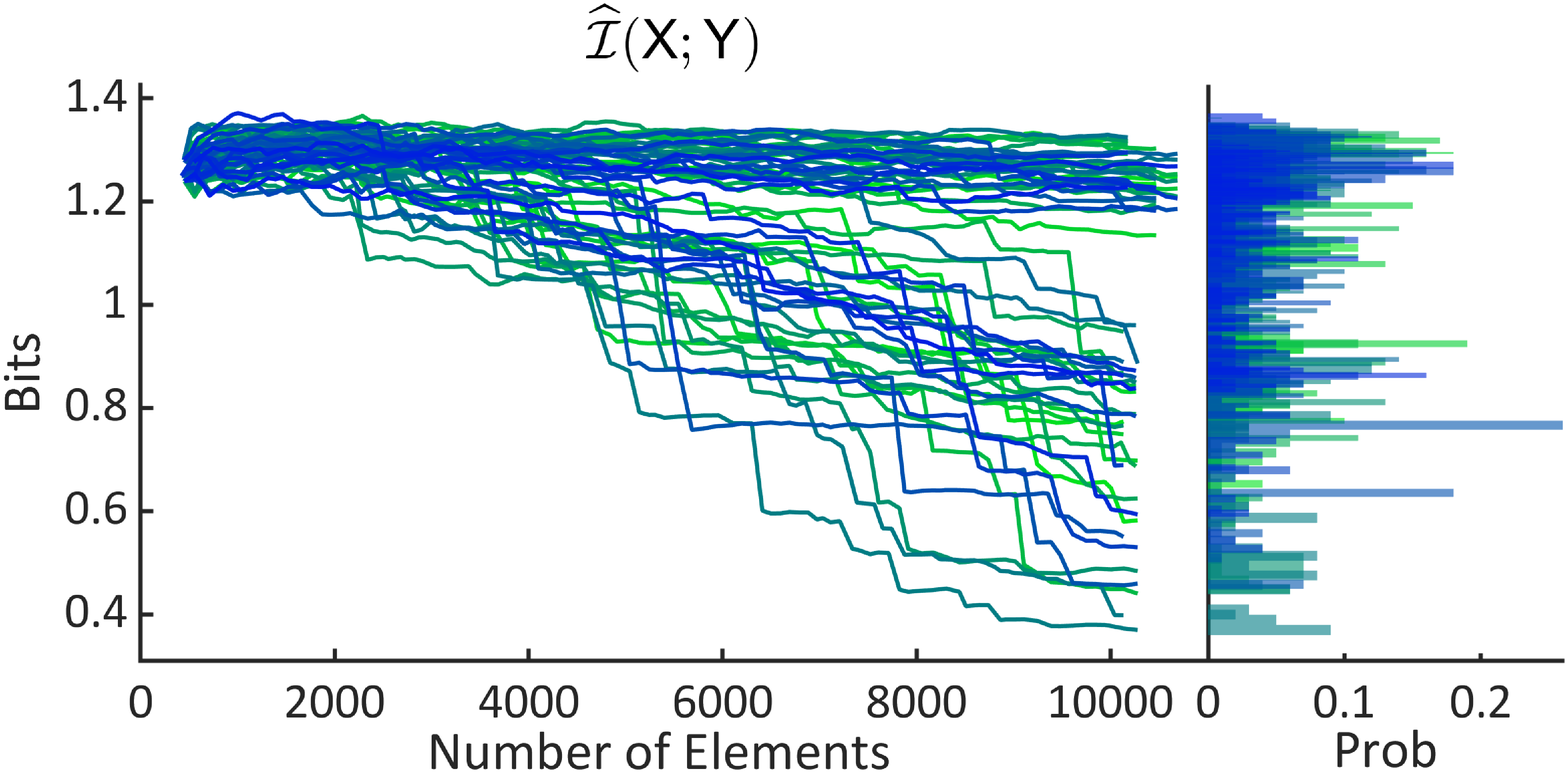}
        \caption{\label{G9a}}
     \end{subfigure}
     \hfill
     \begin{subfigure}[b]{0.48\columnwidth}
         \includegraphics[width=\columnwidth]{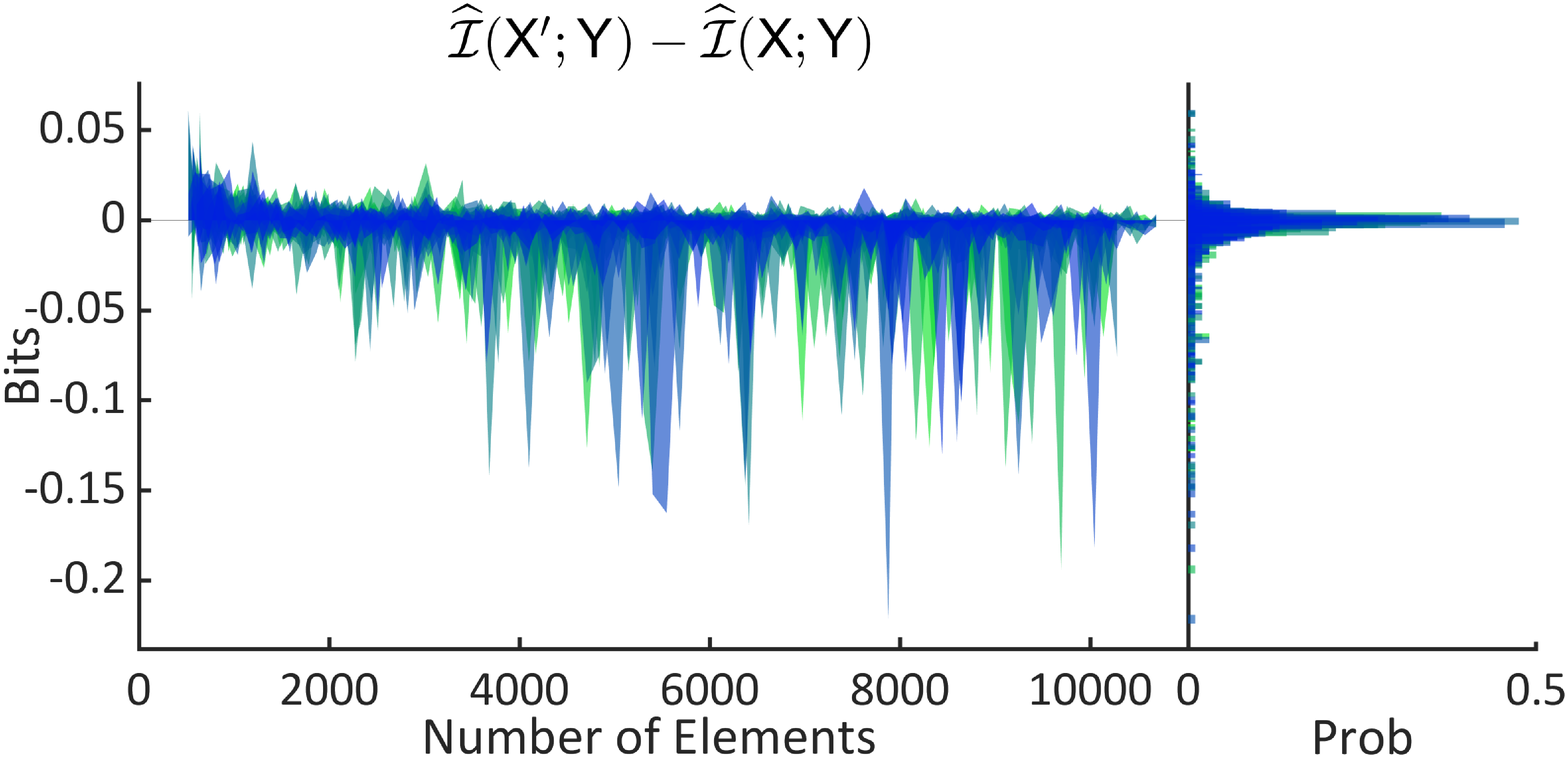}
     \caption{\label{G9b}}
     \end{subfigure}
     \hfill
     \begin{subfigure}[b]{0.48\columnwidth}
         \includegraphics[width=\columnwidth]{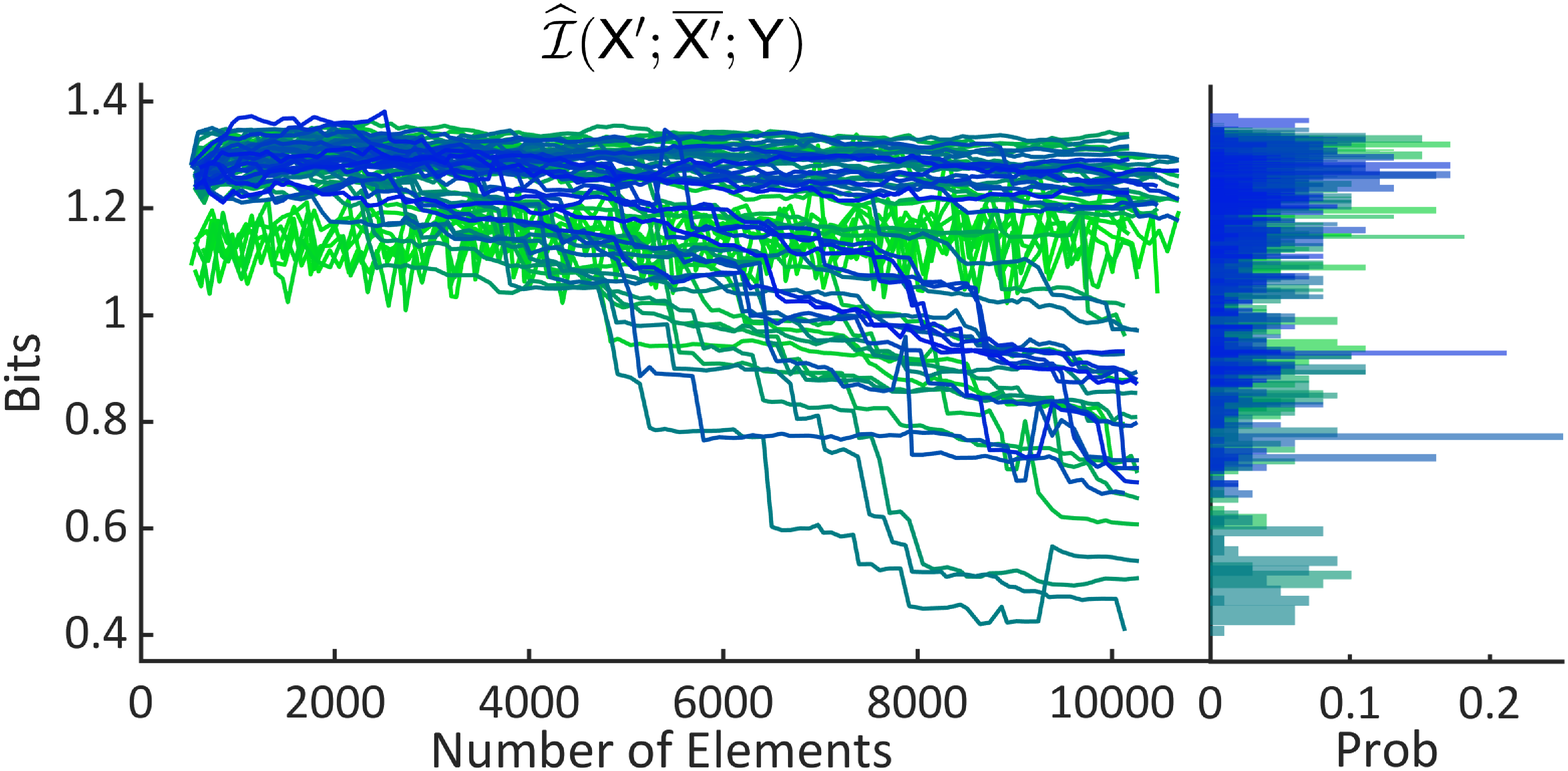}
     \caption{\label{G9c}}
     \end{subfigure}
     \hfill
     \begin{subfigure}[b]{0.48\columnwidth}
         \includegraphics[width=\columnwidth]{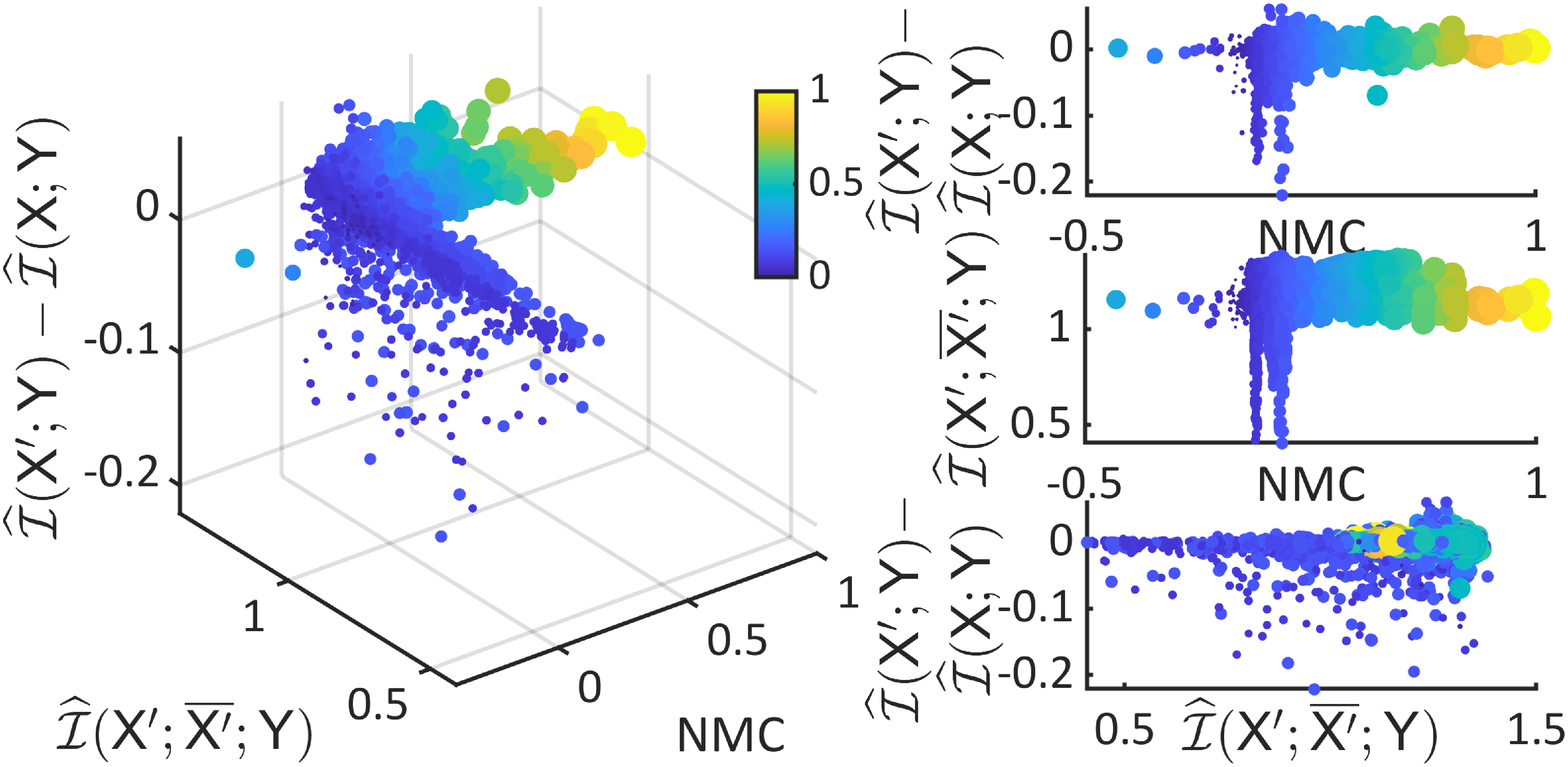}
     \caption{\label{G9d}}
     \end{subfigure}
     \hfill
     \begin{subfigure}[b]{0.48\columnwidth}
         \includegraphics[width=\columnwidth]{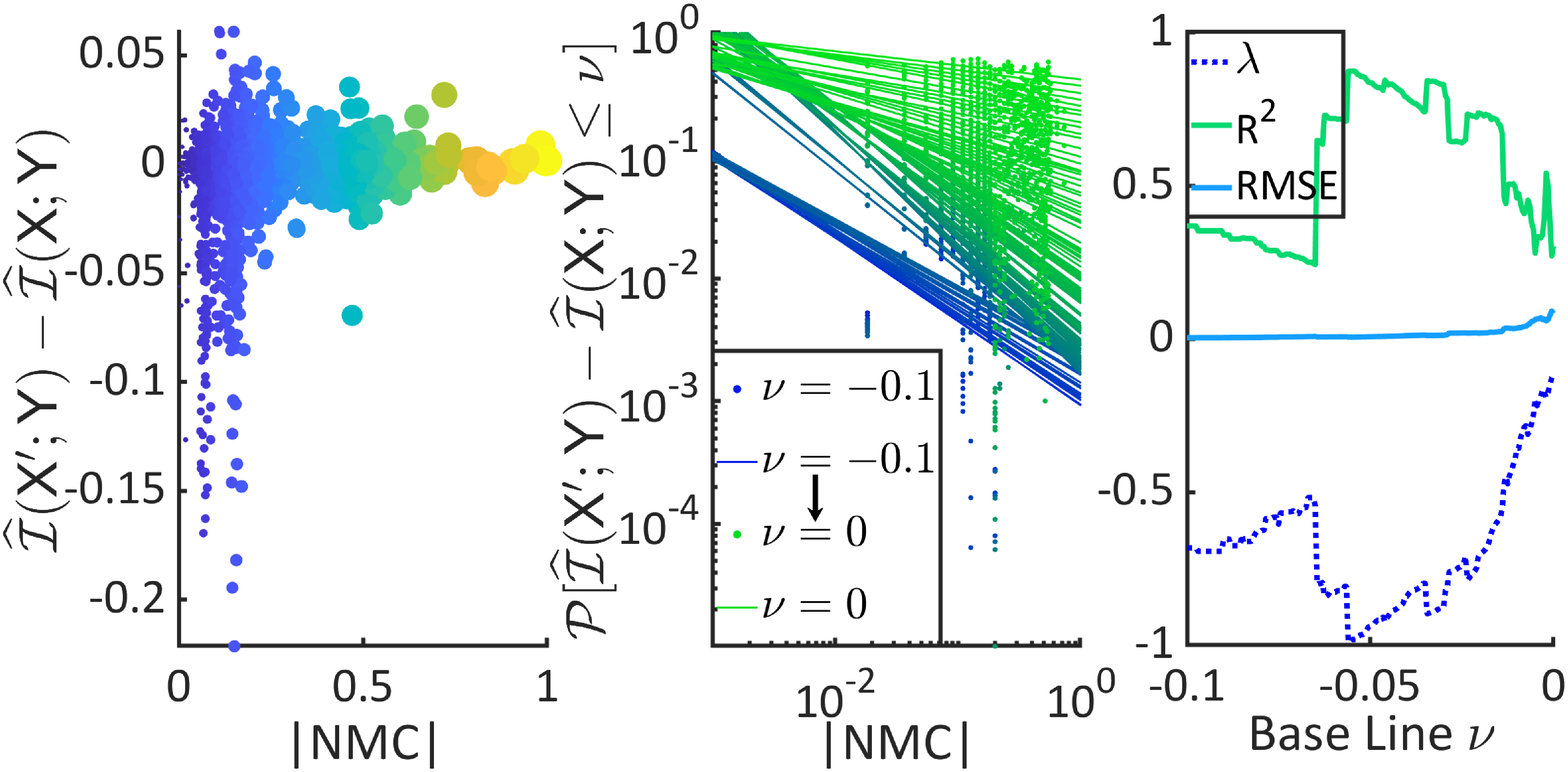}
     \caption{\label{G9e}}
     \end{subfigure}
     \hfill
     \begin{subfigure}[b]{0.48\columnwidth}
         \includegraphics[width=\columnwidth]{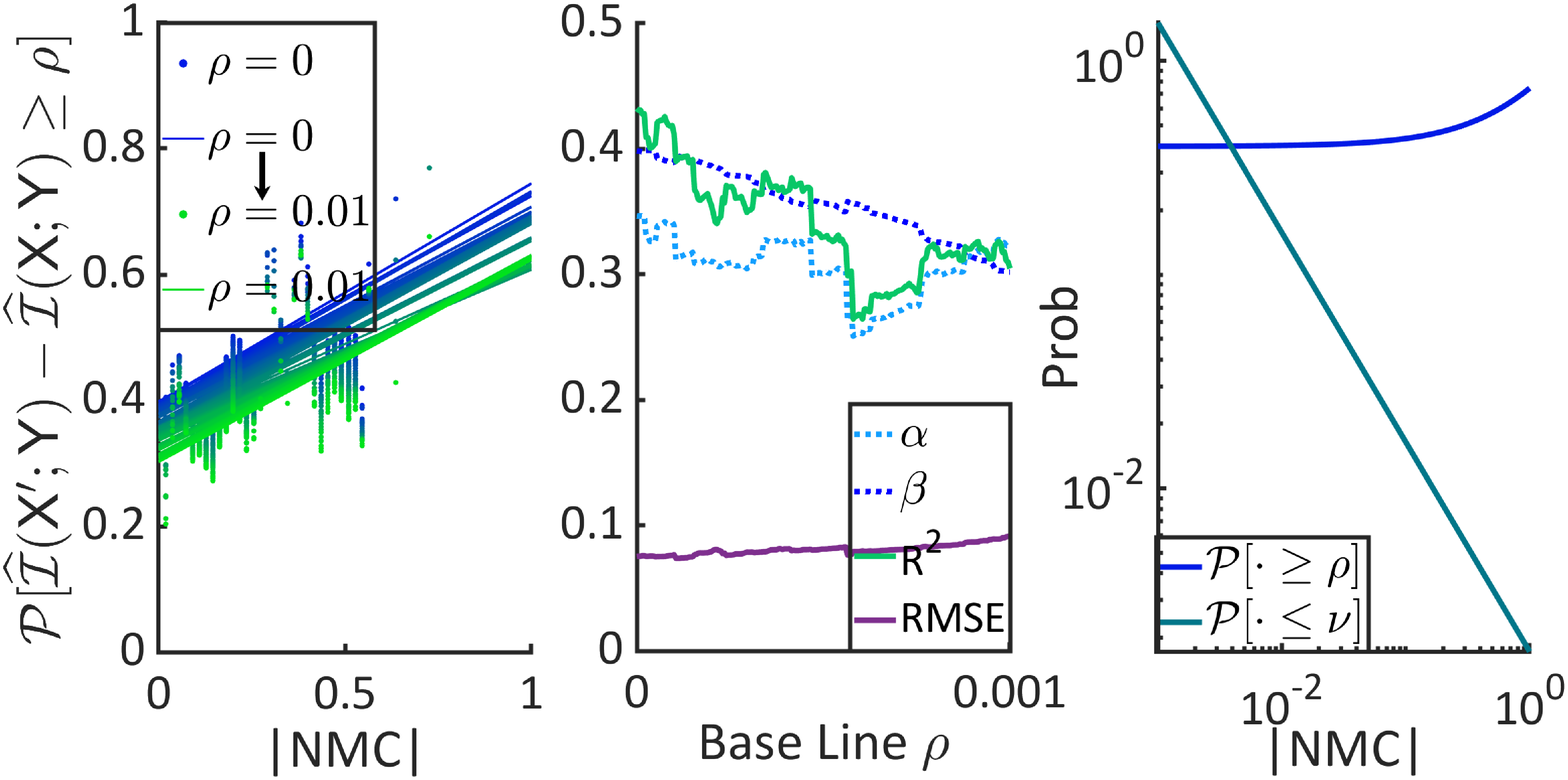}
     \caption{\label{G9f}}
     \end{subfigure}
     \hfill
        \caption{\label{G9} Information thermodynamics in the human brain. (a-c) The estimated information quantities and their probability distributions. (d) The relations between information quantities and the normalized mean correlation (NMC) are shown, where the sizes and colors of data points scale according to NMC. (e) We analyze the information quantity $\widehat{\mathcal{I}}\left(\mathsf{X}_{\left(t\right)}^{\prime};\mathsf{Y}_{\left(t\right)}\right)-\widehat{\mathcal{I}}\left(\mathsf{X}_{\left(t\right)},\mathsf{Y}_{\left(t\right)}\right)$ as a function of the absolute value of NMC (left). By adjusting the base line $\varepsilon$, we can quantify the probability distribution of the case where $\widehat{\mathcal{I}}\left(\mathsf{X}_{\left(t\right)}^{\prime};\mathsf{Y}_{\left(t\right)}\right)-\widehat{\mathcal{I}}\left(\mathsf{X}_{\left(t\right)},\mathsf{Y}_{\left(t\right)}\right)\leq\nu$ under different standards. Our results suggest that this kind of case, irrespective of the selected base line, the probability $\mathcal{P}[\cdot\leq\nu]$ follows a power regression model of $\vert\text{NMC}\vert$ with reasonable fitting accuracy (middle and right). (f) Correspondingly, we set different base line $\rho$ to identify the case where $\widehat{\mathcal{I}}\left(\mathsf{X}_{\left(t\right)}^{\prime};\mathsf{Y}_{\left(t\right)}\right)-\widehat{\mathcal{I}}\left(\mathsf{X}_{\left(t\right)},\mathsf{Y}_{\left(t\right)}\right)\geq\rho$. The probability $\mathcal{P}[\cdot\geq\rho]$ and $\vert\text{NMC}\vert$ feature a linear regression relation with reasonable fitting accuracy, relatively independent of the base line settings (left and middle). Finally, we illustrate a representative instance of the probability distributions $\mathcal{P}[\cdot\leq\nu]$ and $\mathcal{P}[\cdot\geq\rho]$ where $\left(\nu,\rho\right)=\left(-0.0545,0\right)$ (right).}
\end{figure}

 It is computationally costly to calculate information quantities on the whole brain regarding a large number of voxels. Therefore, we implement our analysis utilizing a reverse sequence. Specifically, we randomly initialize a system of $z$ voxels in each brain ($z\in\left[400,600\right]$). Then, we randomly add $r$ voxels into this system in each iteration ($r\in\left[50,150\right]$ and every sequence lasts for 100 iterations). We repeat this random reverse sequence generation in each subject $10$ times. Given these settings, the previous system and the newly added voxel set are complementary sub-systems of the current system (can be referred to as $\mathsf{X}^{\prime}$ and $\overline{\mathsf{X}^{\prime}}$). Fig. \ref{G9a} illustrates the variation trends of the estimated mutual information in these random sequences, suggesting that the system in the previous iteration (includes fewer voxels) might encode more stimulus information ($\mathsf{Y}$) than the current system (includes more voxels). The corresponding differences between the encoded information quantities in previous and current systems are shown in Fig. \ref{G9b}. In Fig. \ref{G9c}, one can see that $\widehat{\mathcal{I}}\left(\mathsf{X}_{\left(t\right)}^{\prime};\overline{\mathsf{X}^{\prime}}_{\left(t\right)};\mathsf{Y}_{\left(t\right)}\right)>0$ always holds across all iterations, revealing the existence of information redundancy. Here we need to emphasize that the iteration is implemented with randomization and therefore frequently adds disharmonious voxels into the system to create redundancy. This result does not deny the possibility of information synergy. 
 
 To relate the above findings with the internal orders of the human brain, we quantify the mean Pearson correlation coefficient between the voxels that previously exist ($\mathsf{X}^{\prime}$) and the newly added voxels ($\overline{\mathsf{X}^{\prime}}$). The correlation coefficients between voxels in functional states, reflecting the relations between neural clusters during neural information processing, usually vary across different cognitive processes and individuals. Moreover, the neural correlation usually concentrates on specific sub-intervals rather than pervades the whole interval of $\left[-1,1\right]$. Therefore, we further normalize the mean Pearson correlation to ensure the generality of our analysis (here, we normalize the coefficients based on their absolute values and maintain their signs). In Fig. \ref{G9d}, we visualize the relations between the estimated information quantities and the normalized mean correlation coefficients (NMC). When NMC approaches $0$, it can be seen that the cases where $\widehat{\mathcal{I}}\left(\mathsf{X}_{\left(t\right)}^{\prime};\mathsf{Y}_{\left(t\right)}\right)\leq\widehat{\mathcal{I}}\left(\mathsf{X}_{\left(t\right)},\mathsf{Y}_{\left(t\right)}\right)$ frequently occur and the $3$-order mutual information drops sharply. This phenomenon inspires us to investigate the variation trend of $\widehat{\mathcal{I}}\left(\mathsf{X}_{\left(t\right)}^{\prime};\mathsf{Y}_{\left(t\right)}\right)-\widehat{\mathcal{I}}\left(\mathsf{X}_{\left(t\right)},\mathsf{Y}_{\left(t\right)}\right)$ as a function of the absolute value of NMC. We implement binning on $\vert\text{NMC}\vert$ following the Freedman-Diaconis approach \cite{freedman1981histogram}, which maintains relative robustness on the non-smooth data. Then, we exclude the bins where no more than $5$ sample exist to obtain the filtered data (covers $99.48\%$ of the raw data and excludes $0.52\%$ of the raw data as outliers). Our results in Fig. \ref{G9e} suggest that the random variable $\widehat{\mathcal{I}}\left(\mathsf{X}_{\left(t\right)}^{\prime};\mathsf{Y}_{\left(t\right)}\right)-\widehat{\mathcal{I}}\left(\mathsf{X}_{\left(t\right)},\mathsf{Y}_{\left(t\right)}\right)\leq\nu$ ($\nu\in\left[-0.1,0\right)$) follows specific power law distributions $\mathcal{P}\left(\cdot\leq \nu\right)\propto\vert\text{NMC}\vert^{-\lambda}$ (here $\lambda\in\left[0.1149,0.9899\right]$, it varies across different $\nu$). Meanwhile, we discover that the random variable $\widehat{\mathcal{I}}\left(\mathsf{X}_{\left(t\right)}^{\prime};\mathsf{Y}_{\left(t\right)}\right)-\widehat{\mathcal{I}}\left(\mathsf{X}_{\left(t\right)},\mathsf{Y}_{\left(t\right)}\right)\geq\rho$ ($\rho\in\left[0,0.001\right]$) follows a probability distribution $\mathcal{P}\left(\cdot\geq \rho\right)=\alpha\vert\text{NMC}\vert+\beta$ (here $\alpha\in\left[0.2507,0.3470\right]$ and $\beta\in\left[0.3019,0.3989\right]$) in Fig. \ref{G9f}. The probability density function fitting of these results obtains reasonable fitting accuracy (please see Fig. \ref{G9e}-\ref{G9f}). In Fig. \ref{G9f}, we illustrate a representative example of the fitted probability distributions $\mathcal{P}\left(\cdot\leq \nu\right)$ and $\mathcal{P}\left(\cdot\geq \nu\right)$. One can see that the probability $\mathcal{P}\left(\cdot\leq \nu\right)$ ($\nu<0$) significantly decreases along with the increasing intra-system coupling strength (quantified by $\vert\text{NMC}\vert$) while the probability $\mathcal{P}\left(\cdot\geq \rho\right)$ ($\rho\geq 0$) shows an opposite pattern. The finding is in consistency with our theory that a system with stronger intra-system coupling more possibly allow specific sub-systems to encode more information than itself.
 
 To this point, we have demonstrated our theoretical results in the brain, a non-isolated system of neurons that has complex intra-system coupling and is coupled with external stimuli. Although we can not calculate (\ref{EQ1}) directly, the indirect analysis demonstrates that the encoded information in a sub-system of neural tissues is not necessarily bound by the irreversible work of the whole system. In the brain, the possibility for a sub-system to encode more stimulus information than the whole system is modified by the neural correlation (the intra-system coupling strength). Suppose we consider the observer perspective discussed in Subsec. \ref{Sec4a}, a similar conclusion can be drawn that a specific amount of encoded information may be hidden in sub-systems.
 
 \section{Discussion}
  The current research pursues to explore the fundamental role of the order inside a non-isolated system in shaping the information thermodynamics coupling between this system and the environment. We begin with a thermodynamic perspective to characterize an arbitrary non-isolated system $\mathsf{X}$ as an information thermodynamics encoder when this system is coupled with an external source $\mathsf{Y}$. The information thermodynamics encoder encodes the information of external source utilizing thermodynamics, providing a unified angle to analyze the inter-system coupling between $\mathsf{X}$ and $\mathsf{Y}$ in term of information and thermodynamics. Rather than stand alone, our idea is rooted in extensive explorations of the physics nature of information (please see \cite{parrondo2015thermodynamics} for a systematic review). Furthermore, similarities and differences coexist between the proposed information thermodynamics encoder and information engines \cite{diana2013finite,mandal2013maxwell,mandal2012work,lu2014engineering,chapman2015autonomous}. Although these two kinds of systems both bridge information and thermodynamics, information thermodynamics encoder mainly focuses on the thermodynamic costs of perceiving external information rather than the potential of information to function as specific thermodynamic fuel \cite{boyd2017correlation,boyd2016maxwell,strasberg2013thermodynamics,barato2013autonomous,hoppenau2014energetics,um2015total,merhav2015sequence}.
  
 To formalize the encoding process with solid physics backgrounds, we try to find an appropriate way to derive the thermodynamics of encoding based on the nonequilibrium second law of thermodynamics. One can find a reverse derivation process from information quantities to the nonequilibrium second law of thermodynamics and fluctuation theorem \cite{sagawa2013role,sagawa2012fluctuation}. Although our proposed encoding process is suggested as a possible generalization of the measurement process \cite{parrondo2015thermodynamics,horowitz2013imitating,sagawa2009minimal,granger2011thermodynamic,sagawa2013role,hasegawa2010generalization,esposito2011second}, we need to emphasize that there exist essential consistency between these two concepts. Given the thermodynamic definition of encoding, a fundamental law is suggested that the mutual information $\mathcal{I}\left(\mathsf{X},\mathsf{Y}\right)$ (the encoded information of $\mathsf{Y}$ by system $\mathsf{X}$ entirety) is bounded by the irreversible work $\mathcal{W}_{\mathsf{irr}}$ from system $\left(\mathsf{X},\mathsf{Y}\right)$ based on (\ref{EQ8}). This upper bound is implied by the nonequilibrium second law of thermodynamics intrinsically \cite{esposito2011second}.
 
 Built on the above foundations, we turn to verify if this bound is followed by the encoded information of $\mathsf{Y}$ in an arbitrary sub-system of $\mathsf{X}$ as well. The necessity to study this problem lies in that it provides an opportunity to glance at the effects of internal orders inside $\mathsf{X}$ on the information thermodynamics link between $\mathsf{X}$ and $\mathsf{Y}$. Despite the substantial progress in exploring the physics of information \cite{parrondo2015thermodynamics}, the critical roles of these internal orders remain elusive. Exploring this problem may help understand information thermodynamics in complex systems, where the diversities of information thermodynamics characteristics might emerge on multiple scales due to heterogeneous elements and intricate internal correlations. As a starting point, the present study is limited at a qualitative level. Combining the $3$-order multiple mutual information and the theory of information synergy and redundancy \cite{mcgill1954multivariate,schneidman2003network}, the proposed thermodynamics of encoding suggests an intrinsic difference between the non-isolated system with internal correlations and the non-isolated system of independent elements when they act as information thermodynamics encoders. Unlike those with independent elements, an encoder $\mathsf{X}$ with internal correlations allows the encoded information of $\mathsf{Y}$ in its sub-systems to exceed the information thermodynamics bound on the joint system $\left(\mathsf{X},\mathsf{Y}\right)$. More specifically, the encoded information of $\mathsf{Y}$ in an arbitrary sub-system $\mathsf{X}^{\prime}$ is not necessarily bound by the irreversible work $\mathcal{W}_{\mathsf{irr}}$ from system $\left(\mathsf{X},\mathsf{Y}\right)$ following (\ref{EQ8}). This difference may originate from the nature of order inside the system $\mathsf{X}$. There is frequently information synergy in the non-isolated system of independent elements; in comparison, there can be either information synergy or information redundancy in the non-isolated system with intra-system coupling. These theoretical findings can be mathematically derived utilizing the Yeung's inequality \cite{yeung1991new}. Furthermore, we have computationally verified them in an Ising model with a random external field and a real data set of the human brain during the perception process. Our analysis demonstrates that the stronger internal correlation inside these systems can create a higher possibility for their sub-systems to encode more information than the global one.
 
 The essential idea that guides our research is to analyze the thermodynamic coupling relation between a non-isolated system and an external source from the information-theoretical perspective. Similar ideas are pervaded in various physics disciplines and become the foundation of information thermodynamics thanks to Szilard's \cite{szilard1929entropieverminderung} and Landauer's \cite{landauer1991information,landauer1996physical} inspiring works. 
  
Recently, this idea has received increasing attention in neuroscience. Neuroscientists have identified the fundamental role of information thermodynamics in supporting brain functions \cite{collell2015brain,street2016neurobiology,street2020upper,sengupta2013information,sartori2014thermodynamic,mlodinow2014relation}. Historically, the physics mechanisms underlying brain functions used to be elusive \cite{dayan2001theoretical}. The brain is a system with complex topology \cite{lynn2019physics}, geometry \cite{zheng2020geometric}, and dynamics \cite{chialvo2010emergent}, which creating numerous obstacles to study how the brain processes information. The application of information thermodynamics in neuroscience may help overcome these obstacles, suggesting a potential direction to understand neural information processing at a physically fundamental level \cite{collell2015brain}. Based on previous studies, our research moves one more step further to explore the information thermodynamics nature of the brain. As a non-isolated system of neurons, the brain is demonstrated to act as an information thermodynamics encoder with intra-system coupling. By deriving the thermodynamics of encoding based on the nonequilibrium second law of thermodynamics, our theory helps reveal the physics foundation of how the brain encodes the information of an external source. This framework might contribute to understanding the intrinsic relations between cognition and the physical brain \cite{collell2015brain,parrondo2015thermodynamics}. Moreover, we combine information thermodynamics with the theory of information synergy and redundancy to implement a unified analysis. We demonstrate that the intra-system coupling within a system allows the encoded information in specific sub-systems to exceed the information thermodynamics bound on the whole system. In other words, more information of the external source can be encoded (or hidden) in certain sub-systems in comparison with the whole system. The emergence probability of this phenomenon is affected by the internal correlation strength inside the brain positively. We suggest that this finding may provide insights into the synergy and labor division among neurons and cortices during neural information processing. Classically, the decomposition of the information encoded by a system of elements is implemented utilizing information synergy and redundancy. The possible significance of synergy and redundancy in the neural system has been discussed for decades, discovering various effects of redundant or synergistic collective dynamics on neural information processing \cite{schneidman2003network,gawne1993independent,brenner2000synergy,palm1988significance}. Our research further connects these previous theoretical findings with the thermodynamics of encoding, revealing the thermodynamic foundation of information synergy and redundancy during the encoding process. This finding may offer a possible explanation for the stimulus-dependent superiority of specific neural or cortical tissues in encoding efficiency comparing with the whole brain. For instance, the superiority of the visual cortex during visual perception \cite{grill2004human} not only means the relative redundancy of other sensory cortices but also corresponds to the larger encoded information quantity that is not restricted by the information thermodynamics bound of the whole brain. In sum, the theory depicted here features the potential to be further applied in neuroscience and related fields, which may help deepen our understanding of the physics foundation of neural information processing.

When we turn to a more general perspective about thermodynamics and information, one might find that limitations and possible insights coexist in our theory. Under the classical framework of information thermodynamics, our theory helps generalize the measurement process \cite{parrondo2015thermodynamics,horowitz2013imitating,sagawa2009minimal,granger2011thermodynamic,sagawa2013role,hasegawa2010generalization,esposito2011second}, one of the basic concepts that bridge information and thermodynamics, to the encoding process. By doing so, we can analyze the origin of mutual information from the nonequilibrium second law of thermodynamics in a more general aspect. Built on this framework, a new perspective is suggested to study the information thermodynamics connection that originates from the coupling between a non-isolated system and its environment. This perspective helps reveal how the intra-system attributes of a non-isolated system shape the information thermodynamics coupling between this system and the environment. The information thermodynamics difference discovered between the non-isolated system with internal correlations and the non-isolated system of independent elements is independent of detailed system definitions and free to be applied to any real system. However, we need to emphasize that our current work remains at a qualitative level. Although the $3$-order multiple mutual information enables us to analyze the internal orders of a non-isolated system in terms of information synergy and redundancy \cite{mcgill1954multivariate,schneidman2003network}, it fails to offer a more precise and quantitative characterization of these orders. This deficiency remains an open challenge for future studies. By overcoming this challenge, our suggested perspective may help explore more information thermodynamics characteristics determined by the nature of order inside the non-isolated system or the inter-system coupling relations. Last but not least, an idea that may interest both physicists and computer scientists relates to the characterized information thermodynamics encoder in our research, which functions as a kind of generalized computer or memory system \cite{parrondo2015thermodynamics}. Given the arbitrariness of non-isolated system selection for defining the encoding process, it is possible for any system with multiple distinguishable states to function as a generalized computer (see similar ideas in thermodynamic computing \cite{hylton2021vision,parrondo2015thermodynamics,still2012thermodynamics}). 

To conclude, the information thermodynamics link that originates from system coupling is explored in our research. We pursue a thermodynamics theory of encoding and propose a novel idea to analyze an arbitrary non-isolated system as an information thermodynamics encoder. This perspective indicates the critical role of the nature of order inside the non-isolated system in shaping information thermodynamics coupling. The presented theoretical findings might lay a cornerstone for the future exploration of various related topics.

\section{Acknowledgements}
The authors are grateful to Prof. Oren Raz for his inspiring suggestion on the research and valuable comments on the manuscript. Y.T appreciates the support of the YutChun Program. Moreover, the authors appreciate Mr. Ben Hou and Mr. Tianyi Wang for their proofreading on mathematics. 

This research is supported by the Artificial and General Intelligence Research Program of Guo Qiang Research Institute at Tsinghua University (2020GQG1017) as well as the Tsinghua University Initiative Scientific Research Program. 
 \begin{appendix}
 \section{The relative entropy and the mutual information}\label{ASec1}
Here we show that for each $\tau$
 \begin{align}
    &\mathcal{D}\big[\mathcal{P}\left(\mathsf{X}_{\left(\tau\right)},\mathsf{Y}_{\left(\tau\right)}\right)\big\Vert\widehat{\mathcal{P}}\left(\mathsf{X}_{\left(\tau\right)}\right)\widehat{\mathcal{P}}\left(\mathsf{Y}_{\left(\tau\right)}\right)\big]\notag\\
    =&\sum_{\Omega\left(\mathsf{X}\right)\times\Omega\left(\mathsf{Y}\right)}\mathcal{P}\left(\mathsf{X}_{\left(\tau\right)},\mathsf{Y}_{\left(\tau\right)}\right)\log\left(\frac{\mathcal{P}\left(\mathsf{X}_{\left(\tau\right)},\mathsf{Y}_{\left(\tau\right)}\right)}{\mathcal{P}\left(\mathsf{X}_{\left(\tau\right)}\right)\mathcal{P}\left(\mathsf{Y}_{\left(\tau\right)}\right)}\right)+\notag\\
    &\sum_{\Omega\left(\mathsf{X}\right)\times\Omega\left(\mathsf{Y}\right)}\mathcal{P}\left(\mathsf{X}_{\left(\tau\right)},\mathsf{Y}_{\left(\tau\right)}\right)\log\left(\frac{\mathcal{P}\left(\mathsf{X}_{\left(\tau\right)}\right)\mathcal{P}\left(\mathsf{Y}_{\left(\tau\right)}\right)}{\widehat{\mathcal{P}}\left(\mathsf{X}_{\left(\tau\right)}\right)\widehat{\mathcal{P}}\left(\mathsf{Y}_{\left(\tau\right)}\right)}\right)\label{A1}\\
    =&\mathcal{I}\left(\mathsf{X}_{\left(\tau\right)};\mathsf{Y}_{\left(\tau\right)}\right)+\sum_{\Omega\left(\mathsf{X}\right)\times\Omega\left(\mathsf{Y}\right)}\mathcal{P}\left(\mathsf{X}_{\left(\tau\right)},\mathsf{Y}_{\left(\tau\right)}\right)\log\left(\frac{\mathcal{P}\left(\mathsf{X}_{\left(\tau\right)}\right)\mathcal{P}\left(\mathsf{Y}_{\left(\tau\right)}\right)}{\widehat{\mathcal{P}}\left(\mathsf{X}_{\left(\tau\right)}\right)\widehat{\mathcal{P}}\left(\mathsf{Y}_{\left(\tau\right)}\right)}\right),\label{A2}
\end{align}
 Then, one can reformulate
 \begin{align}
    &\sum_{\Omega\left(\mathsf{X}\right)\times\Omega\left(\mathsf{Y}\right)}\mathcal{P}\left(\mathsf{X}_{\left(\tau\right)},\mathsf{Y}_{\left(\tau\right)}\right)\log\left(\frac{\mathcal{P}\left(\mathsf{X}_{\left(\tau\right)}\right)\mathcal{P}\left(\mathsf{Y}_{\left(\tau\right)}\right)}{\widehat{\mathcal{P}}\left(\mathsf{X}_{\left(\tau\right)}\right)\widehat{\mathcal{P}}\left(\mathsf{Y}_{\left(\tau\right)}\right)}\right)\notag\\
    =&-\sum_{\Omega\left(\mathsf{X}\right)\times\Omega\left(\mathsf{Y}\right)}\mathcal{P}\left(\mathsf{X}_{\left(\tau\right)},\mathsf{Y}_{\left(\tau\right)}\right)\log\left(\frac{\widehat{\mathcal{P}}\left(\mathsf{X}_{\left(\tau\right)}\right)\widehat{\mathcal{P}}\left(\mathsf{Y}_{\left(\tau\right)}\right)}{\mathcal{P}\left(\mathsf{X}_{\left(\tau\right)}\right)\mathcal{P}\left(\mathsf{Y}_{\left(\tau\right)}\right)}\right)\label{A3},
\end{align}
and prove the following derivations 
 \begin{align}
    &-\sum_{\Omega\left(\mathsf{X}\right)\times\Omega\left(\mathsf{Y}\right)}\mathcal{P}\left(\mathsf{X}_{\left(\tau\right)},\mathsf{Y}_{\left(\tau\right)}\right)\log\left(\frac{\widehat{\mathcal{P}}\left(\mathsf{X}_{\left(\tau\right)}\right)\widehat{\mathcal{P}}\left(\mathsf{Y}_{\left(\tau\right)}\right)}{\mathcal{P}\left(\mathsf{X}_{\left(\tau\right)}\right)\mathcal{P}\left(\mathsf{Y}_{\left(\tau\right)}\right)}\right)\notag\\
    =&-\sum_{\Omega\left(\mathsf{X}\right)\times\Omega\left(\mathsf{Y}\right)}\mathcal{P}\left(\mathsf{X}_{\left(\tau\right)},\mathsf{Y}_{\left(\tau\right)}\right)\log\left(\frac{\widehat{\mathcal{P}}\left(\mathsf{X}_{\left(\tau\right)}\right)}{\mathcal{P}\left(\mathsf{X}_{\left(\tau\right)},\mathsf{Y}_{\left(\tau\right)}\right)}\right)-\sum_{\Omega\left(\mathsf{X}\right)\times\Omega\left(\mathsf{Y}\right)}\mathcal{P}\left(\mathsf{X}_{\left(\tau\right)},\mathsf{Y}_{\left(\tau\right)}\right)\log\widehat{\mathcal{P}}\left(\mathsf{Y}_{\left(\tau\right)}\right)-\notag\\&\sum_{\Omega\left(\mathsf{X}\right)\times\Omega\left(\mathsf{Y}\right)}\mathcal{P}\left(\mathsf{X}_{\left(\tau\right)},\mathsf{Y}_{\left(\tau\right)}\right)\log\left(\frac{\mathcal{P}\left(\mathsf{X}_{\left(\tau\right)},\mathsf{Y}_{\left(\tau\right)}\right)}{\mathcal{P}\left(\mathsf{X}_{\left(\tau\right)}\right)\mathcal{P}\left(\mathsf{Y}_{\left(\tau\right)}\right)}\right)\label{A4}.
\end{align}
Combining (\ref{A3}) and (\ref{A4}), we can apply the Jensen inequality \cite{cover1999elements} to obtain
\begin{align}
 &\sum_{\Omega\left(\mathsf{X}\right)\times\Omega\left(\mathsf{Y}\right)}\mathcal{P}\left(\mathsf{X}_{\left(\tau\right)},\mathsf{Y}_{\left(\tau\right)}\right)\log\left(\frac{\mathcal{P}\left(\mathsf{X}_{\left(\tau\right)}\right)\mathcal{P}\left(\mathsf{Y}_{\left(\tau\right)}\right)}{\widehat{\mathcal{P}}\left(\mathsf{X}_{\left(\tau\right)}\right)\widehat{\mathcal{P}}\left(\mathsf{Y}_{\left(\tau\right)}\right)}\right)\notag\\
\geq&-\log\sum_{\Omega\left(\mathsf{X}\right)\times\Omega\left(\mathsf{Y}\right)}\widehat{\mathcal{P}}\left(\mathsf{X}_{\left(\tau\right)}\right)-\sum_{\Omega\left(\mathsf{X}\right)\times\Omega\left(\mathsf{Y}\right)}\mathcal{P}\left(\mathsf{X}_{\left(\tau\right)},\mathsf{Y}_{\left(\tau\right)}\right)\log\widehat{\mathcal{P}}\left(\mathsf{Y}_{\left(\tau\right)}\right)-\notag\\&\sum_{\Omega\left(\mathsf{X}\right)\times\Omega\left(\mathsf{Y}\right)}\mathcal{P}\left(\mathsf{X}_{\left(\tau\right)},\mathsf{Y}_{\left(\tau\right)}\right)\log\left(\frac{\mathcal{P}\left(\mathsf{X}_{\left(\tau\right)},\mathsf{Y}_{\left(\tau\right)}\right)}{\mathcal{P}\left(\mathsf{X}_{\left(\tau\right)}\right)\mathcal{P}\left(\mathsf{Y}_{\left(\tau\right)}\right)}\right).\label{A5}
\end{align}
One can immediately find that the first term in (\ref{A5}) satisfies $-\log\sum_{\Omega\left(\mathsf{X}\right)\times\Omega\left(\mathsf{Y}\right)}\widehat{\mathcal{P}}\left(\mathsf{X}_{\left(\tau\right)}\right)=0$. Therefore, we have 
\begin{align}
 &\sum_{\Omega\left(\mathsf{X}\right)\times\Omega\left(\mathsf{Y}\right)}\mathcal{P}\left(\mathsf{X}_{\left(\tau\right)},\mathsf{Y}_{\left(\tau\right)}\right)\log\left(\frac{\mathcal{P}\left(\mathsf{X}_{\left(\tau\right)}\right)\mathcal{P}\left(\mathsf{Y}_{\left(\tau\right)}\right)}{\widehat{\mathcal{P}}\left(\mathsf{X}_{\left(\tau\right)}\right)\widehat{\mathcal{P}}\left(\mathsf{Y}_{\left(\tau\right)}\right)}\right)\notag\\
\geq&-\sum_{\Omega\left(\mathsf{X}\right)\times\Omega\left(\mathsf{Y}\right)}\mathcal{P}\left(\mathsf{X}_{\left(\tau\right)},\mathsf{Y}_{\left(\tau\right)}\right)\log\widehat{\mathcal{P}}\left(\mathsf{Y}_{\left(\tau\right)}\right)-\sum_{\Omega\left(\mathsf{X}\right)\times\Omega\left(\mathsf{Y}\right)}\mathcal{P}\left(\mathsf{X}_{\left(\tau\right)},\mathsf{Y}_{\left(\tau\right)}\right)\log\left(\frac{\mathcal{P}\left(\mathsf{X}_{\left(\tau\right)},\mathsf{Y}_{\left(\tau\right)}\right)}{\mathcal{P}\left(\mathsf{X}_{\left(\tau\right)}\right)\mathcal{P}\left(\mathsf{Y}_{\left(\tau\right)}\right)}\right)\label{A6}\\
\geq&\zeta-\mathcal{I}\left(\mathsf{X}_{\left(\tau\right)};\mathsf{Y}_{\left(\tau\right)}\right)\label{A7},
\end{align}
where we mark
\begin{align}
 \zeta=-\sum_{\Omega\left(\mathsf{X}\right)\times\Omega\left(\mathsf{Y}\right)}\mathcal{P}\left(\mathsf{X}_{\left(\tau\right)},\mathsf{Y}_{\left(\tau\right)}\right)\log\widehat{\mathcal{P}}\left(\mathsf{Y}_{\left(\tau\right)}\right).\label{A8}
\end{align}
To analyze the value range of $\zeta$, we define a measure $\mu\left[\Omega\left(\mathsf{X}\right)\right]=m\left[\Omega\left(\mathsf{X}\right)\right]m^{-1}\left[\Omega\left(\mathsf{X}\right)\times\Omega\left(\mathsf{Y}\right)\right]$, where $m\left(\cdot\right)$ denotes the area (the area quantities of discrete sets can be obtained by element counting). Based on this measure, one can see that $\sum_{\Omega\left(\mathsf{X}\right)\times\Omega\left(\mathsf{Y}\right)}\widehat{\mathcal{P}}\left(\mathsf{Y}_{\left(\tau\right)}\right)=\mu\left[\Omega\left(\mathsf{X}\right)\right]$. Then, we can reformulate (\ref{A8}) as
\begin{align}
 \zeta=&-\sum_{\Omega\left(\mathsf{X}\right)\times\Omega\left(\mathsf{Y}\right)}\mathcal{P}\left(\mathsf{X}_{\left(\tau\right)},\mathsf{Y}_{\left(\tau\right)}\right)\log\Bigg\lbrace\frac{\widehat{\mathcal{P}}\left(\mathsf{Y}_{\left(\tau\right)}\right)}{\mu\left[\Omega\left(\mathsf{X}\right)\right]}\mu\left[\Omega\left(\mathsf{X}\right)\right]\Bigg\rbrace\label{A9}\\
 =&-\sum_{\Omega\left(\mathsf{X}\right)\times\Omega\left(\mathsf{Y}\right)}\mathcal{P}\left(\mathsf{X}_{\left(\tau\right)},\mathsf{Y}_{\left(\tau\right)}\right)\log\frac{\widehat{\mathcal{P}}\left(\mathsf{Y}_{\left(\tau\right)}\right)}{\mu\left[\Omega\left(\mathsf{X}\right)\right]}-\sum_{\Omega\left(\mathsf{X}\right)\times\Omega\left(\mathsf{Y}\right)}\mathcal{P}\left(\mathsf{X}_{\left(\tau\right)},\mathsf{Y}_{\left(\tau\right)}\right)\log\mu\left[\Omega\left(\mathsf{X}\right)\right].\label{A10}
\end{align}
It is trivial to know that $\sum_{\Omega\left(\mathsf{X}\right)\times\Omega\left(\mathsf{Y}\right)}\frac{\widehat{\mathcal{P}}\left(\mathsf{Y}_{\left(\tau\right)}\right)}{\mu\left[\Omega\left(\mathsf{X}\right)\right]}=1$. Therefore, we can apply the Gibbs inequality \cite{cover1999elements} on the first term of (\ref{A10}) to prove that
\begin{align}
 \zeta\geq& -\sum_{\Omega\left(\mathsf{X}\right)\times\Omega\left(\mathsf{Y}\right)}\mathcal{P}\left(\mathsf{X}_{\left(\tau\right)},\mathsf{Y}_{\left(\tau\right)}\right)\log\mathcal{P}\left(\mathsf{X}_{\left(\tau\right)},\mathsf{Y}_{\left(\tau\right)}\right)-\sum_{\Omega\left(\mathsf{X}\right)\times\Omega\left(\mathsf{Y}\right)}\mathcal{P}\left(\mathsf{X}_{\left(\tau\right)},\mathsf{Y}_{\left(\tau\right)}\right)\log\mu\left[\Omega\left(\mathsf{X}\right)\right]\label{A11}\\\geq&\mathcal{S}\left(\mathsf{X}_{\left(\tau\right)},\mathsf{Y}_{\left(\tau\right)}\right)-\sum_{\Omega\left(\mathsf{X}\right)\times\Omega\left(\mathsf{Y}\right)}\mathcal{P}\left(\mathsf{X}_{\left(\tau\right)},\mathsf{Y}_{\left(\tau\right)}\right)\log\mu\left[\Omega\left(\mathsf{X}\right)\right].\label{A12}
\end{align}
As for the second term of (\ref{A12}), its non-negativity is trivial because $\mu\left[\Omega\left(\mathsf{X}\right)\right]\leq 1$ (the equality is satisfied if and only if $m\left[\Omega\left(\mathsf{X}\right)\right]=m\left[\Omega\left(\mathsf{X}\right)\times\Omega\left(\mathsf{Y}\right)\right]$). Therefore, we can know
\begin{align}
 \zeta\geq\mathcal{S}\left(\mathsf{X}_{\left(\tau\right)},\mathsf{Y}_{\left(\tau\right)}\right).\label{A13}
\end{align}
Based on (\ref{A7}) and (\ref{A13}), one can derive
\begin{align}
 &\sum_{\Omega\left(\mathsf{X}\right)\times\Omega\left(\mathsf{Y}\right)}\mathcal{P}\left(\mathsf{X}_{\left(\tau\right)},\mathsf{Y}_{\left(\tau\right)}\right)\log\left(\frac{\mathcal{P}\left(\mathsf{X}_{\left(\tau\right)}\right)\mathcal{P}\left(\mathsf{Y}_{\left(\tau\right)}\right)}{\widehat{\mathcal{P}}\left(\mathsf{X}_{\left(\tau\right)}\right)\widehat{\mathcal{P}}\left(\mathsf{Y}_{\left(\tau\right)}\right)}\right)\notag\\
\geq&\mathcal{S}\left(\mathsf{X}_{\left(\tau\right)},\mathsf{Y}_{\left(\tau\right)}\right)-\mathcal{I}\left(\mathsf{X}_{\left(\tau\right)};\mathsf{Y}_{\left(\tau\right)}\right)\label{A14}\\\geq&0.\label{A15}
\end{align}
To conclude, we can ensure that the term in (\ref{A2}) is non-negative. This result further implies that $\mathcal{D}\big[\mathcal{P}\left(\mathsf{X}_{\left(t\right)},\mathsf{Y}_{\left(t\right)}\right)\big\Vert\widehat{\mathcal{P}}\left(\mathsf{X}_{\left(t\right)}\right)\widehat{\mathcal{P}}\left(\mathsf{Y}_{\left(t\right)}\right)\big]\geq\mathcal{I}\big(\mathsf{X}_{\left(t\right)};\mathsf{Y}_{\left(t\right)}\big)$ (please see (\ref{EQ8}) in our main text).

\section{The \textbf{UCE Condition} and the mutual information}\label{ASec2}
Assume that the thermodynamics of encoding defined in (\ref{EQ7}-\ref{EQ8}) is independent of the \textbf{UCE Condition}, then $\mathcal{D}\big[\mathcal{P}\left(\mathsf{X}_{\left(\tau\right)},\mathsf{Y}_{\left(\tau\right)}\right)\big\Vert\widehat{\mathcal{P}}\left(\mathsf{X}_{\left(\tau\right)},\mathsf{Y}_{\left(\tau\right)}\right)\big]$ is not necessarily equivalent to $\mathcal{D}\big[\mathcal{P}\left(\mathsf{X}_{\left(\tau\right)},\mathsf{Y}_{\left(\tau\right)}\right)\big\Vert\widehat{\mathcal{P}}\left(\mathsf{X}_{\left(\tau\right)}\right)\widehat{\mathcal{P}}\left(\mathsf{Y}_{\left(\tau\right)}\right)\big]$. The derivations of $\mathcal{D}\big[\mathcal{P}\left(\mathsf{X}_{\left(t\right)},\mathsf{Y}_{\left(t\right)}\right)\big\Vert\widehat{\mathcal{P}}\left(\mathsf{X}_{\left(t\right)}\right)\widehat{\mathcal{P}}\left(\mathsf{Y}_{\left(t\right)}\right)\big]\geq\mathcal{I}\big(\mathsf{X}_{\left(t\right)};\mathsf{Y}_{\left(t\right)}\big)$ do not necessarily hold under this condition, remaining for further verification.  

Following the idea in appendix \ref{ASec2}, we know that for each $\tau$
 \begin{align}
    &\mathcal{D}\big[\mathcal{P}\left(\mathsf{X}_{\left(\tau\right)},\mathsf{Y}_{\left(\tau\right)}\right)\big\Vert\widehat{\mathcal{P}}\left(\mathsf{X}_{\left(\tau\right)},\mathsf{Y}_{\left(\tau\right)}\right)\big]\notag\\
    =&\mathcal{I}\left(\mathsf{X}_{\left(\tau\right)};\mathsf{Y}_{\left(\tau\right)}\right)-\sum_{\Omega\left(\mathsf{X}\right)\times\Omega\left(\mathsf{Y}\right)}\mathcal{P}\left(\mathsf{X}_{\left(\tau\right)},\mathsf{Y}_{\left(\tau\right)}\right)\log\left(\frac{\widehat{\mathcal{P}}\left(\mathsf{X}_{\left(\tau\right)},\mathsf{Y}_{\left(\tau\right)}\right)}{\mathcal{P}\left(\mathsf{X}_{\left(\tau\right)}\right)\mathcal{P}\left(\mathsf{Y}_{\left(\tau\right)}\right)}\right),\label{B1}
\end{align}
where the second term of (\ref{B1}) is equivalent to
 \begin{align}
    &-\sum_{\Omega\left(\mathsf{X}\right)\times\Omega\left(\mathsf{Y}\right)}\mathcal{P}\left(\mathsf{X}_{\left(\tau\right)},\mathsf{Y}_{\left(\tau\right)}\right)\log\left(\frac{\widehat{\mathcal{P}}\left(\mathsf{X}_{\left(\tau\right)},\mathsf{Y}_{\left(\tau\right)}\right)}{\mathcal{P}\left(\mathsf{X}_{\left(\tau\right)}\right)\mathcal{P}\left(\mathsf{Y}_{\left(\tau\right)}\right)}\right)\notag\\
    =&-\sum_{\Omega\left(\mathsf{X}\right)\times\Omega\left(\mathsf{Y}\right)}\mathcal{P}\left(\mathsf{X}_{\left(\tau\right)},\mathsf{Y}_{\left(\tau\right)}\right)\log\left(\frac{\widehat{\mathcal{P}}\left(\mathsf{X}_{\left(\tau\right)},\mathsf{Y}_{\left(\tau\right)}\right)}{\mathcal{P}\left(\mathsf{X}_{\left(\tau\right)},\mathsf{Y}_{\left(\tau\right)}\right)}\right)-\notag\\&\sum_{\Omega\left(\mathsf{X}\right)\times\Omega\left(\mathsf{Y}\right)}\mathcal{P}\left(\mathsf{X}_{\left(\tau\right)},\mathsf{Y}_{\left(\tau\right)}\right)\log\left(\frac{\mathcal{P}\left(\mathsf{X}_{\left(\tau\right)},\mathsf{Y}_{\left(\tau\right)}\right)}{\mathcal{P}\left(\mathsf{X}_{\left(\tau\right)}\right)\mathcal{P}\left(\mathsf{Y}_{\left(\tau\right)}\right)}\right)\label{B2}\\
    =&-\sum_{\Omega\left(\mathsf{X}\right)\times\Omega\left(\mathsf{Y}\right)}\mathcal{P}\left(\mathsf{X}_{\left(\tau\right)},\mathsf{Y}_{\left(\tau\right)}\right)\log\left(\frac{\widehat{\mathcal{P}}\left(\mathsf{X}_{\left(\tau\right)},\mathsf{Y}_{\left(\tau\right)}\right)}{\mathcal{P}\left(\mathsf{X}_{\left(\tau\right)},\mathsf{Y}_{\left(\tau\right)}\right)}\right)-\mathcal{I}\left(\mathsf{X}_{\left(\tau\right)};\mathsf{Y}_{\left(\tau\right)}\right)\label{B3}\\
     \geq&-\log\sum_{\Omega\left(\mathsf{X}\right)\times\Omega\left(\mathsf{Y}\right)}\widehat{\mathcal{P}}\left(\mathsf{X}_{\left(\tau\right)},\mathsf{Y}_{\left(\tau\right)}\right)-\mathcal{I}\left(\mathsf{X}_{\left(\tau\right)};\mathsf{Y}_{\left(\tau\right)}\right)\label{B4}\\
    \geq&0-\mathcal{I}\left(\mathsf{X}_{\left(\tau\right)};\mathsf{Y}_{\left(\tau\right)}\right)\label{B5}.
\end{align}
Here (\ref{B4}) can be derived utilizing the Jensen inequality \cite{cover1999elements}. Combining (\ref{B1}) and (\ref{B5}), what we can prove is
 \begin{align}
    \mathcal{D}\big[\mathcal{P}\left(\mathsf{X}_{\left(\tau\right)},\mathsf{Y}_{\left(\tau\right)}\right)\big\Vert\widehat{\mathcal{P}}\left(\mathsf{X}_{\left(\tau\right)},\mathsf{Y}_{\left(\tau\right)}\right)\big]\geq 0\label{B6}
\end{align}
rather than the inequality (\ref{EQ8}) in our theory. The inequality (\ref{B6}) is trivial since the non-negativity of the relative entropy has already be known. To derive the stronger conclusion in (\ref{EQ8}), the \textbf{UCE Condition} should be included in our theory.

 \section{Simulation setting for Ising model}\label{ASec3}
 The following are the parameters used in our simulation for system $\left(\bm{\sigma},h\right)$.
    \begin{table}[ht]
\centering
\caption{Parameter settings in simulation}
\begin{tabular}{ccc}
\hline
\textbf{Parameter} & \textbf{Value} & \textbf{Meaning} \\
\hline
$t$	& $10000$ & Duration length\\
$\vert\bm{\sigma}\vert$	& $8$ & Number of spins\\
$m$	& $5$ & Maximum external field\\
$K_{J}$	& $10$ & Number of coupling strength conditions\\
$K_{T}$	& $10$ & Number of temperature conditions\\
$K_{\theta}$ & $10$ & Number of transition rate conditions\\
$\xi$	& $10\times 2^{\vert\bm{\sigma}\vert}$ & Times of repetition\\
\hline
\end{tabular}
\end{table}
Specifically, the coupling strength $J$, temperature $T$, and transition rate $\theta^{-1}$ conditions are set as following
\begin{table}[ht]
\centering
\caption{Condition settings in simulation}
\begin{tabular}{cc}
\hline
\textbf{Parameter} & \textbf{Value} \\
\hline
$J$	& $J\in\{0.1,0.5,1,1.5,2,2.5,3,3.5,4,4.5\}$ \\
$T$	& $T\in\{0.1,0.5,1,1.5,2,2.5,3,3.5,4,4.5\}$ \\
$\theta^{-1}$	& $\theta^{-1}\in\{1,\frac{1}{5},\frac{1}{10},\frac{1}{20},\frac{1}{50},\frac{1}{100},\frac{1}{200},\frac{1}{500},\frac{1}{1000},\frac{1}{2000}\}$ \\
\hline
\end{tabular}
\end{table}
Please note that the parameter $\theta$ measures the interval between two times of transitions of the external field. Its inverse $\theta^{-1}$ quantifies the transition speed.

In our experiment, we simulate the Ising model under each combination $\left(J,T,\theta^{-1}\right)$ to obtain the results. Moreover, we can treat $J$, $T$, and $\theta^{-1}$ as three parameter directions, respectively. By averaging experiment results in arbitrary two parameter directions, we can further analyze these results as the functions of the remaining parameter direction. 

\section{fMRI data acquisition and pre-processing}\label{ASec4}
The fMRI data set used in the present study is acquired from \cite{ds000105}, which has been used in several neuroscience studies \cite{haxby2001distributed,hanson2004combinatorial,o2005partially}. Neural responses, reflected by the fMRI BOLD signals, are measured from 6 subjects (5 females and 1 male) with gradient echo echoplanar imaging on a GE 3T scanner (General Electric, Milwaukee, WI). Repetition time (TR) is 2500 ms, echo time (TE) is 30 ms, the flip angle is 90, and the field of view (FOV) is 240 mm. In each time of repetition, 40 slices of 3.5-mm-thick sagittal images are obtained. T1-weighted spoiled gradient recall (SPGR) signals (124 slices of 1.2-mm-thick sagittal images) are obtained as the high-resolution information of detailed anatomy (please note that our analysis is implemented on the BOLD signals rather than the T1 signals). Stimuli are gray-scale images of 8 types of visual objects (faces, houses, cats, bottles, scissors, shoes, chairs, and the nonsense patterns that are generated as the phase-scrambled images of intact objects).  More details about the stimulus sequences in visual perception experiments can be seen in \cite{ds000105}. In our analysis, these stimuli are represented by the indices of their types.

Pre-processing of the BOLD signals is performed based on SPM12 (a MATLAB toolbox designed for brain imaging data analysis, one can see \url{https://www.fil.ion.ucl.ac.uk/spm/} for details), including slice-timing correction and motion correction. These corrections are standard workflows in BOLD signal pre-processing and have no significant influence on the observed phenomena in Sec. \ref{Sec4b}. As for the T1 signals, we have not implemented further pre-processing on them since the anatomy information is not necessary for our study.  

In general, our analysis in Sec. \ref{Sec4b} does not rely on the data set and pre-processing techniques critically. One can implement the same analysis on any neural data set involving the perception of multiple stimuli.
 \end{appendix}

\bibliographystyle{plain}
\bibliography{references}  

\begin{thebibliography}{100}

\bibitem{abeles1991corticonics}
Moshe Abeles.
\newblock {\em Corticonics: Neural circuits of the cerebral cortex}.
\newblock Cambridge University Press, 1991.

\bibitem{abreu2011extracting}
David Abreu and Udo Seifert.
\newblock Extracting work from a single heat bath through feedback.
\newblock {\em EPL (Europhysics Letters)}, 94(1):10001, 2011.

\bibitem{acharyya1998nonequilibrium}
Muktish Acharyya.
\newblock Nonequilibrium phase transition in the kinetic ising model: Dynamical
  symmetry breaking by randomly varying magnetic field.
\newblock {\em Physical Review E}, 58(1):174, 1998.

\bibitem{andrieux2008nonequilibrium}
David Andrieux and Pierre Gaspard.
\newblock Nonequilibrium generation of information in copolymerization
  processes.
\newblock {\em Proceedings of the National Academy of Sciences},
  105(28):9516--9521, 2008.

\bibitem{barato2013autonomous}
A~Cardoso Barato and Udo Seifert.
\newblock An autonomous and reversible maxwell's demon.
\newblock {\em EPL (Europhysics Letters)}, 101(6):60001, 2013.

\bibitem{barato2014efficiency}
Andre~C Barato, David Hartich, and Udo Seifert.
\newblock Efficiency of cellular information processing.
\newblock {\em New Journal of Physics}, 16(10):103024, 2014.

\bibitem{berut2012experimental}
Antoine B{\'e}rut, Artak Arakelyan, Artyom Petrosyan, Sergio Ciliberto, Raoul
  Dillenschneider, and Eric Lutz.
\newblock Experimental verification of landauer’s principle linking
  information and thermodynamics.
\newblock {\em Nature}, 483(7388):187--189, 2012.

\bibitem{berut2013detailed}
Antoine B{\'e}rut, Artyom Petrosyan, and Sergio Ciliberto.
\newblock Detailed jarzynski equality applied to a logically irreversible
  procedure.
\newblock {\em EPL (Europhysics Letters)}, 103(6):60002, 2013.

\bibitem{bhanot1988metropolis}
Gyan Bhanot.
\newblock The metropolis algorithm.
\newblock {\em Reports on Progress in Physics}, 51(3):429, 1988.

\bibitem{boyd2016maxwell}
Alexander~B Boyd and James~P Crutchfield.
\newblock Maxwell demon dynamics: Deterministic chaos, the szilard map, and the
  intelligence of thermodynamic systems.
\newblock {\em Physical review letters}, 116(19):190601, 2016.

\bibitem{boyd2017correlation}
Alexander~B Boyd, Dibyendu Mandal, and James~P Crutchfield.
\newblock Correlation-powered information engines and the thermodynamics of
  self-correction.
\newblock {\em Physical Review E}, 95(1):012152, 2017.

\bibitem{brandner2016periodic}
Kay Brandner and Udo Seifert.
\newblock Periodic thermodynamics of open quantum systems.
\newblock {\em Physical Review E}, 93(6):062134, 2016.

\bibitem{brenner2000synergy}
Naama Brenner, Steven~P Strong, Roland Koberle, William Bialek, and Rob R de
  Ruyter~van Steveninck.
\newblock Synergy in a neural code.
\newblock {\em Neural computation}, 12(7):1531--1552, 2000.

\bibitem{callen1998thermodynamics}
Herbert~B Callen.
\newblock Thermodynamics and an introduction to thermostatistics, 1998.

\bibitem{campisi2009thermodynamics}
Michele Campisi, Peter Talkner, and Peter H{\"a}nggi.
\newblock Thermodynamics and fluctuation theorems for a strongly coupled open
  quantum system: an exactly solvable case.
\newblock {\em Journal of Physics A: Mathematical and Theoretical},
  42(39):392002, 2009.

\bibitem{cao2015thermodynamics}
Yuansheng Cao, Zongping Gong, and HT~Quan.
\newblock Thermodynamics of information processing based on enzyme kinetics: An
  exactly solvable model of an information pump.
\newblock {\em Physical Review E}, 91(6):062117, 2015.

\bibitem{chapman2015autonomous}
Adrian Chapman and Akimasa Miyake.
\newblock How an autonomous quantum maxwell demon can harness correlated
  information.
\newblock {\em Physical Review E}, 92(6):062125, 2015.

\bibitem{chialvo2010emergent}
Dante~R Chialvo.
\newblock Emergent complex neural dynamics.
\newblock {\em Nature physics}, 6(10):744--750, 2010.

\bibitem{churchland2010stimulus}
Mark~M Churchland, M~Yu Byron, John~P Cunningham, Leo~P Sugrue, Marlene~R
  Cohen, Greg~S Corrado, William~T Newsome, Andrew~M Clark, Paymon Hosseini,
  Benjamin~B Scott, et~al.
\newblock Stimulus onset quenches neural variability: a widespread cortical
  phenomenon.
\newblock {\em Nature neuroscience}, 13(3):369--378, 2010.

\bibitem{cipra1987introduction}
Barry~A Cipra.
\newblock An introduction to the ising model.
\newblock {\em The American Mathematical Monthly}, 94(10):937--959, 1987.

\bibitem{collell2015brain}
Guillem Collell and Jordi Fauquet.
\newblock Brain activity and cognition: a connection from thermodynamics and
  information theory.
\newblock {\em Frontiers in psychology}, 6:818, 2015.

\bibitem{cover1999elements}
Thomas~M Cover.
\newblock {\em Elements of information theory}.
\newblock John Wiley \& Sons, 1999.

\bibitem{dayan2001theoretical}
Peter Dayan and Laurence~F Abbott.
\newblock {\em Theoretical neuroscience: computational and mathematical
  modeling of neural systems}.
\newblock Computational Neuroscience Series, 2001.

\bibitem{diana2013finite}
Giovanni Diana, G~Baris Bagci, and Massimiliano Esposito.
\newblock Finite-time erasing of information stored in fermionic bits.
\newblock {\em Physical Review E}, 87(1):012111, 2013.

\bibitem{donald1987free}
Matthew~J Donald.
\newblock Free energy and the relative entropy.
\newblock {\em Journal of statistical physics}, 49(1):81--87, 1987.

\bibitem{dopfer2007general}
Kurt Dopfer and Jason Potts.
\newblock {\em The general theory of economic evolution}.
\newblock Routledge, 2007.

\bibitem{esposito2010entropy}
Massimiliano Esposito, Katja Lindenberg, and Christian Van~den Broeck.
\newblock Entropy production as correlation between system and reservoir.
\newblock {\em New Journal of Physics}, 12(1):013013, 2010.

\bibitem{esposito2011second}
Massimiliano Esposito and Christian Van~den Broeck.
\newblock Second law and landauer principle far from equilibrium.
\newblock {\em EPL (Europhysics Letters)}, 95(4):40004, 2011.

\bibitem{freedman1981histogram}
David Freedman and Persi Diaconis.
\newblock On the histogram as a density estimator: L 2 theory.
\newblock {\em Zeitschrift f{\"u}r Wahrscheinlichkeitstheorie und verwandte
  Gebiete}, 57(4):453--476, 1981.

\bibitem{gao2017estimating}
Weihao Gao, Sreeram Kannan, Sewoong Oh, and Pramod Viswanath.
\newblock Estimating mutual information for discrete-continuous mixtures.
\newblock In {\em Advances in neural information processing systems}, pages
  5986--5997, 2017.

\bibitem{gawne1993independent}
Timothy~J Gawne and Barry~J Richmond.
\newblock How independent are the messages carried by adjacent inferior
  temporal cortical neurons?
\newblock {\em Journal of Neuroscience}, 13(7):2758--2771, 1993.

\bibitem{gelin2009thermodynamics}
Maxim~F Gelin and Michael Thoss.
\newblock Thermodynamics of a subensemble of a canonical ensemble.
\newblock {\em Physical Review E}, 79(5):051121, 2009.

\bibitem{granger2011thermodynamic}
L{\'e}o Granger and Holger Kantz.
\newblock Thermodynamic cost of measurements.
\newblock {\em Physical Review E}, 84(6):061110, 2011.

\bibitem{grill2004human}
Kalanit Grill-Spector and Rafael Malach.
\newblock The human visual cortex.
\newblock {\em Annu. Rev. Neurosci.}, 27:649--677, 2004.

\bibitem{hanson2004combinatorial}
Stephen~Jos{\'e} Hanson, Toshihiko Matsuka, and James~V Haxby.
\newblock Combinatorial codes in ventral temporal lobe for object recognition:
  Haxby (2001) revisited: is there a “face” area?
\newblock {\em Neuroimage}, 23(1):156--166, 2004.

\bibitem{hasegawa2010generalization}
H-H Hasegawa, J~Ishikawa, K~Takara, and DJ~Driebe.
\newblock Generalization of the second law for a nonequilibrium initial state.
\newblock {\em Physics Letters A}, 374(8):1001--1004, 2010.

\bibitem{haxby2001distributed}
James~V Haxby, M~Ida Gobbini, Maura~L Furey, Alumit Ishai, Jennifer~L Schouten,
  and Pietro Pietrini.
\newblock Distributed and overlapping representations of faces and objects in
  ventral temporal cortex.
\newblock {\em Science}, 293(5539):2425--2430, 2001.

\bibitem{ds000105}
J.V. Haxby, M.I. Gobbini, M.L. Furey, A.~Ishai, J.L. Schouten, and P.~Pietrini.
\newblock "visual object recognition", 2016.

\bibitem{haynie2001biological}
Donald~T Haynie.
\newblock {\em Biological thermodynamics}.
\newblock Cambridge University Press, 2001.

\bibitem{heeger2002does}
David~J Heeger and David Ress.
\newblock What does fmri tell us about neuronal activity?
\newblock {\em Nature Reviews Neuroscience}, 3(2):142--151, 2002.

\bibitem{hoppenau2014energetics}
Johannes Hoppenau and Andreas Engel.
\newblock On the energetics of information exchange.
\newblock {\em EPL (Europhysics Letters)}, 105(5):50002, 2014.

\bibitem{horowitz2013imitating}
Jordan~M Horowitz, Takahiro Sagawa, and Juan~MR Parrondo.
\newblock Imitating chemical motors with optimal information motors.
\newblock {\em Physical review letters}, 111(1):010602, 2013.

\bibitem{hylton2021vision}
Todd Hylton, Thomas~M Conte, and Mark~D Hill.
\newblock A vision to compute like nature: thermodynamically.
\newblock {\em Communications of the ACM}, 64(6):35--38, 2021.

\bibitem{ito2018stochastic}
Sosuke Ito.
\newblock Stochastic thermodynamic interpretation of information geometry.
\newblock {\em Physical review letters}, 121(3):030605, 2018.

\bibitem{ito2013information}
Sosuke Ito and Takahiro Sagawa.
\newblock Information thermodynamics on causal networks.
\newblock {\em Physical review letters}, 111(18):180603, 2013.

\bibitem{jarzynski1999microscopic}
C~Jarzynski.
\newblock Microscopic analysis of clausius--duhem processes.
\newblock {\em Journal of statistical physics}, 96(1):415--427, 1999.

\bibitem{jarzynski2004nonequilibrium}
Chris Jarzynski.
\newblock Nonequilibrium work theorem for a system strongly coupled to a
  thermal environment.
\newblock {\em Journal of Statistical Mechanics: Theory and Experiment},
  2004(09):P09005, 2004.

\bibitem{jarzynski2008thermodynamics}
Christopher Jarzynski.
\newblock The thermodynamics of writing a random polymer.
\newblock {\em Proceedings of the National Academy of Sciences},
  105(28):9451--9452, 2008.

\bibitem{jarzynski2017stochastic}
Christopher Jarzynski.
\newblock Stochastic and macroscopic thermodynamics of strongly coupled
  systems.
\newblock {\em Physical Review X}, 7(1):011008, 2017.

\bibitem{jorgensen2004towards}
Sven~Erik Jorgensen and Yuri~M Svirezhev.
\newblock {\em Towards a thermodynamic theory for ecological systems}.
\newblock Elsevier, 2004.

\bibitem{jun2014high}
Yonggun Jun, Mom{\v{c}}ilo Gavrilov, and John Bechhoefer.
\newblock High-precision test of landauer’s principle in a feedback trap.
\newblock {\em Physical review letters}, 113(19):190601, 2014.

\bibitem{katchalsky1962thermodynamics}
A~Katchalsky and O~Kedem.
\newblock Thermodynamics of flow processes in biological systems.
\newblock {\em Biophysical journal}, 2(2 Pt 2):53, 1962.

\bibitem{kawai2007dissipation}
Ryoichi Kawai, Juan~MR Parrondo, and Christian Van~den Broeck.
\newblock Dissipation: The phase-space perspective.
\newblock {\em Physical review letters}, 98(8):080602, 2007.

\bibitem{kraskov2004estimating}
Alexander Kraskov, Harald St{\"o}gbauer, and Peter Grassberger.
\newblock Estimating mutual information.
\newblock {\em Physical review E}, 69(6):066138, 2004.

\bibitem{kuhl2003mass}
Ellen Kuhl and Paul Steinmann.
\newblock Mass--and volume--specific views on thermodynamics for open systems.
\newblock {\em Proceedings of the Royal Society of London. Series A:
  Mathematical, Physical and Engineering Sciences}, 459(2038):2547--2568, 2003.

\bibitem{landauer1996physical}
Rolf Landauer.
\newblock The physical nature of information.
\newblock {\em Physics letters A}, 217(4-5):188--193, 1996.

\bibitem{landauer1991information}
Rolf Landauer et~al.
\newblock Information is physical.
\newblock {\em Physics Today}, 44(5):23--29, 1991.

\bibitem{lebowitz1972uniqueness}
Joel~L Lebowitz and Anders Martin-L{\"o}f.
\newblock On the uniqueness of the equilibrium state for ising spin systems.
\newblock {\em Communications in Mathematical Physics}, 25(4):276--282, 1972.

\bibitem{leff2014maxwell}
Harvey~S Leff and Andrew~F Rex.
\newblock {\em Maxwell's demon: entropy, information, computing}.
\newblock Princeton University Press, 2014.

\bibitem{lu2014engineering}
Zhiyue Lu, Dibyendu Mandal, and Christopher Jarzynski.
\newblock Engineering maxwell’s demon.
\newblock {\em Physics Today}, 67(8):60--61, 2014.

\bibitem{lynn2019physics}
Christopher~W Lynn and Danielle~S Bassett.
\newblock The physics of brain network structure, function and control.
\newblock {\em Nature Reviews Physics}, 1(5):318--332, 2019.

\bibitem{mandal2012work}
Dibyendu Mandal and Christopher Jarzynski.
\newblock Work and information processing in a solvable model of maxwell’s
  demon.
\newblock {\em Proceedings of the National Academy of Sciences},
  109(29):11641--11645, 2012.

\bibitem{mandal2013maxwell}
Dibyendu Mandal, HT~Quan, and Christopher Jarzynski.
\newblock Maxwell’s refrigerator: an exactly solvable model.
\newblock {\em Physical review letters}, 111(3):030602, 2013.

\bibitem{mcgill1954multivariate}
William McGill.
\newblock Multivariate information transmission.
\newblock {\em Transactions of the IRE Professional Group on Information
  Theory}, 4(4):93--111, 1954.

\bibitem{merhav2015sequence}
Neri Merhav.
\newblock Sequence complexity and work extraction.
\newblock {\em Journal of Statistical Mechanics: Theory and Experiment},
  2015(6):P06037, 2015.

\bibitem{mlodinow2014relation}
Leonard Mlodinow and Todd~A Brun.
\newblock Relation between the psychological and thermodynamic arrows of time.
\newblock {\em Physical Review E}, 89(5):052102, 2014.

\bibitem{o2005partially}
Alice~J O'toole, Fang Jiang, Herv{\'e} Abdi, and James~V Haxby.
\newblock Partially distributed representations of objects and faces in ventral
  temporal cortex.
\newblock {\em Journal of cognitive neuroscience}, 17(4):580--590, 2005.

\bibitem{palm1988significance}
G~Palm, AMHJ Aertsen, and GL~Gerstein.
\newblock On the significance of correlations among neuronal spike trains.
\newblock {\em Biological cybernetics}, 59(1):1--11, 1988.

\bibitem{parrondo2015thermodynamics}
Juan~MR Parrondo, Jordan~M Horowitz, and Takahiro Sagawa.
\newblock Thermodynamics of information.
\newblock {\em Nature physics}, 11(2):131--139, 2015.

\bibitem{philbin2016thermal}
Thomas~G Philbin and Janet Anders.
\newblock Thermal energies of classical and quantum damped oscillators coupled
  to reservoirs.
\newblock {\em Journal of Physics A: Mathematical and Theoretical},
  49(21):215303, 2016.

\bibitem{pucci2013entropy}
Lorenzo Pucci, Massimiliano Esposito, and Luca Peliti.
\newblock Entropy production in quantum brownian motion.
\newblock {\em Journal of Statistical Mechanics: Theory and Experiment},
  2013(04):P04005, 2013.

\bibitem{qian2001relative}
Hong Qian.
\newblock Relative entropy: Free energy associated with equilibrium
  fluctuations and nonequilibrium deviations.
\newblock {\em Physical Review E}, 63(4):042103, 2001.

\bibitem{qian2005nonequilibrium}
Hong Qian and Timothy~C Reluga.
\newblock Nonequilibrium thermodynamics and nonlinear kinetics in a cellular
  signaling switch.
\newblock {\em Physical review letters}, 94(2):028101, 2005.

\bibitem{ruelle1972use}
David Ruelle.
\newblock On the use of “small external fields” in the problem of symmetry
  breakdown in statistical mechanics.
\newblock {\em Annals of Physics}, 69(2):364--374, 1972.

\bibitem{ruth2013integrating}
Matthias Ruth.
\newblock {\em Integrating economics, ecology and thermodynamics}, volume~3.
\newblock Springer Science \& Business Media, 2013.

\bibitem{sagawa2014thermodynamic}
Takahiro Sagawa.
\newblock Thermodynamic and logical reversibilities revisited.
\newblock {\em Journal of Statistical Mechanics: Theory and Experiment},
  2014(3):P03025, 2014.

\bibitem{sagawa2009minimal}
Takahiro Sagawa and Masahito Ueda.
\newblock Minimal energy cost for thermodynamic information processing:
  measurement and information erasure.
\newblock {\em Physical review letters}, 102(25):250602, 2009.

\bibitem{sagawa2010generalized}
Takahiro Sagawa and Masahito Ueda.
\newblock Generalized jarzynski equality under nonequilibrium feedback control.
\newblock {\em Physical review letters}, 104(9):090602, 2010.

\bibitem{sagawa2012fluctuation}
Takahiro Sagawa and Masahito Ueda.
\newblock Fluctuation theorem with information exchange: Role of correlations
  in stochastic thermodynamics.
\newblock {\em Physical review letters}, 109(18):180602, 2012.

\bibitem{sagawa2013role}
Takahiro Sagawa and Masahito Ueda.
\newblock Role of mutual information in entropy production under information
  exchanges.
\newblock {\em New Journal of Physics}, 15(12):125012, 2013.

\bibitem{sartori2014thermodynamic}
Pablo Sartori, L{\'e}o Granger, Chiu~Fan Lee, and Jordan~M Horowitz.
\newblock Thermodynamic costs of information processing in sensory adaptation.
\newblock {\em PLoS Comput Biol}, 10(12):e1003974, 2014.

\bibitem{sawyer2005social}
R~Keith Sawyer and Robert Keith~Sawyer Sawyer.
\newblock {\em Social emergence: Societies as complex systems}.
\newblock Cambridge University Press, 2005.

\bibitem{schneidman2003network}
Elad Schneidman, Susanne Still, Michael~J Berry, William Bialek, et~al.
\newblock Network information and connected correlations.
\newblock {\em Physical review letters}, 91(23):238701, 2003.

\bibitem{schnitzer2003multineuronal}
Mark~J Schnitzer and Markus Meister.
\newblock Multineuronal firing patterns in the signal from eye to brain.
\newblock {\em Neuron}, 37(3):499--511, 2003.

\bibitem{schwarz1995thermodynamics}
RB~Schwarz and AG~Khachaturyan.
\newblock Thermodynamics of open two-phase systems with coherent interfaces.
\newblock {\em Physical review letters}, 74(13):2523, 1995.

\bibitem{seifert2012stochastic}
Udo Seifert.
\newblock Stochastic thermodynamics, fluctuation theorems and molecular
  machines.
\newblock {\em Reports on progress in physics}, 75(12):126001, 2012.

\bibitem{seifert2016first}
Udo Seifert.
\newblock First and second law of thermodynamics at strong coupling.
\newblock {\em Physical review letters}, 116(2):020601, 2016.

\bibitem{sekimoto2010stochastic}
Ken Sekimoto.
\newblock {\em Stochastic energetics}, volume 799.
\newblock Springer, 2010.

\bibitem{sengupta2013information}
Biswa Sengupta, Martin~B Stemmler, and Karl~J Friston.
\newblock Information and efficiency in the nervous system—a synthesis.
\newblock {\em PLoS Comput Biol}, 9(7):e1003157, 2013.

\bibitem{still2012thermodynamics}
Susanne Still, David~A Sivak, Anthony~J Bell, and Gavin~E Crooks.
\newblock Thermodynamics of prediction.
\newblock {\em Physical review letters}, 109(12):120604, 2012.

\bibitem{strasberg2013thermodynamics}
Philipp Strasberg, Gernot Schaller, Tobias Brandes, and Massimiliano Esposito.
\newblock Thermodynamics of a physical model implementing a maxwell demon.
\newblock {\em Physical review letters}, 110(4):040601, 2013.

\bibitem{street2016neurobiology}
Sterling Street.
\newblock Neurobiology as information physics.
\newblock {\em Frontiers in systems neuroscience}, 10:90, 2016.

\bibitem{street2020upper}
Sterling Street.
\newblock Upper limit on the thermodynamic information content of an action
  potential.
\newblock {\em Frontiers in Computational Neuroscience}, 14:37, 2020.

\bibitem{svirezhev2000thermodynamics}
Yuri~M Svirezhev.
\newblock Thermodynamics and ecology.
\newblock {\em Ecological Modelling}, 132(1-2):11--22, 2000.

\bibitem{szilard1929entropieverminderung}
Leo Szilard.
\newblock {\"U}ber die entropieverminderung in einem thermodynamischen system
  bei eingriffen intelligenter wesen.
\newblock {\em Zeitschrift f{\"u}r Physik}, 53(11-12):840--856, 1929.

\bibitem{talkner2016open}
Peter Talkner and Peter H{\"a}nggi.
\newblock Open system trajectories specify fluctuating work but not heat.
\newblock {\em Physical Review E}, 94(2):022143, 2016.

\bibitem{talkner2020colloquium}
Peter Talkner and Peter H{\"a}nggi.
\newblock Colloquium: Statistical mechanics and thermodynamics at strong
  coupling: Quantum and classical.
\newblock {\em Reviews of Modern Physics}, 92(4):041002, 2020.

\bibitem{ting1962amount}
HU~Kuo Ting.
\newblock On the amount of information.
\newblock {\em Theory of Probability \& Its Applications}, 7(4):439--447, 1962.

\bibitem{toyabe2010experimental}
Shoichi Toyabe, Takahiro Sagawa, Masahito Ueda, Eiro Muneyuki, and Masaki Sano.
\newblock Experimental demonstration of information-to-energy conversion and
  validation of the generalized jarzynski equality.
\newblock {\em Nature physics}, 6(12):988--992, 2010.

\bibitem{um2015total}
Jaegon Um, Haye Hinrichsen, Chulan Kwon, and Hyunggyu Park.
\newblock Total cost of operating an information engine.
\newblock {\em New Journal of Physics}, 17(8):085001, 2015.

\bibitem{watanabe1960information}
Satosi Watanabe.
\newblock Information theoretical analysis of multivariate correlation.
\newblock {\em IBM Journal of research and development}, 4(1):66--82, 1960.

\bibitem{yarrow2012fisher}
Stuart Yarrow, Edward Challis, and Peggy Seri{\`e}s.
\newblock Fisher and shannon information in finite neural populations.
\newblock {\em Neural computation}, 24(7):1740--1780, 2012.

\bibitem{yeung1991new}
Raymond~W Yeung.
\newblock A new outlook on shannon's information measures.
\newblock {\em IEEE transactions on information theory}, 37(3):466--474, 1991.

\bibitem{zhang2016critical}
Yirui Zhang and Andre~C Barato.
\newblock Critical behavior of entropy production and learning rate: Ising
  model with an oscillating field.
\newblock {\em Journal of Statistical Mechanics: Theory and Experiment},
  2016(11):113207, 2016.

\bibitem{zheng2020geometric}
Muhua Zheng, Antoine Allard, Patric Hagmann, Yasser Alem{\'a}n-G{\'o}mez, and
  M~{\'A}ngeles Serrano.
\newblock Geometric renormalization unravels self-similarity of the multiscale
  human connectome.
\newblock {\em Proceedings of the National Academy of Sciences},
  117(33):20244--20253, 2020.

\end{thebibliography}






\end{document}